\documentclass[3p,sort&compress,review]{elsarticle}

\usepackage{xcolor}
\usepackage{amsmath}
\usepackage{amssymb}
\usepackage{amsfonts}
\usepackage{amsthm}
\usepackage{fixmath}
\usepackage{braket}
\usepackage{slashed}
\usepackage{fontenc}
\usepackage{bm}
\usepackage[mathscr]{euscript}

\newcommand{\be}{\begin{equation}}
\newcommand{\ee}{\end{equation}}
\newcommand{\bea}{\begin{eqnarray}}
\newcommand{\eea}{\end{eqnarray}}
\newcommand{\nn}{\nonumber}
\newcommand{\nnnl}{\nonumber\\}

\newcommand{\al}[1]{\vskip-3ex\begin{align}#1\end{align}}
\newcommand{\Eqref}[1]{\mbox{Eq.}~(\ref{#1})}
\newcommand{\Eqsref}[1]{\mbox{Eqs.}~(\ref{#1})}
\newcommand{\Figref}[1]{\mbox{Fig.}~\ref{#1}}
\newcommand{\Tabref}[1]{\mbox{Tab.}~\ref{#1}}
\newcommand{\Secref}[1]{\mbox{Sec.}~\ref{#1}}

\renewcommand{\bm}{\mathbold}
\newcommand{\mr}{\mathrm}

\newcommand{\mc}{\mathcal}

\def\ri{\mr{i}} 
\def\tr{\mr{tr}~}

\def\ga{\gamma}

\def\Nc{N_c}
\def\Nf{N_f}
\def\Nt{N_\tau}

\def\TF{T_F}
\def\CF{C_F}
\def\CA{C_A}

\def\MS{\overline{\rm MS}}
\def\lQ{\Lambda_{\rm QCD}}
\def\lMSbt{{\Lambda_{\MS}^{\Nf=3}}}

\def\als{\alpha_s} 
\def\as{\alpha_s} 
\newcommand{\qbq}{{q\bar q}}
 
\def\av{\alpha_{V}} 
\def\cdf{\chi^2/{\rm d.o.f.}}

\def\Vs{V_s}
\def\Vo{V_o}
\def\FS{F_S}

\def\cf{\mbox{cf.~}}
\def\eg{\mbox{e.g.~}}
\def\ie{\mbox{i.e.~}}

\newcommand{\chpt}{$\chi$PT}
\newcommand{\order}{\ensuremath{\text{O}}}

\newcommand{\GeV}{\text{GeV}}
\newcommand{\MeV}{\text{MeV}} 

\newcommand{\mbar}{\overline{m}}
\newcommand{\MRS}{{\text{MRS}}}

\newcommand{\msb}{{\ensuremath{\overline{\text{MS}}}}}
\newcommand{\fpiPDG}{\ensuremath{f_{\pi,\text{PDG}}}}

\newcommand{\Lambdabar}{{\ensuremath{\bar{\Lambda}}}}

\definecolor{orange}{rgb}{1.0,0.5,0.0}
\definecolor{darkblue}{rgb}{0.0,0.1,0.7}
\definecolor{brown}{rgb}{0.6,0.1,0.0}
\definecolor{grey}{rgb}{0.6,0.6,0.6}

\newcommand{\strike}[1]{}			

\usepackage{lineno}

\begin{document}

\begin{frontmatter}

\title{ \vspace{1cm} Strong coupling constant and quark masses from lattice QCD}

\author[3]{Javad~Komijani}
\ead{jkomijani@ut.ac.ir}
\author[1]{Peter~Petreczky\corref{cor1}}
\ead{petreczk@quark.phy.bnl.gov}
\author[2]{Johannes~Heinrich~Weber}
\ead{weberjo8@msu.edu}

\address[3]{%
Department of Physics, University of Tehran, Tehran 1439955961, Iran
}
\address[1]{%
Physics Department, Brookhaven National Laboratory, Upton, NY 11973, USA
}
\address[2]{%
Department of Computational Mathematics, Science and Engineering, and 
Department of Physics and Astronomy,
Michigan State University, East Lansing, MI 48824, USA
}

\cortext[cor1]{Corresponding author}

\begin{abstract}
We review lattice determinations of the charm and bottom quark masses and 
the strong coupling constant obtained by different methods. 
We explain how effective field theory approaches, such as Non-Relativistic 
QCD (NRQCD), potential Non-Relativistic QCD (pNRQCD), Heavy Quark Effective 
Theory (HQET) and Heavy Meson rooted All-Staggered Chiral Perturbation 
Theory (HMrAS$\chi$PT) can help in these determinations. 
After critically reviewing different lattice results we determine
lattice world averages for the strong coupling constant, 
$\alpha_s(M_Z,\Nf{=}5)=0.11803^{+0.00047}_{-0.00068}$,
as well as for the charm quark mass, 
$m_c(m_c,\Nf{=}4)=1.2735(35)$ GeV, and the bottom quark mass,
$m_b(m_b,\Nf{=}5)=4.188(10)$ GeV. 
The above determinations are more precise than the ones
obtained by Particle Data Group (PDG).
\end{abstract}

\end{frontmatter}

\tableofcontents

\section{Introduction}
The strong coupling constant and quark masses are important parameters of the 
Standard Model (SM) and thus their knowledge is required for its testing. 
Heavy quark masses are needed to test the Higgs mechanism of mass 
generation~\cite{Lepage:2014fla} and accurate determination of
the SM parameters \cite{Erler:2017knj, Erler:2000nx, Erler:2006vu}.
The accurate determination of the strong coupling constant is needed for 
predictions of
the Higgs boson branching ratios \cite{Lepage:2014fla, Dawson:2013bba} 
and stability of the SM vacuum \cite{Buttazzo:2013uya, Espinosa:2013lma}.

Traditionally the quark masses and the strong coupling constant $\als$ 
have been obtained from a comparison of perturbative QCD calculations to the 
experimental data involving a hard scale. 
For example, $\als$ can be extracted from jets~\cite{Abazov:2009nc, 
Malaescu:2012ts, Chatrchyan:2013txa, CMS:2014mna, Khachatryan:2014waa},
$e^{+}e^{-}$~annihilation \cite{Dissertori:2009ik, Dissertori:2009qa, 
OPAL:2011aa, Bethke:2008hf, Davison:2008vx, Abbate:2010xh, Gehrmann:2012sc, 
Hoang:2015hka}, 
deep inelastic scattering~\cite{Blumlein:2006be, Alekhin:2012ig, 
JimenezDelgado:2008hf, Martin:2009bu, Harland-Lang:2015nxa, Ball:2011us},
or $\tau$-decay \cite{Baikov:2008jh, Pich:2013lsa, Davier:2013sfa, 
Boito:2014sta}. 
The charm and bottom quark masses can be extracted from the 
$e^{+}e^{-}$ ~annihilation data above the open charm threshold 
\cite{Kuhn:2001dm, Dehnadi:2015fra} or from deep inelastic 
scattering~\cite{Alekhin:2012vu}.

Lattice QCD calculations play an increasingly important role in the 
determination of the quark masses and $\als$ because of the increase 
in available computation resources and algorithmic improvements. 
The basis of lattice QCD is the formulation of the QCD path integral on 
a discretized Euclidean space-time lattice with lattice spacing $a$. 
This formulation allows for ultraviolet regularization of QCD at the level 
of the path integral and not only at the level of Feynman diagrams. 
The path integral can be calculated non-perturbatively using 
Monte-Carlo simulations. 
Finally, to recover the continuum physics<, the limit $a \to 0$ 
should be taken. There are different ways to discretize the gauge and 
fermions fields on a lattice, i.e. different lattice QCD actions. 
For the gauge degrees of freedom one can use the standard Wilson 
plaquette action or the improved L\"uscher-Weisz action. 
For quark fields different improved actions are used, such as 
the clover-improved Wilson quark action, 
AsqTad staggered quark action~\cite{Orginos:1999cr}, 
Highly Improved Staggered Quark (HISQ) action~\cite{Follana:2006rc},
twisted mass quark action~\cite{Frezzotti:2000nk}, 
or any domain wall fermion (DWF) action \cite{Kaplan:1992bt}.
For a recent review on the basics and current status of lattice QCD we 
refer the reader to \mbox{Ref.}~\cite{Bazavov:2009bb}.

Phenomenological determinations of the strong coupling constant typically 
give $\als$ defined at large energy scales.
The only low energy determination of $\als$ comes from the $\tau$ decay. 
For some applications it is important to know the running of the strong 
coupling constant at low energy scales. 
An example is the testing the applicability of the weak-coupling approach 
to QCD thermodynamics through comparison to lattice QCD results 
\cite{Bazavov:2018wmo, Bazavov:2017dsy, Bazavov:2016uvm, Berwein:2015ayt, 
Ding:2015fca, Haque:2014rua, Bazavov:2013uja}. 
Here one needs $\als$ at relatively low scale $\mu \simeq \pi T$, 
with $T$ being the temperature.
As we will see, lattice QCD calculations are well suited to determine the 
running of $\als$ at low energy scales down to the charm quark mass.

In lattice QCD the quark masses are fixed by setting the masses of a set of 
hadrons composed of those quarks to their physical values. 
This procedure gives the so-called bare quark masses in a specific lattice 
scheme. 
To obtain the quark masses in the commonly used $\MS$ scheme
one has to calculate the renormalization constant that converts the 
lattice scheme to the $\MS$ scheme.
There are several methods to do this which will be discussed in this review.

The lattice determination of $\als$ is based on the idea that one calculates 
the quantity $O(\nu)$, that depends on some physical energy scale $\nu$
(not to be confused with the renormalization scale) non-perturbatively on a 
lattice.
Then one compares the result with the perturbative calculation of $O(\nu)$ 
as an expansion in powers of $\als(\nu)$.
From this one can obtain the value of $\als(\nu)$ provided that the 
truncated power series in $\als(\nu)$ yields a sufficiently accurate 
description of $O(\nu)$. 
One of the big challenges in the lattice determination of the quark masses
and $\als$ is the so-called window problem. 
The scale $\nu$ has to be much smaller than the lattice cutoff ($a^{-1}$) 
to avoid large discretization effects (lattice artifacts), and at the same 
time sufficiently large to make the perturbative expansion accurate, \ie
\begin{equation}
\Lambda_{QCD} \ll \nu \ll \frac{1}{a}.
\end{equation} 

This means one has to perform the calculations at small values of the 
lattice spacing, which is computationally very demanding and is limited by 
the critical slowing down of Monte-Carlo algorithms. 
Controlling discretization effects is the biggest challenge for an accurate 
determination of $\als$ and the quark masses.
We will discuss this issue in detail for calculations of $\als$ from 
the energy of a static quark anti-quark ($\qbq$) pair and the moments of 
quarkonium correlators as examples. 
As we will see, the interplay between lattice QCD calculations and
effective field theory methods will play an important role in addressing 
this issue.

The rest of this review is organized as follows. 
In the next section we summarize some key aspects of lattice calculations 
including a discussion on determination of the light quark masses.
In \Secref{sec:static} we will discuss the determination of $\als$
from the energy of static $\qbq$ states. 
In \Secref{sec:moments} we discuss the determination of $\als$ and 
heavy quark masses from the moments of quarkonium correlators. 
In \Secref{sec:heavylight} we present the determination of the quark 
masses using a combined effective field theory and lattice QCD calculations. 
The discussion of other lattice methods for the determination of $\als$
as well as the comparison of different results will be given 
in \Secref{sec:otherals}.
Overview of different determinations of heavy quark masses will be given in 
\Secref{sec:otherhqm}. 
Readers not interested in technical details of the lattice calculations 
and perturbation theory can skip to \Secref{sec:exec} for the main results, 
where we present an executive summary of the present status. 
Finally, \Secref{sec:concl} contains our conclusions. 
The lattice determination of the strong coupling constant and the quark 
masses is also reviewed by the 
Flavor Lattice Averaging Group (FLAG)~\cite{Aoki:2019cca}. 
The FLAG findings will be incorporated into our discussions where 
appropriate.

\section{General remarks on lattice calculations and 
determinations of the light and strange quark masses}
\label{sec:lattice}

Out of the six quarks present in the SM the up, down, and strange quarks have 
masses that are smaller or comparable to the nonperturbative QCD scale $\lQ$.
As a result, their effects cannot be taken into account by perturbation 
theory unlike the heavy quarks.
Therefore, the effects of the $u$, $d$, and $s$ quarks have to be included 
directly in any lattice QCD calculations. 
Since isospin symmetry is believed to be a good symmetry of QCD, most 
lattice calculations set $m_u=m_d$ and denote their masses by $m_l$. 
Therefore, lattice QCD calculations involving $u$, $d$, and $s$ quarks 
are usually referred to as (2+1)-flavor or $N_f=2+1$ calculations. 
The charm-quark mass is much larger than $\lQ$, and therefore its effect 
can be reasonably well approximated by perturbation theory. This was also
confirmed in a recent lattice study~\cite{Athenodorou:2018wpk}. 
Nevertheless many lattice QCD calculations nowadays include the effect 
of dynamical charm quarks. 
These calculations are referred to as (2+1+1)-flavor or 
$N_f=2+1+1$ calculations. 

In order to perform the continuum ($a \to 0$) limit of a quantity of 
interest the lattice spacing has to be varied, which is performed by 
varying the bare gauge coupling $g_0$ in the lattice QCD action. 
It is customary to parametrize $g_0$ in terms of $\beta=6/g_0^2$ or 
$\beta=10/g_0^2$ if an improved gauge action is used, see 
\eg Ref.~\cite{Aubin:2004wf}. 
To fix the lattice spacing we need to calculate a physical quantity, 
for example the pion decay constant $f_{\pi}$, in lattice units%
\footnote{A quantity in lattice units refers to a dimensionless number 
obtained by multiplying the quantity by an appropriate power of the 
lattice spacing $a$.
For instance, the pion decay constant $f_{\pi}$ in lattice units is 
$af_\pi$.}.
Using the experimental value of the pion decay constant from the 
Particle Data Group~\cite{PDG18} one can then determine the lattice 
spacing in physical units.
It is also possible to determine the lattice spacing in terms of 
quantities that are not measured experimentally. 
Examples for such a quantity are the Sommer scale $r_0$~\cite{Sommer:1993ce} 
and the $r_1$~\cite{Aubin:2004wf} and $r_2$~\cite{Bazavov:2017dsy}~scales.
These scales are defined in terms of the energy of a static 
quark-antiquark pair, $E(r)$, as
\al{
 &\left.\frac{r^2}{a^2}\frac{d [aE(r/a)]}{d [r/a]}\right|_{r=r_i}
 =\left.r^2\frac{d E(r)}{dr}\right|_{r=r_i} = c_i,&
 & \left\{\begin{array}{rlll} 
 r_i ~\to & r_0 & r_1 & r_2 \\
 c_i ~\to & 1.65 & 1.0 & 0.5 \\
 \end{array}\right..
}
To use for example $r_1$ for scale setting one has to determine it in physical 
units. 
To do this one first calculates $r_1/a$ and also $f_{\pi} a$ (or any other 
experimentally measured quantity) 
at several lattice spacings and performs the continuum extrapolation of the 
dimensionless quantity $r_1 f_{\pi}$. 
Using the known value of $f_{\pi}$ one obtains $r_1$ in physical units. 
This value can be used to set the lattice spacing in a different lattice 
calculation. 
The same applies to the $r_0$~or the $r_2$ scales.
Other similar quantities that can be used for scale setting are the 
gradient flow parameters $t_0$~\cite{Luscher:2010iy} and 
$w_0$~\cite{Borsanyi:2012zs}. 
The values of $r_0$, $r_1$, $r_2$, $w_0$ and $t_0$ are now well 
established. 
In \mbox{Secs.}~\ref{sec:static} and \ref{sec:moments} we use the value 
$r_1=0.3106(8)(14)(4)\,\mr{fm}$~\cite{Bazavov:2010hj} in the case of 
(2+1)-flavor QCD with HISQ action.
The contributions to the uncertainty of $r_1$ are due to the continuum 
extrapolation, the chiral extrapolation, and the experimental uncertainty 
of $f_\pi$. 
The calculations of the static energy using DWF formulation, discussed in
\mbox{Sec.}~\ref{sec:static} use
$r_1=0.311(2)\,\mr{fm}$~\cite{Sommer:2014mea}, which is based on the same 
lattice calculation of $r_1$ in terms of $f_\pi$. 

As we vary the lattice spacing or equivalently $\beta$ we also need to 
vary the quark masses such that the physical hadron masses remain constant 
in physical units. 
This procedure defines the line of constant physics $m_l=m_l(\beta)$, 
$m_s=m_s(\beta)$ and $m_c=m_c(\beta)$. 
The computational costs of lattice QCD calculations increase with decreasing 
quark masses and could be very large for physical values of $m_l$.  
For this reason many lattice QCD calculations fix the charm- and 
strange-quark masses to their physical values, while $m_l$ is taken to be 
larger than the physical value.
Performing calculations at several values of $m_l$ that are larger than 
the physical one and performing chiral extrapolations the physical limit 
can be recovered.
Currently many lattice calculations are performed directly at or very 
close to the physical value of $m_l$ and no extrapolation is needed.
Once the values of $m_l(\beta)$, $m_s(\beta)$ and $m_c(\beta)$ are 
determined in lattice calculations one has to calculate the 
renormalization factor that converts the quark masses in the lattice 
scheme to the $\MS$ scheme at a certain scale $\mu$, which is 
conventionally set to $2$ GeV. 
This can be done directly in perturbation theory on the lattice or
indirectly by first converting 
the lattice quark masses to some intermediate scheme, like RI-MOM, and then 
matching the RI-MOM masses to the $\MS$ masses. 

To determine the light quark masses in the isospin symmetric limit 
one has to consider the isospin symmetric hadron masses. 
Typically pseudoscalar meson masses are used to fix the quark masses and 
chiral perturbation theory is used to estimate the isospin symmetric masses. 
The status of the lattice determination of $m_s$ and $m_l$ is reviewed in 
detail by FLAG~\cite{Aoki:2019cca} 
and therefore we will only summarize their main findings here. 
The quark mass determinations in FLAG are grouped according to some 
quality criteria and only the lattice determinations that pass certain 
quality criteria are used to calculate the averaged values of the 
quark masses.
The results from averaging $N_f=2+1$ and $N_f=2+1+1$ lattice calculations 
are reported separately. 
FLAG finds for the $\MS$ masses at $\mu=2$ GeV \cite{Aoki:2019cca}:
\begin{eqnarray}
&
m_l=3.364(41)~ {\rm MeV}, ~ m_s=92.03(88) ~ {\rm MeV}, ~~N_f=2+1,\\[0.3mm]
&
m_l=3.410(43)~ {\rm MeV}, ~ m_s=93.44(68) ~ {\rm MeV}, ~~N_f=2+1+1.
\end{eqnarray}
The quark mass ratios are scheme and scale independent, 
and thus can be evaluated directly from ratios of the bare masses.
For the ratio $m_s/m_l$ FLAG obtains \cite{Aoki:2019cca}:
\begin{eqnarray}
&
m_s/m_l=27.42(12),~~N_f=2+1,\\[0.3mm]
&
m_s/m_l=27.23(10),~~N_f=2+1+1.
\end{eqnarray}
We see that the $N_f=2+1$ and $N_f=2+1+1$ determinations agree well with 
each other implying that the effects of dynamical charm quarks are small.

Calculating for instance charmonium ground state masses on a lattice 
one can obtain $m_c(\beta)$. 
Combining this with $m_s=m_s(\beta)$ and performing continuum extrapolation 
one obtains $m_c/m_s$; see for example Ref. \cite{Maezawa:2016vgv} for $N_f=2+1$, which finds 
\begin{equation}
m_c/m_s=11.877(91).
\end{equation}
This agrees well with the FLAG average of $m_c/m_s=11.82(16), N_f=2+1$, 
though it was excluded from the FLAG averaging procedure. 
For $N_f=2+1+1$ FLAG finds: $m_c/m_s=11.768(33)$.

To obtain $m_u$ and $m_d$ masses isospin breaking effects including QED 
effects have to be considered.
The latter implies that one has to perform lattice calculations in QED + QCD. 
Many of the present day calculations consider QED effects only in the valence 
sector, i.e. use electro-quenched approximation. 
This approach has the advantage that no new lattice gauge configurations have 
to be generated. 
The lattice determination of $m_u$ and $m_d$ is also reviewed by FLAG in great 
detail and here we only give the summary of their findings:
\begin{eqnarray}
&
m_u=2.27(9)~ {\rm MeV}, ~ m_d=4.67(9) ~ {\rm MeV},~ m_u/m_d=0.485(19),~~N_f=2+1,\\[0.3mm]
&
m_u=2.50(17)~ {\rm MeV}, ~ m_d=4.88(20) ~ {\rm MeV},~ m_u/m_d=0.513(31),~~N_f=2+1+1.
\end{eqnarray}
The above lattice results for the up-, down-, and strange-quark masses are 
much more precise than the values quoted by PDG \cite{PDG18}. 
The determination of the charm and bottom quark mass may potentially suffer 
from large discretization effects and will be discussed separately in the 
following sections.

\section{Static $\qbq$ energy}
\label{sec:static}

The QCD static energy $ E $ of a quark-antiquark pair is an observable 
in QCD up to an additive constant. 
As an analog of positronium in QED, the QCD static energy is the most 
simple bound state problem of QCD. 
Due to the infinite mass of the static quark and antiquark the two color charges 
are strictly immobile, and their distance is a well-defined quantum number.  
The static energy $ E(r) $ is defined in terms of the large time limit 
of the expectation value of the Wilson loop $W_S(r,t)$ as 
\al{\label{eq:Emink}
 E(r) 
 &=
 \lim\limits_{t \to \infty} 
 \ri\frac{1}{t} \Braket{ \ln W_S(\bm r,t) }
 =
 \lim\limits_{t \to \infty} 
 \ri\frac{\partial}{\partial t} \Braket{ \ln W_S(\bm r,t) }. 
}
The Wilson loop is a smooth path-ordered contour in the continuum 
\al{\label{eq:WS}
 W_S(\bm r,t)
 &=
 e^{ \ri g \oint_{\bm r,t} dz^\mu A_\mu }.
}
The Wilson loop has a self-energy divergence proportional to its 
circumference.
The static energy cannot depend on the gauge fields at infinite time 
separation due to the cluster decomposition. 
Hence, the static energy may be defined as well by evaluating the 
correlation function of two temporal Wilson lines in a suitable 
gauge, \eg Coulomb gauge, 
\al{\label{eq:CS}
 C_S(\bm r,t)
 &=
 e^{\ri g \int_{0}^{t} d\tilde{t} A_0(\bm 0,\tilde{t}) }
 e^{-\ri g \int_{0}^{t} d\tilde{t} A_0(\bm r,\tilde{t}) }.
}
The definition in terms of the Wilson line correlation function avoids 
the self-energy divergence associated with the spatial distance between 
the two Wilson lines. 
The static energy $ E $ is a function of the fixed distance $ r $ between 
the static quark and antiquark, the QCD coupling $\als=g^2/4\pi$, and 
the masses of the $\Nf$ dynamical quarks in the sea. 
In particular, the static energy depends on $\als$ already at the tree-level, 
where it is up to a trivial color factor $\CF$ and the different coupling 
constant formally the same as the QED Coulomb potential.

\subsection{Static energy in the weak-coupling approach}

The QCD static energy has been calculated in the weak-coupling approach 
since the earliest days of QCD~\cite{Appelquist:1977tw, Fischler:1977yf, 
Billoire:1979ih, Schroder:1998vy, Brambilla:1999qa} and is currently known 
at the next-to-next-to-next-to-leading log ($\mr{N^3LL}$) 
level~\cite{Brambilla:2009bi, Anzai:2009tm, Smirnov:2009fh, Lee:2016cgz, 
Lee:2016lvq}. 
At each loop order there is a one-to-one correspondence between the 
coupling $\als$ in the weak-coupling expression for the static energy 
and the QCD Lambda parameter. 
Since the masses of the up, down, and strange sea quarks are much smaller
than the QCD Lambda parameter, their non-vanishing masses play only a 
negligible role for the running of the coupling at scales much higher 
than the Lambda parameter. 
For this reason, these sea quarks are generally taken to be in the massless 
limit in perturbation theory.
Although there is an explicit dependence on the masses of the dynamical 
sea quark flavors as well, such contributions to the static energy are 
suppressed by positive powers of the distance between the quark and 
antiquark and of the coupling $\als$ itself~\cite{Eiras:2000rh}. 
These small contributions to the static energy can be neglected safely. 

\subsubsection{Static energy at $\mr{N^3LO}$}

The QCD static energy $E$ is a {multi-scale problem}, \ie $E$ depends on 
the scales $1/r$ and $\als/r$, where $r$ is the distance between the 
static quark and antiquark. 
It is a convention to call the former the soft scale, $\nu \sim 1/r$, 
and the latter the ultrasoft scale, $\mu_\mr{us} \sim \als/r$.
Such a multi-scale problem is ideally approached in an effective field 
theory approach, \ie in potential nonrelativistic QCD 
(pNRQCD)~\cite{Brambilla:1999xf, Brambilla:1999qa}, which permits 
integrating out the higher scale, $1/r$, while solving the more simple 
problem at the lower scale, $\als/r$. 
In pNRQCD the system is described in terms of the interactions among a 
heavy quark-antiquark pair at the relative distance $r$ in the 
color-singlet or the color-octet configurations. 
These are the nonrelativistic degrees of freedom, whereas the light quarks 
and gluons, which are at the even lower scale $\lQ$, are included in terms 
of nonperturbative matrix elements. 
The Wilson coefficients of pNRQCD have an explicit dependence on the 
distance $r$ and represent various contributions to the quark-antiquark 
potential in QCD. 
The static energy has been calculated in the weak-coupling approach in 
the limit of $\Nf$ massless sea quarks. 
The static energy can be decomposed at the next-to-next-to-next-to-leading 
order into 
\al{\label{eq:E3L}
 E(r)&=
 \Vs(r,\mu_\mr{cut}) + \Lambda_s+ \delta E_{us}(r,\mu_\mr{cut}).
}
Here, $\mu_\mr{cut}$ is a cutoff scale between the soft and ultrasoft scales, 
$\nu \sim 1/r \gg \mu_\mr{cut} \gtrsim \mu_\mr{us}\sim \als/r$. 
The first term, $\Vs$, is the color singlet potential of pNRQCD. 
Like its color octet counterpart $\Vo$ it is a Coulomb potential 
up to the trivial color factors at the tree-level: 
$\Vs^\mr{tree}(r)=-\CF~\tfrac{\als}{r}$, 
$\Vo^\mr{tree}(r)=+\frac{1}{2\Nc}~\frac{\als}{r}$.
The singlet potential can be obtained through the Fourier transform 
\al{\label{eq:VsaV}
 \Vs(r,\mu_\mr{cut}) = 
 -4\pi\CF \int_{q^2\ge \mu_{us}^2} 
 \frac{d^3q}{(2\pi)^3}
 ~e^{\ri\bm q\cdot \bm r}
 ~\frac{\av(q^2)}{q^2},
}
where the coupling in the V scheme (also called the $\qbq$ scheme) 
is given by 
\al{\label{eq:aV}
 \alpha_{V}(q^2) = 
 \als(\nu^2) \sum\limits_{n=0}^{L} 
 \left\{ p_n[\ln(\nu^2/q^2)] + d_n[\ln(\nu^2/q^2)] \right\} 
 \left( \frac{\als(\nu^2)}{4\pi} \right)^n.
}
The $n$-th order polynomials $p_n$ and $d_n$ are known at the 
next-to-next-to-next-to-leading order~\cite{Fischler:1977yf, 
Billoire:1979ih, Schroder:1998vy,Anzai:2009tm, Smirnov:2009fh, Lee:2016cgz, 
Lee:2016lvq, Brambilla:1999qa}. 
In particular, $p_n[0]=a_n$, where the constants $a_n$ are reproduced to 
this order in \ref{app:force}. 
The latter ($d_n$) contain the IR divergent contributions 
that are regularized by the cutoff at or above the {ultrasoft scale} 
$\mu_\mr{cut} \gtrsim \mu_\mr{us} \sim \als/r$. 
The first nontrivial contribution $d_n$ occurs for $n=3$, 
\ie it contributes at the next-to-next-to-next-to-leading order. 
The singlet potential $\Vs(r,\mu_\mr{cut})$ is affected by an $r$-independent 
renormalon%
\footnote{For a discussion on renormalons see Ref.~\cite{Beneke:1998ui}.
For a brief discussion in the context of the pole mass see 
\Secref{sec:EFT:func:MRS}.}.
The second term $\Lambda_s$ in \Eqref{eq:E3L} is the residual mass, which 
is also affected by a renormalon; see~\mbox{Ref.}~\cite{Appelquist:1977tw}.  This renormalon exactly cancels the constant renormalon of the 
singlet potential~\cite{Aglietti:1995tg}. 
The last term $\delta E_\mr{us}(r,\mu_\mr{cut})$ in \Eqref{eq:E3L} is called the 
{ultrasoft contribution}. 
It is the time integral of the chromoelectric correlation function 
regularized at the {ultrasoft scale} $\mu_\mr{cut}$,
\al{
 \delta E_\mr{us}(r,\mu_\mr{cut}) =
 -\ri\frac{g^2}{\Nc} V_A^2 \int\limits_0^\infty dt
 ~e^{-\ri t[\Vo-\Vs](r)}
 ~\braket{\tr{\bm r\cdot \bm E(t) ~\bm r \cdot \bm E(0)}}.
}
$\delta E_\mr{us}(r,\mu_\mr{cut})$ can be understood in pNRQCD as being 
due to the intermediate fluctuation of the static quark and antiquark 
into the color-octet configurations. 
The chromoelectric fields can be separated as the sum of the 
perturbative contribution $\bm E_\mr{us}(t)$ at the ultrasoft scale, 
$\mu_\mr{us} \sim \als/r$, and the nonperturbative contribution 
$\bm E_\mr{np}(t)$ at the scale $\lQ \ll \mu_\mr{us}$, where the 
latter undergoes a very slow variation with time. 
Hence, the integral splits into two contributions\footnote{
A mixed term is forbidden due to the rotational invariance of the problem.}, 
where
\al{
 -\ri\frac{g^2}{\Nc} V_A^2 \int\limits_0^\infty dt
 ~e^{-\ri t[\Vo-\Vs](r)}
 \braket{\tr{\bm r\cdot \bm E_\mr{np}(0) ~\bm r \cdot \bm E_\mr{np}(0)}}
 \sim
  -\frac{\frac{g^2}{\Nc}}{\frac{\Nc}{2}\frac{\als}{r}} V_A^2 ~
 r^2\braket{\tr{\bm E_\mr{np}^2(0)}}
 \sim
  -\frac{\pi}{\Nc} ~
 r^3 \braket{G^2}
}
is the nonperturbative contribution from the dimension four 
gluon condensate with $r^3$ power-law form, while 
\al{
 -\ri\frac{g^2}{\Nc} V_A^2 \int\limits_0^\infty dt
 ~e^{-\ri t[\Vo-\Vs](r)}
 \braket{\tr{\bm r\cdot \bm E_\mr{us}(t) ~\bm r \cdot \bm E_\mr{us}(0)}}
}
can be evaluated in perturbation theory with the UV cutoff $\mu_\mr{cut}$.  
Here, $\mu_\mr{us}\sim \als/r\sim(\Vo-\Vs) \sim 1/t$ acts as an IR cutoff, 
namely, contributions from larger times $ t \gg 1/\mu_\mr{us}$ are 
suppressed due to the rapid oscillations. 
At the level of the leading logarithms, the UV divergent contributions 
from this part of $\delta E_\mr{us}(r,\mu_\mr{cut})$ at the UV cutoff 
$\mu_\mr{cut}$ cancel against the IR divergent contributions to 
$\Vs(r,\mu_\mr{cut})$ at the same IR cutoff $\mu_\mr{cut}$. 
\vskip1ex
The renormalon of $\Lambda_s$ may be evaded by instead considering the 
force $F=dE/dr$~\cite{Necco:2001xg}, which is given at the 
next-to-next-to-next-to-leading order ($\mr{N^3LO}$) as 
\al{\label{eq:F3L}
 F(r,\nu) 
 &= \frac{\CF}{r^2}\als(\nu)\Bigg[1
 +\frac{\als(\nu)}{4\pi}\Big(\tilde{a}_1-2\beta_0\Big)
 +\frac{\als^2(\nu)}{(4\pi)^2}\Big(\tilde{a}_2-4\tilde{a}_1\beta_0-2\beta_1\Big)
 \nn\\&
 \phantom{=\frac{\CF}{r^2}\als(\nu)\Bigg[}
 +\frac{\als^3(\nu)}{(4\pi)^3}\Big(
 \tilde{a}_3-6\tilde{a}_2\beta_0-4\tilde{a}_1\beta_1-2\beta_2
 +a_3^L\ln\frac{C_A\als(\nu)}{2}\Big)+\mathcal{O}(\als^4,\als^4\ln^2\als)\Bigg].
}
The first three terms correspond to the tree-level, the one-loop, and 
the two-loop order, respectively. 
Both parts of the fourth term correspond to the three-loop order. 
Here and in the following, the coupling $\als(\nu)$ is understood as 
renormalized in the $\MS$ scheme at the {soft scale} $\nu$, and can 
be traded for the corresponding QCD Lambda parameter, \ie at three flavors 
in the $\MS$ scheme $\lMSbt$. 
Large logarithms $\ln(r\nu)$ and their cancellations between different terms 
can be avoided, if the {soft scale} $\nu$ is chosen proportional to the 
inverse distance, \eg a standard choice is $\nu=1/r$, which we use hereafter. 
Integrating \Eqref{eq:F3L} one obtains a renormalon-free expression for the 
static energy up to an arbitrary but finite integration constant. 
The coefficients in \mbox{Eq.}~\eqref{eq:F3L} are defined 
in~\cite{Tormo:2013tha} and reproduced in~\ref{app:force}. 
At the three-loop order there is a contribution proportional to $\ln \als(1/r)$
as well as a non-logarithmic term. 
The origin of the former can be understood in pNRQCD~\cite{Brambilla:1999qa} 
as the interplay between the singlet potential $\Vs$ and the last term in 
\Eqref{eq:E3L}. 
The intermediate scale $\mu_\mr{us} = \CA \als(1/r)/(2r) = \Vo-\Vs$ is the 
physical IR cutoff for the contribution from the singlet potential, $\Vs$, and 
the UV cutoff for the {ultrasoft contribution}. 
For this reason, and to distinguish it from the {soft scale} $\nu$, 
$\mu_\mr{us}$ is called the {ultrasoft scale}. 
Note that there are alternative conventions to use an $r$-independent 
value for the {ultrasoft scale}, \eg 
$0.8\,\mr{GeV}$~\cite{Bazavov:2012ka, Bazavov:2014soa}, or 
$3\,\lMSbt$ or $4\,\lMSbt$~\cite{Takaura:2018vcy, Takaura:2018lpw} (using the PDG 
value of $\lMSbt$). 

\subsubsection{Resummation of the ultrasoft logarithms}

\begin{figure}[tb]
\begin{center}
\begin{minipage}[t]{16.5 cm}
\begin{minipage}[t]{8 cm}
\includegraphics[scale=0.48]{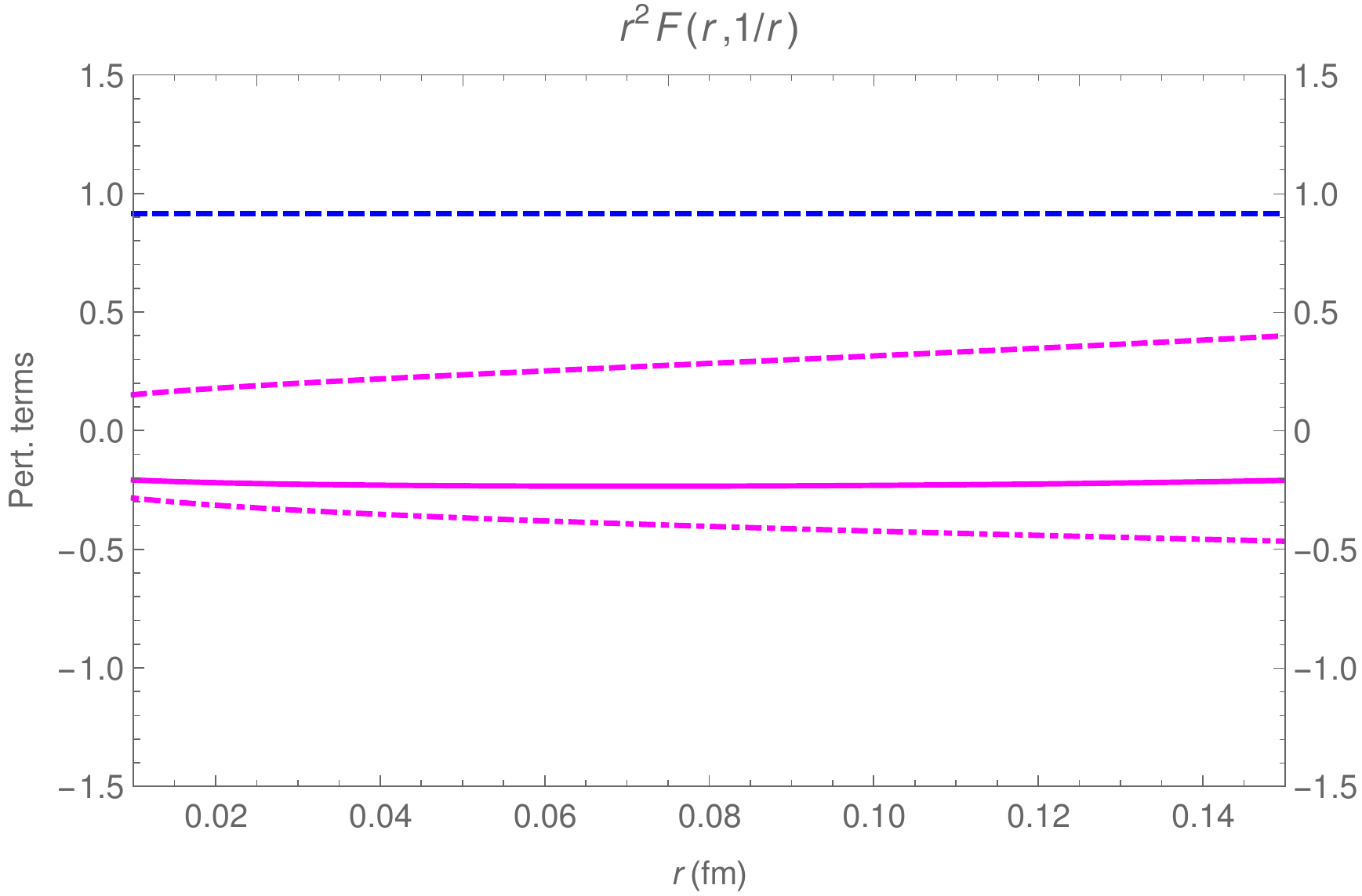}
\end{minipage}
\hspace{0.4cm}
\begin{minipage}[t]{8 cm}
\includegraphics[scale=0.45]{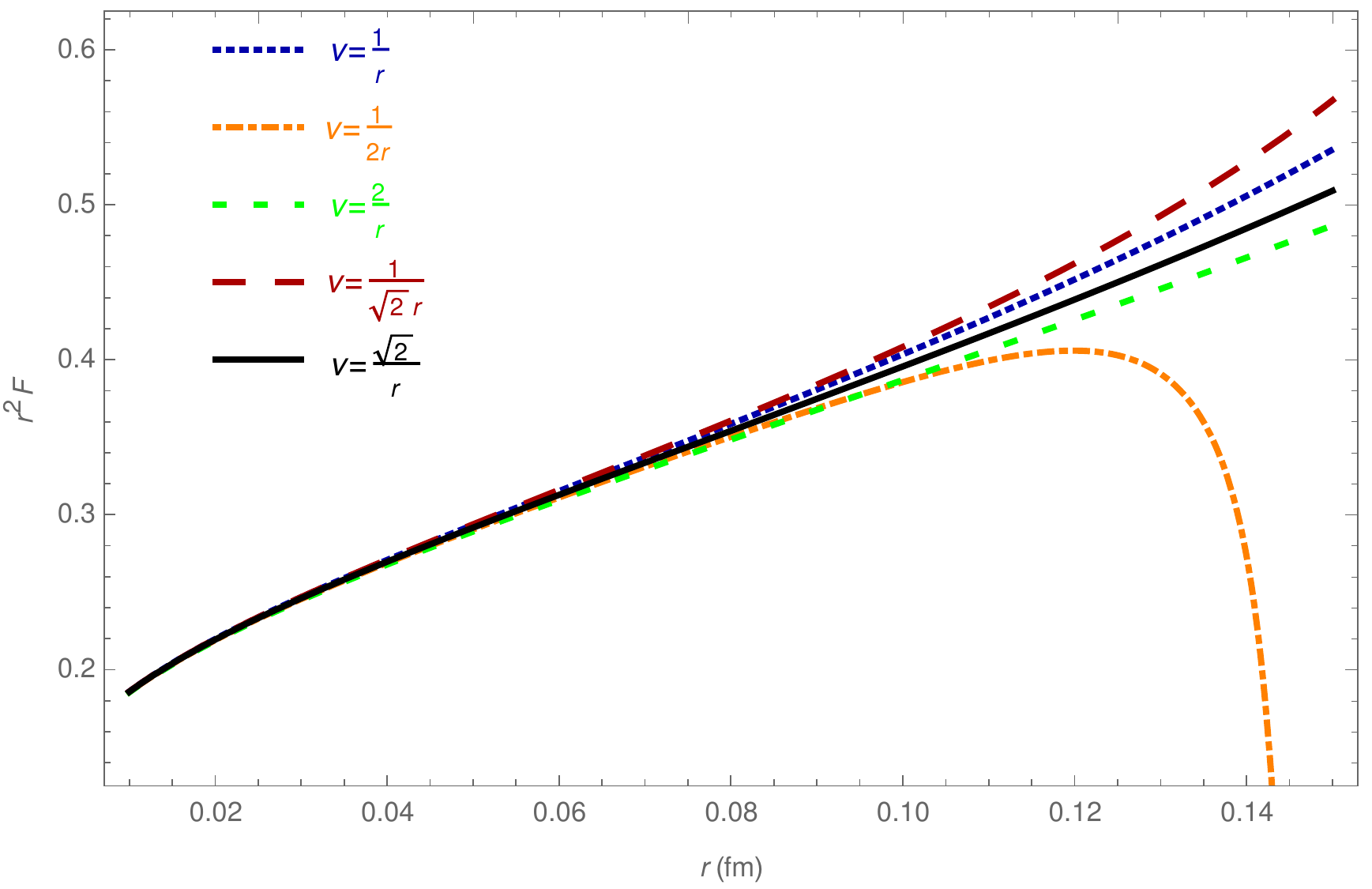}
\end{minipage}
\end{minipage}
\begin{minipage}[t]{16.5 cm}
\caption{
(left) The magenta lines represent three-loop contributions:
dashed ({soft term}, the first term in the last line of 
\Eqref{eq:F3L}), solid ({ultrasoft term}, the second term 
in the last line of \Eqref{eq:F3L}), and dot-dashed ({resummed 
ultrasoft term}, the terms in the second line of 
\Eqref{eq:FN3LL}).
The blue dashed line represents the two-loop contribution 
(last term in the first line of \Eqref{eq:F3L}).
A factor \(\als^3(1/r)\) has been dropped from all terms. 
(right) 
The {three-loop force with resummed leading ultrasoft logarithms} 
according to \Eqref{eq:FN3LL} at different values of the 
{soft scale} $\nu$. 
The dependence on the {soft scale} is non-monotonic and has a 
minimum for $\nu\approx 1/(\sqrt{2}r)$. 
(both)
The plots use $\lMSbt=315\,\mr{MeV}$ and the standard value 
$\mu_{us}=\CA\als(1/r)/(2r)$ for the {ultrasoft} scale.
\label{fig:F3L}
}
\end{minipage}
\end{center}
\end{figure}

Eventually, for sufficiently small values of the coupling $\als(1/r)$, 
\ie for short enough distances $r$, the {ultrasoft logarithm} 
$\ln \als(1/r)$ becomes very large. 
Hence, one could resum the terms $\als^{3+n} \ln^n(\als)$ or 
$\als^{4+n} \ln^n(\als)$ that contribute to the force. 
This resummation is accomplished using the renormalization group equations 
in pNRQCD~\cite{Pineda:2000gza, Brambilla:2004jw, Brambilla:2009bi}. 
In this case, it is not immediately clear, whether the naive power 
counting $\ln{\als} \sim \mc{O}(\als^0)$ could be appropriate. 
One may attempt a resummation of these contributions by treating the 
logarithm as $\ln{\als} \sim \mc{O}(\als^{-n})$, where 
$n=0,~1,~2,~\ldots$. 
We show the different terms at the two- and three-loop orders 
of \Eqref{eq:F3L} in \Figref{fig:F3L} for distances that are accessible 
in state-of-the-art lattice simulations.
Inspecting the magnitude of the terms, \ie first noting that the two-loop 
term is systematically larger than each three-loop term as demanded by 
the naive power counting, and second noting that all three-loop terms are 
of a similar magnitude, it is evident that the counting of the {ultrasoft 
logarithms} as of a lower order, is not justified in this distance range. 
%
%
Nevertheless, using the naive power counting 
$\ln{\als} \sim \mc{O}(\als^0)$ one may still resum the leading 
{ultrasoft logarithms} as~\cite{Bazavov:2014soa}
\al{\label{eq:FN3LL}
 F(r,\nu) 
 &=  \frac{C_F}{r^2}\als(\nu)\Bigg[1
 +\frac{\als(\nu)}{4\pi}\Big(\tilde{a}_1-2\beta_0\Big)
 +\frac{\als^2(\nu)}{(4\pi)^2}\Big(\tilde{a}_2-4\tilde{a}_1\beta_0-2\beta_1\Big)
 \nn\\& 
 \phantom{=\frac{\CF}{r^2}\als(\nu)\Bigg[}
 -\frac{\als^2(\nu)}{(4\pi)^2}\frac{a_3^L}{2\beta_0}\ln\frac{\als(\mu_\mr{us})}{\als(\nu)}
 +\frac{\als^2(\nu)\als(\mu_\mr{us})}{(4\pi)^3}a_3^L\ln\frac{C_A\als(\nu)}{2\mu_\mr{us}/\nu}
 \nn\\&
  \phantom{=\frac{\CF}{r^2}\als(\nu)\Bigg[}
 +\frac{\als^3(\nu)}{(4\pi)^3}\Big(
 \tilde{a}_3-6\tilde{a}_2\beta_0-4\tilde{a}_1\beta_1-2\beta_2\Big)
 +\mathcal{O}(\als^4)\Bigg].
}
The second term of the last line of \Eqref{eq:F3L} is recovered from the 
second line of \Eqref{eq:FN3LL} in the limit $\mu_\mr{us} \to \nu$. 
Figure \ref{fig:F3L} shows that the {resummed ultrasoft term} partially 
cancels the $r$ dependence of the {soft term} at three loops, which may 
be accidental, \ie this may fail to happen at higher loop orders. 
For this reason, it is conservative to consider the perturbative 
uncertainties due to the {soft} or {ultrasoft terms} 
as independent uncertainties. 
In the following we refer to the static energy $E(r)$ obtained from 
integrating $F(r,1/r)$ in \Eqref{eq:FN3LL} as the 
{three-loop result with resummed leading ultrasoft logarithms}. 

\subsubsection{Perturbative uncertainty}
\label{sec:Eperterr}

The perturbative uncertainty can be estimated by varying the details of 
the perturbative calculation. 
On the one hand, one may estimate the influence of unknown higher order 
terms by multiplying the {soft term} at the highest known order 
($\mr{N^3LO}$) by $\pm\als(\nu)$, and include this in the weak-coupling 
result. 
On the other hand, one may estimate the influence of unknown higher order 
terms by varying the {soft scale} $\nu$ in the calculation. 
It is customary to vary the {soft scale} by factor two, \ie between 
$ \nu=1/(2r)$ and $\nu=2/r$ for the standard choice of $\nu=1/r$. 
Figure~\ref{fig:F3L} shows that the {soft scale dependence} is 
non-monotonic and becomes rather large for the variation by a factor of
two, if distances larger than $r \gtrsim 0.1\,\mr{fm}$ are considered. 
The perturbative uncertainty due to the {soft term} is then 
given by the larger of these two estimates, where the upper error is 
always found to be determined in terms of the scale variation. 
Adopting this approach an asymmetric uncertainty 
due to the soft term could be obtained~\cite{Bazavov:2019qoo}. 
Moreover, the uncertainty originating in the {ultrasoft contribution} 
could be estimated similarly (by varying the {ultrasoft scale}). 
Alternatively, one could vary the resummation scheme for the 
{ultrasoft term} to estimate the related uncertainty, \ie 
consider the difference between the $\mr{N^3LO}$ result using \Eqref{eq:F3L} 
or the {three-loop result with resummed leading ultrasoft logarithms} 
using \Eqref{eq:FN3LL}. 
This approach has been adopted to obtain a symmetric uncertainty due to 
the {ultrasoft term}~\cite{Bazavov:2019qoo}. 

\subsubsection{OPE calculation of the renormalon subtracted singlet potential}

An alternative procedure to addressing the renormalon of the singlet 
potential $\Vs$ in the weak-coupling approach relies on a very particular 
version of an operator product expansion 
(OPE)~\cite{Sumino:2005cq, Mishima:2016vna}. 
For small distances $r$~the exponential $e^{\ri qr}$ in \Eqref{eq:VsaV} 
can be multipole expanded into real and imaginary terms, 
and the momentum integral can be evaluated with the 
lower momentum cutoff $\nu_f \ll \sqrt{q^2}$. 
Using a deformed contour integration the singlet potential $\Vs$ can be 
separated into a renormalon-free singlet potential $\Vs^\mr{RF}(r)$ and 
a renormalon contribution $\mr{R}(r,\nu_f)$ as 
\al{\label{eq:VsRF}
 &\Vs(r,\nu_f) = \Vs^\mr{RF}(r) + \mr{R}(r,\nu_f),&
 &\Vs^\mr{RF}(r) = V_C(r) + \mc{C}_1 r,&
 &\mr{R}(r,\nu_f) = \mc{C}_0(\nu_f) + \mc{C}_2(\nu_f) r^2
}
up to higher orders in the multipole expansion. 
It is not immediately clear, how the coefficients $\mc{C}_1$ or $\mc{C}_2$, 
which would represent condensates of mass dimension two or three, 
respectively, could be related to the usual QCD condensates of at least 
mass dimension four. 
Here, the renormalon-free terms are independent of the momentum cutoff, 
\ie 
\al{
 &\mc{C}_{-1} 
 = \frac{2\CF}{2\pi \ri} \int_{C_{\lQ}} \frac{dq}{q} \av(q^2),
 \quad 
 \mc{C}_{1} 
 = -\frac{\CF}{2\pi \ri} \int_{C_{\lQ}} \frac{dq}{q} q^2 \av(q^2), \\
 &V_C(r) 
 = -\frac{1}{r} \left\{
 \frac{2\CF}{\pi} \int\limits_0^\infty 
 \frac{dq}{q}~e^{-qr}~\mr{Im}~{\av(q^2+\ri 0)-\mc{C}_{-1} }
 \right\},
}
where $C_{\lQ}$ is a closed contour around the singularity 
of the coupling $\av(q^2)$. $V_C(r)$ is defined via an 
integral whose contour has been rotated to the line 
$\ri q$ with a real positive $q$.
Explicit dependence on the momentum cutoff $\nu_f$ appears only 
through the imaginary terms in the multipole expansion, 
\al{
 \frac{2\CF}{\pi r} \mr{Im}~\int_{C_\mr{upper}(\nu_f)} \frac{dq}{q}
 ~\left\{ \ri qr -\frac{\ri(qr)^3}{3!} \right\} 
 \av(q^2) 
 = \mc{C}_0(\nu_f) + \mc{C}_2(\nu_f)~r^2, 
}
where $C_\mr{upper}(\nu_f)$ is a contour from $0$ to $\nu_f$ in the upper 
half-plane above the singularity of the coupling $\av(q^2)$. 
The coefficients $\mc{C}_0(\nu_f)$ and $\mc{C}_2(\nu_f)$ have to be 
determined nonperturbatively. 
The second order renormalon proportional to $r^2$ can only be resolved at 
sufficiently large distances.

\subsection{Static energy determination on the lattice}

The static energy is also amenable to lattice-QCD calculations. 
For nonperturbative lattice calculations one is forced to use the 
imaginary-time formalism. 
After the Wick rotation $ t \to -\ri\tau$ the static energy can 
be obtained as 
\al{\label{eq:Eeucl}
 &aE\left(\frac{r}{a}\right) 
 =
 \lim\limits_{\tau \to \infty} 
 -\frac{a}{\tau} 
 \Braket{ \ln W \left( \frac{\bm r}{a},\frac{\tau}{a} \right) }
 =
 \lim\limits_{t \to \infty} 
 -\frac{\partial}{\partial (\tau/a)} 
 \Braket{ \ln W \left( \frac{\bm r}{a},\frac{\tau}{a} \right) }.
}
$W$ can be any lattice correlation function suitable for 
discretizing the Wilson loop $W_S$. 
The correlation function in the lattice approach is constructed as a 
product of gauge link variables (elementary parallel transporters 
by a single lattice step). 
Each of the gauge links has the same self-energy divergence, which is 
related to the charge renormalization factor, and results in an 
additive constant contribution to the QCD static energy in the lattice 
approach that diverges in the continuum limit. 
Since the lattice has a reduced symmetry in terms of the cubic group $W_3$ 
instead of the full rotation symmetry group $O(3)$, there is only a finite 
set of displacements between the static quark and antiquark that are 
geometrically equivalent. 
The static energy is accessible in the lattice approach at distances 
$ r = \sqrt{n_x^2+n_y^2+n_z^2} a $, 
where $n_x,~n_y,~n_z \in [0,~1,~2,~\ldots]$.
In order to resolve very small distances $r$, simulations with very fine 
lattice spacings are required. 
Moreover, for distances of the order of the lattice spacing, $r \sim a$, 
lattice calculations are affected by severe discretization artifacts. 
The origin of these discretization artifacts is apparent from the fact 
that the paths connecting the static quark and antiquark belong to different 
representations of the symmetry group. 
On the other hand, at distances larger than $r/a \sim 5$ the discretization 
artifacts of the QCD static energy are usually of a similar size as the 
statistical errors, and thus cannot be resolved clearly. 

\subsubsection{Interpolating operator dependence}
\label{sec:opdep}

\begin{figure}[tb]
\begin{center}
\begin{minipage}[t]{16.5 cm}
\begin{minipage}[t]{8 cm}
\includegraphics[scale=0.7]{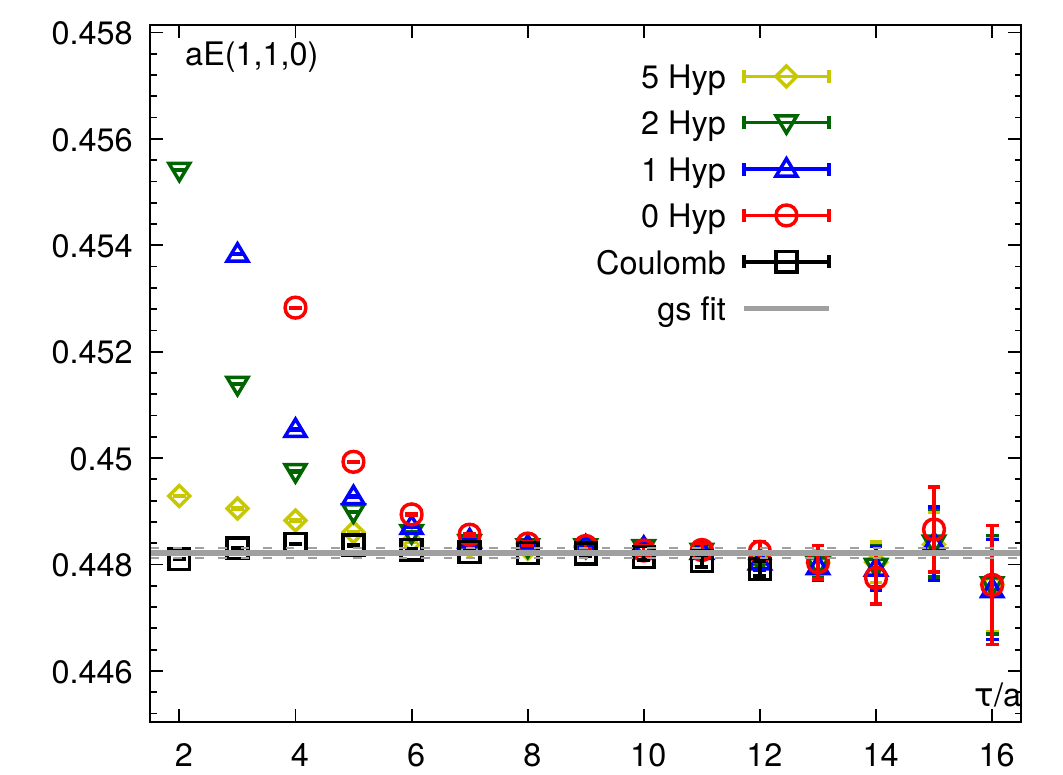}
\end{minipage}
\begin{minipage}[t]{8 cm}
\includegraphics[scale=0.7]{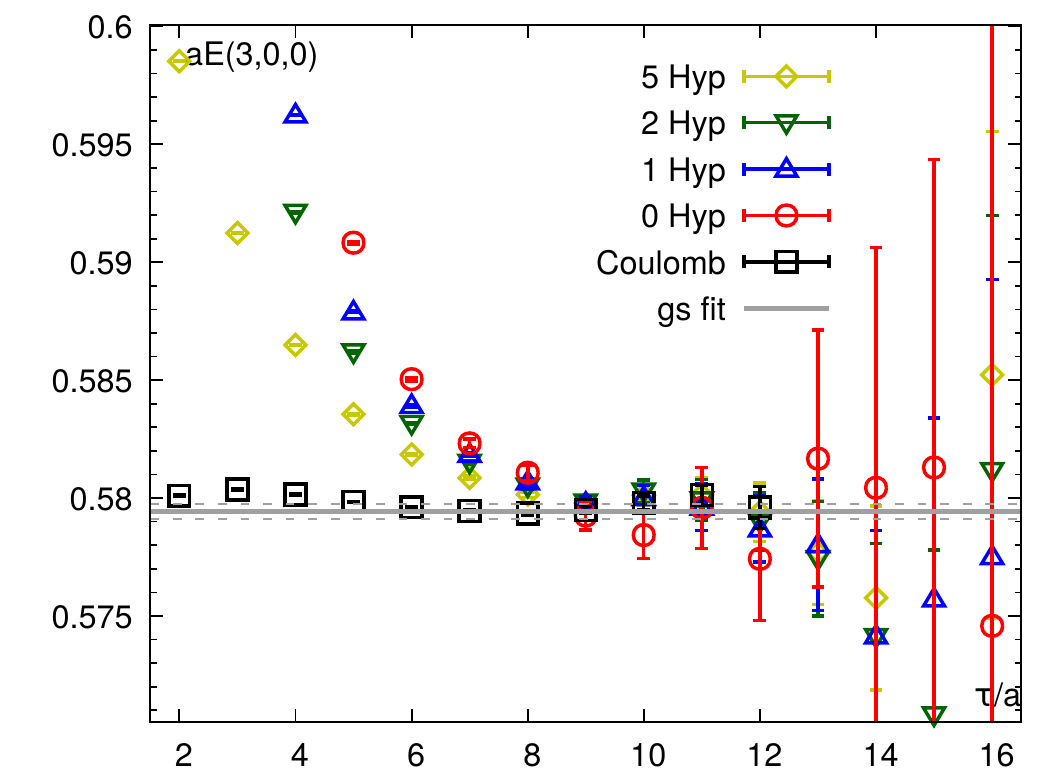}
\end{minipage}
\end{minipage}
\begin{minipage}[t]{16.5 cm}
\caption{
The effective masses
$aE(\bm r) = -\ln{\left(W(\bm r,\tau)/W(\bm r,\tau+a)\right)}$
of the Wilson line correlator in Coulomb gauge (black) and the Wilson
loop with spatially (only cubic HYP) smeared spatial Wilson lines
reach the same plateau value at large times.
At larger distances larger amounts of smearing are required for the
same extent of excited state suppression.
The data are obtained in (2+1)-flavor QCD at $\beta=7.825$ with the HISQ
action. 
\label{fig:WvsC}
}
\end{minipage}
\end{center}
\end{figure}

A lattice regularized Wilson loop cannot be realized as a smooth contour, 
but has to be implemented as a {rectangular Wilson loop}. 
This {rectangular Wilson loop} introduces an additional problem of receiving 
large contributions from cusp divergences due to its four corners, whose 
precise cancellation requires very high statistical accuracy. 
Due to the UV noise associated with the cusp divergences the statistical 
error of the {rectangular Wilson loops} tends to be too large for obtaining 
a useful signal without any smoothing of the links.
Typical link smoothing algorithms are the APE smearing~\cite{Albanese:1987ds}, 
the HYP (hypercubic) smearing~\cite{Hasenfratz:2001hp}, 
stout smearing~\cite{Morningstar:2003gk} or the 
Wilson flow~\cite{Luscher:2010iy, Luscher:2011bx}. 
The spatial Wilson lines of the {rectangular Wilson loop} are path dependent, 
and bring about the linear self-energy divergences, which are suppressed by 
applying smoothing techniques for the gauge fields along the paths. 
Moreover, for the {rectangular Wilson loops} there are different levels of 
cusp divergences associated with non-straight paths.
Nevertheless, as more and more smoothing is applied to the spatial links the 
signal-to-noise ratio of the Wilson loops steadily improves and the cusp 
divergences can be suppressed until they become numerically irrelevant. 
Alternatively, due to the cluster decomposition, the static energy may 
be defined in a suitably fixed gauge by evaluating \Eqref{eq:CS}, which 
evades all difficulties associated with the spatial paths. 
Whereas the two definitions have different levels of excited state 
contaminations, both correlators yield the same large-time behavior, and 
become independent of the details of the smearing, \ie both reach the same 
ground state, which is the static energy in units of the lattice spacing, 
$aE$; see \Figref{fig:WvsC}. 

\subsubsection{Discretization artifacts at short distances}

\begin{figure}[tb]
\begin{center}
\begin{minipage}[t]{16.5 cm}
\begin{minipage}[t]{8 cm}
\includegraphics[scale=0.7]{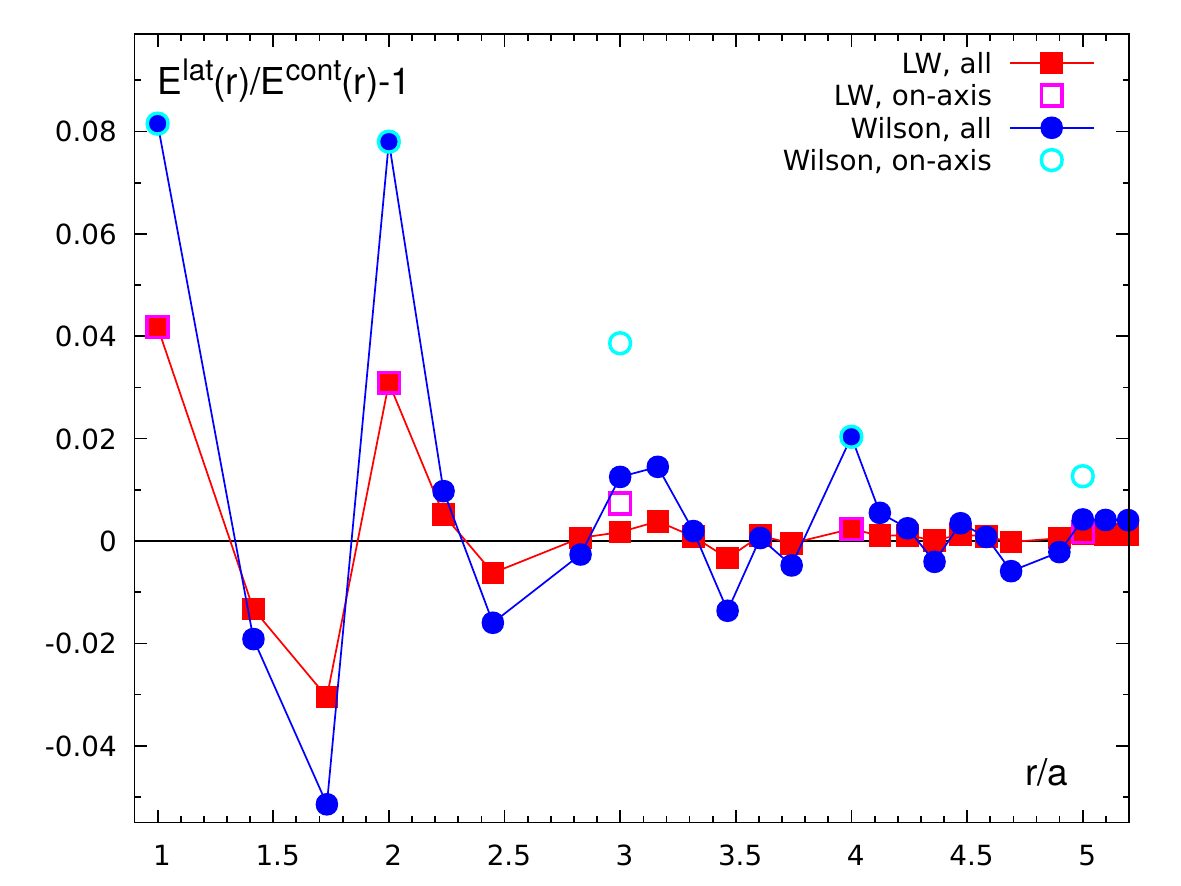}
\end{minipage}
\begin{minipage}[t]{8 cm}
\includegraphics[scale=0.77]{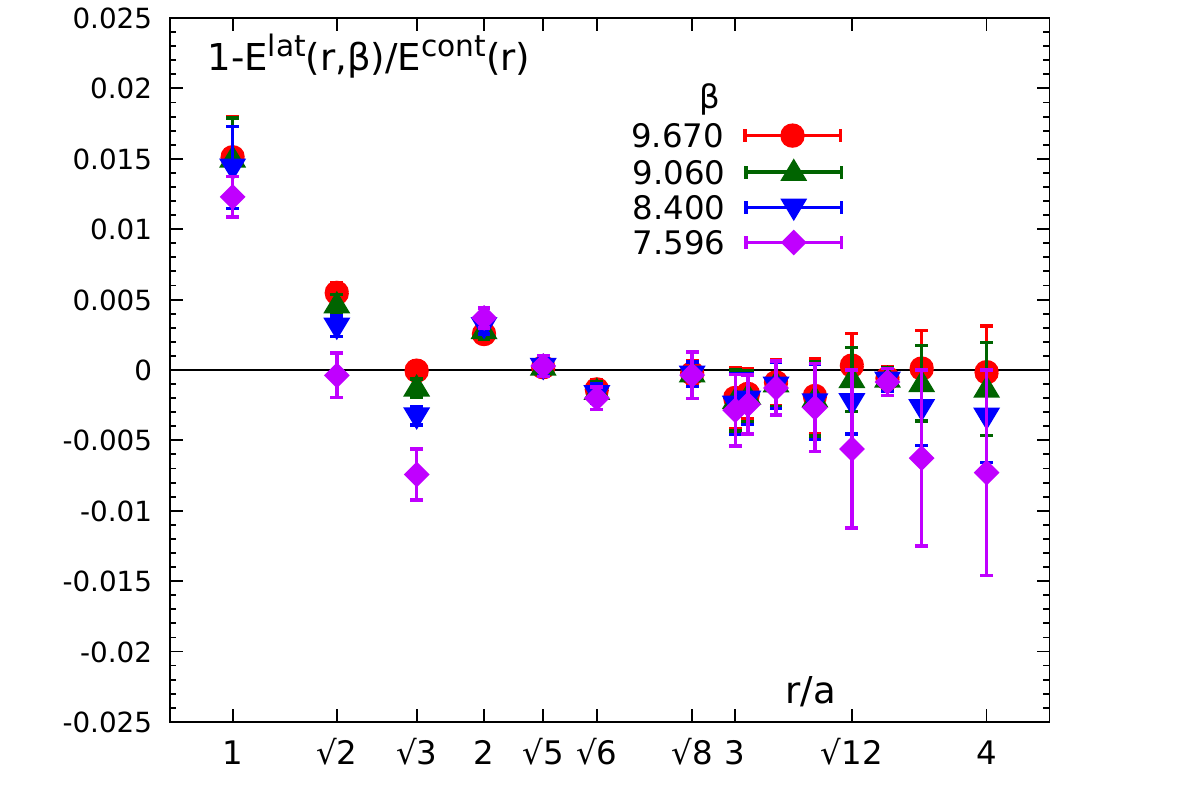}
\end{minipage}
\end{minipage}
\begin{minipage}[t]{16.5 cm}
\caption{
The ratio between the static energy on the lattice and in the continuum,
$E^\mr{lat}/E^\mr{cont}$, shows a distinctive, non-smooth pattern of 
$ r/a $-dependent discretization artifacts. 
(left) The discretization artifacts of the static energy in the lattice 
approach at the tree-level have a distinct pattern at the percent level 
as a function of $r/a$. 
Use of an improved gauge action reduces these discretization artifacts 
significantly, such that they become almost negligible (permille level) 
for $r/a > \sqrt{6}$. 
(right) The discretization artifacts of the QCD static energy in the lattice 
approach after the {tree-level correction} show a similar pattern with 
a reduced size and opposite sign.
\label{fig:disc}
}
\end{minipage}
\end{center}
\end{figure}

At very short distances that are most relevant to the comparison of the 
static energy calculated on a lattice and in the weak-coupling expansion, 
the static quark-antiquark correlators are beset by severe discretization 
artifacts. 
These discretization artifacts can be understood to a large extent in 
terms of the tree-level calculation, 
\al{
 E^\mr{lat,tree}(\bm r) = 
 -\CF g^2 \int \frac{d^3k}{(2\pi)^3}
 ~D_{00}(\bm k,k_0=0) 
 ~e^{\ri \bm k \cdot \bm r} ,
}
\ie one-gluon exchange between a static 
quark-antiquark pair without a running coupling. 
The discretization artifacts are due the lattice gluon propagator 
\al{
 D_{00}(\bm k,k_0=0) = \left(\sum\limits_{j=1}^3 
 \sin^2\left(\frac{ak_j}{2}\right)
 +c_w \sin^4\left(\frac{ak_j}{2}\right) \right)^{-1},
}
where $c_w=0$ for the (unimproved) Wilson gauge action and $c_w=1/3$ 
for the (improved) L\"uscher-Weisz action~\cite{Luscher:1985zq}. 
In the continuum this calculation yields 
$ E^\mr{cont,tree}(r) = \Vs^\mr{tree}(r) = -\CF\as/r$. 
A tree-level {improved distance} $r_\mr{I}$ is defined by equating 
\al{
 E^\mr{lat,tree}(\bm r) \equiv -\CF\as/r_\mr{I} = 
 E^\mr{cont,tree}(r_\mr{I}),
}
where $r_\mr{I}$ depends on the details of the path geometry, \ie 
paths with the same length but geometries that belong to different 
representations of the cubic group $W_3$ correspond to different 
tree-level {improved distances} $r_\mr{I}$. 
The assignment of the lattice data to the tree-level 
{improved distances} is called the {tree-level correction}.
These {improved distances} $r_\mr{I}$ differ from the naive (bare) 
distances by up to 8\% for the unimproved gauge action and by up to 
4\% for the improved gauge action (at $r/a=1$), 
see \Figref{fig:disc} (left). 
\begin{figure}[tb]
\begin{center}
\begin{minipage}[t]{16.5 cm}
\begin{minipage}[t]{8 cm}
\includegraphics[scale=0.7]{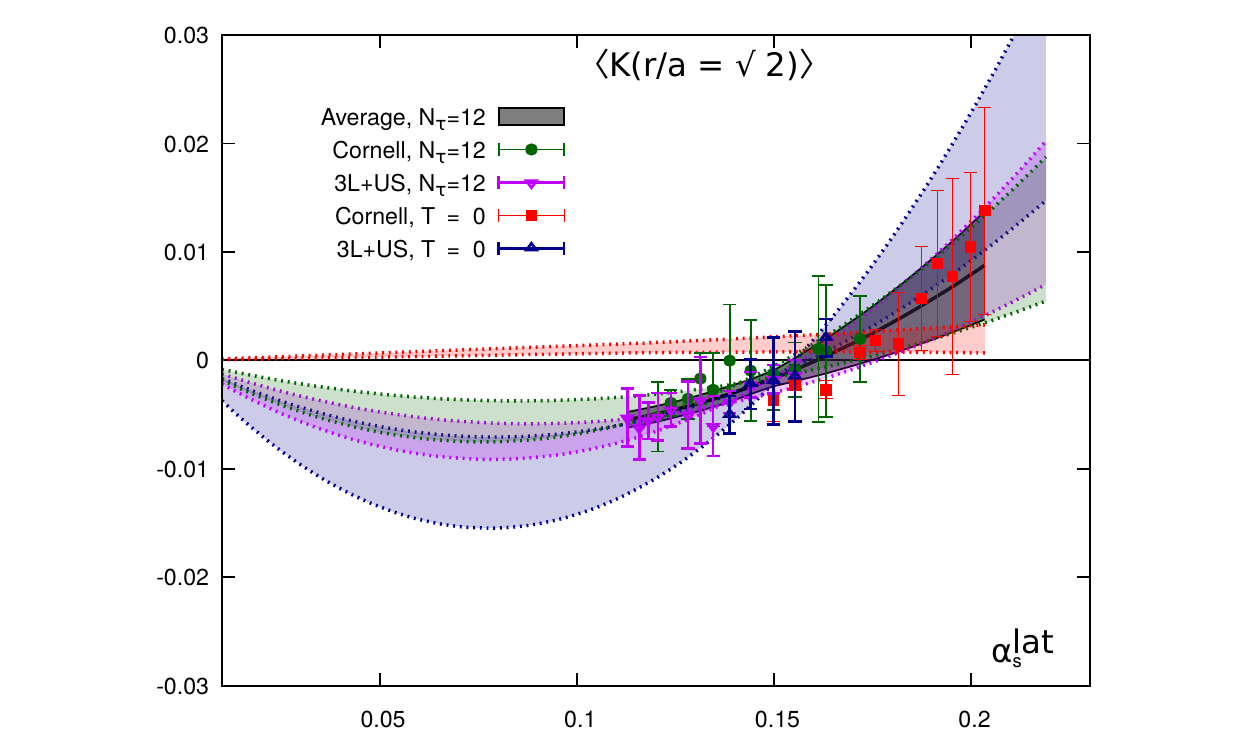}
\end{minipage}
\begin{minipage}[t]{8 cm}
\includegraphics[scale=0.7]{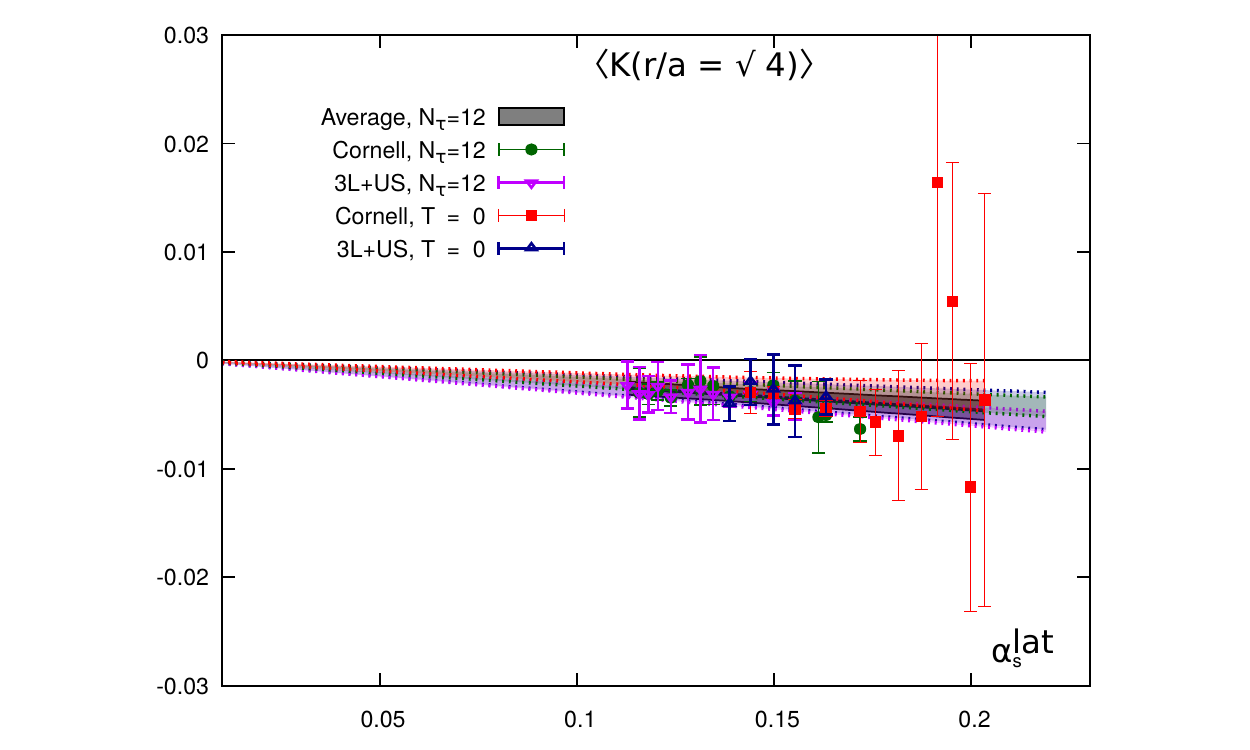}
\end{minipage}
\end{minipage}
\begin{minipage}[t]{16.5 cm}
\caption{
Nonperturbative correction of the {tree-level corrected} static 
energy in (2+1)-flavor QCD with improved gauge action and HISQ action. 
Results obtained using a Cornell Ansatz for interpolating 
the data on fine lattices, or using the integral of the
{three-loop force with resummed leading ultrasoft logarithms} 
are numerically consistent. 
Substituting the QCD static energy (at $T=0$) by the QCD singlet free 
energy at $T>0$, and restricted to short enough distances, a similar 
result can be obtained at much smaller values of the gauge coupling. 
\label{fig:npc}
}
\end{minipage}
\end{center}
\end{figure}
On the one hand, the {tree-level correction} reduces the size of the 
residual discretization artifacts of the QCD static energy 
$E^\mr{lat}(\bm r)$ to the level of the typical statistical 
errors in QCD lattice simulations at distances $r/a \gtrsim 3$. 
On the other hand, the {tree-level correction} accounts for the 
largest part of the discretization artifacts of the QCD static energy 
$E^\mr{lat}(\bm r)$ even at distances $r/a \lesssim 3$ , since the QCD 
gauge coupling is small at such short distances. 
Nevertheless, a similar pattern of discretization artifacts remains even 
after applying the {tree-level correction}, 
see \Figref{fig:disc} (right).
There are two possible approaches to account for these residual 
discretization artifacts. 
One may either calculate the corrections order by order in $g^2$ using 
the lattice perturbation theory, or one may use a continuum estimate for 
the QCD static energy, and calculate the necessary 
nonperturbative correction beyond the tree-level correction. 
This has been achieved in two schemes that ultimately yield quantitatively 
consistent answers. 

First, the same distance $r$ corresponds to different $r/a$ on fine or 
coarse lattices. 
Thus, the static energy result on fine lattices (with the improved gauge 
action and after the {tree-level correction}) at distances where 
the discretization artifacts are statistically irrelevant may serve
as a continuum estimate for determining the discretization artifacts in 
the static energy on coarser lattices. 
This approach introduces a systematic uncertainty, because this requires 
interpolating the data on the fine lattices to the distances where data 
are available on the coarse lattices. 
The other drawback of this correction scheme is that it does not provide 
accurate information on the {nonperturbative correction} for small 
$r/a$ on fine lattices, which are most important for comparing to the 
weak-coupling result. 
Second, one may compare to the weak-coupling calculation directly to 
estimate the {nonperturbative correction}. 
In this scheme there are a few important ingredients.
One has to ensure that the weights of the data where corrections are 
needed are small enough that these do not dominate the weighted 
residues in the comparison. 
Moreover, one has to marginalize over the details of the weak-coupling 
result used in the comparison in order to not introduce a bias towards 
a specific value of the QCD Lambda parameter. 
Lastly, the comparison has to be restricted to the distance range where 
the weak-coupling result and the lattice data are expected to agree in 
the continuum limit, \ie following~\cite{Bazavov:2014soa} for 
$r \lesssim 0.5\,r_1$ (or $r \lesssim 0.15\,\mr{fm}$). 
Due to the $\mc{O}(a^2)$ or $\mc{O}(\als a^2,a^4)$ forms of the dominant 
discretization errors of the Wilson or L\"uscher-Weisz actions, the
discretization artifacts of the tree-level corrected QCD static energy 
have to be of the form $\als^n~a^2/r^2$, where $n \ge 1$. 
Namely, the discretization artifacts for the fixed $r/a$ are 
polynomials in the bare gauge coupling $\als^\mr{bare}$. 
The {nonperturbative corrections} from either of these estimates may be extrapolated in the (tadpole-improved%
\footnote{Here,the ``average'' link variable $u_0$ in the tadpole 
improvement is defined using the expectation value of the plaquette.
An alternative approach would be to directly measure it in a smooth 
gaugelike Landau gauge. 
The tadpole improvement is used to enhance the convergent of the 
perturbative expansion of the discretized theory. 
See Ref.~\cite{Lepage:1992xa} for details.})
gauge coupling $\als^\mr{lat}=\als^\mr{bare}/u_0^4$ towards the 
continuum limit. 
This extrapolation indicates that corrections of the order 
$(\als^\mr{lat})^2$ are required for $r/a<2$, while the order 
$\als^\mr{lat}$ is sufficient for $2\le r/a <\sqrt{8}$ at the 
present numerical accuracy, see \Figref{fig:npc} for two typical cases. 
After the {nonperturbative correction} lattice results for the 
static energy at different lattice spacings are statistically 
consistent with each other up to the divergent, additive constant.

\subsubsection{Sea quark effects in lattice simulations}

\begin{figure}[tb]
\begin{center}
\begin{minipage}[t]{16.5 cm}
\begin{minipage}[t]{8 cm}
\includegraphics[scale=0.6]{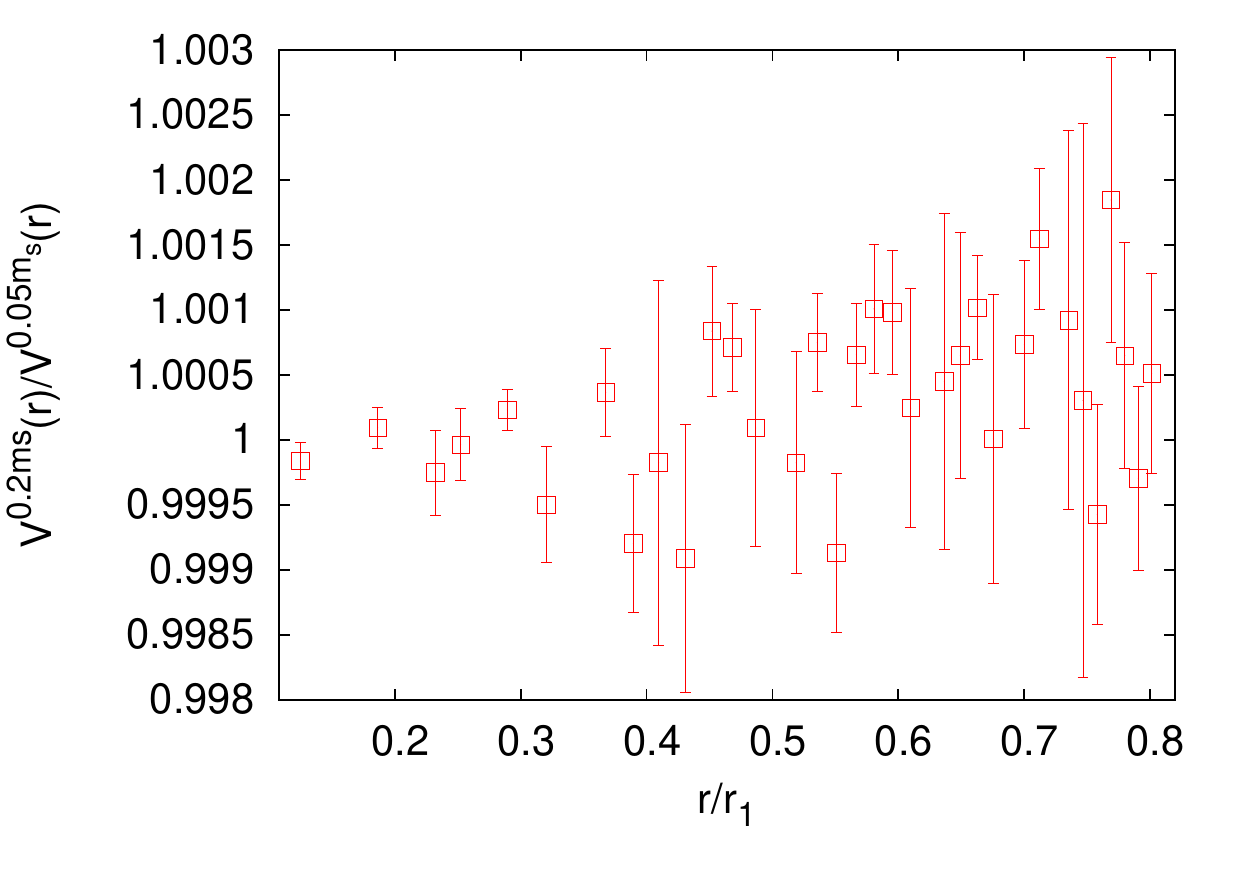}
\end{minipage}
\begin{minipage}[t]{8 cm}
\includegraphics[scale=0.6]{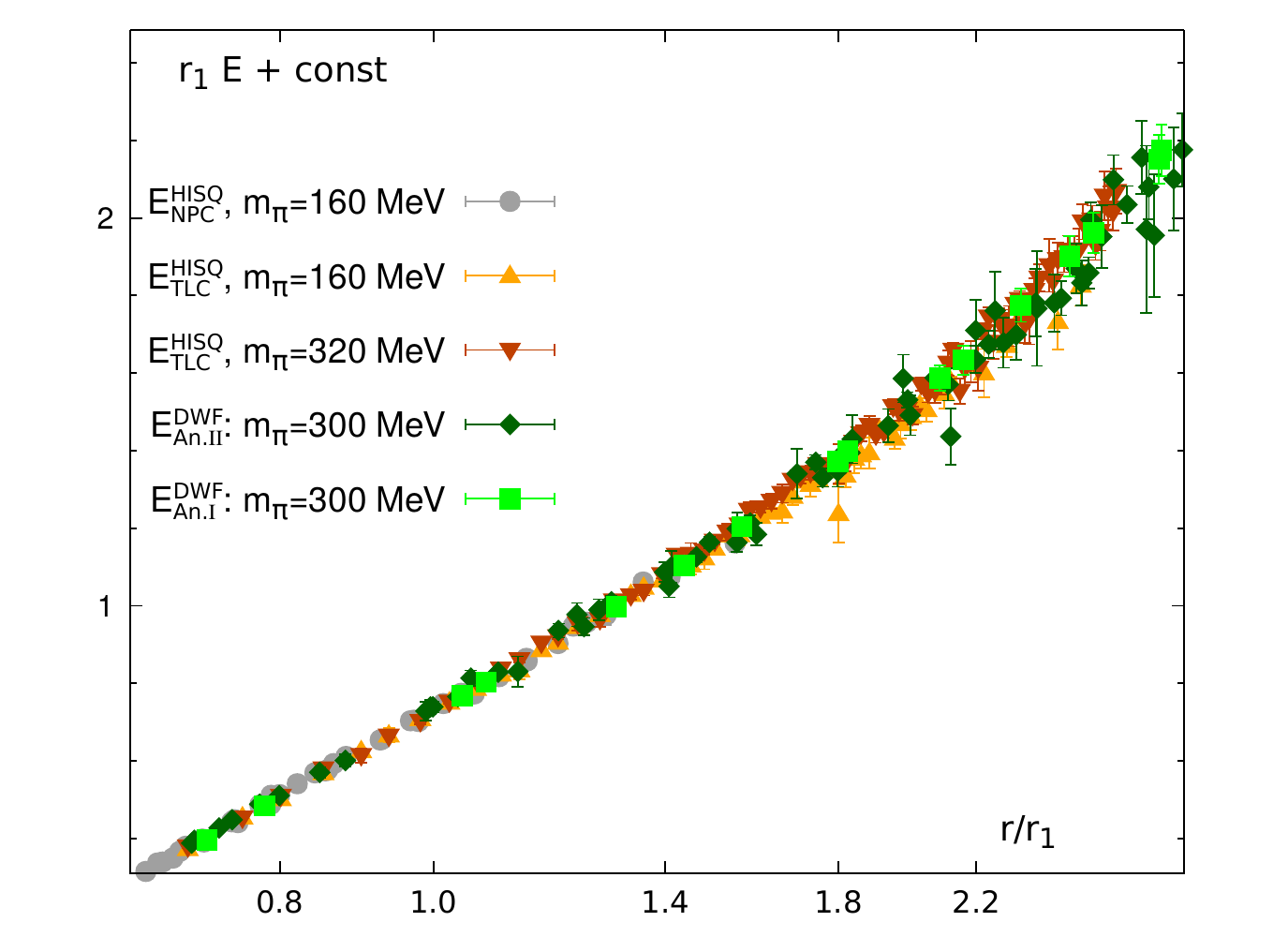}
\end{minipage}
\end{minipage}
\begin{minipage}[t]{16.5 cm}
\caption{
The sea quark mass dependence of the QCD static energy is negligible 
at short distances, $r \ll r_1$, but becomes numerically significant 
at larger distances. 
(left) The ratio of the QCD static energy for different sea quark 
masses corresponding to a pion mass of $160\,\mr{MeV}$ or $320\,\mr{MeV}$ 
in the continuum limit. 
The data are obtained in (2+1)-flavor QCD at $\beta=7.825$ with the HISQ 
action~\cite{Bazavov:2017dsy, Weber:2018bam}. 
(right) The QCD static energy shows the clearly visible sea quark mass 
dependence at larger distances $r \gtrsim r_1$, which is quite 
independent of the discretization used for the sea quarks. 
The DWF data, which were obtained using spatially smeared Wilson loops in 
(2+1)-flavor QCD with a pion mass of $300\,\mr{MeV}$ were extrapolated to 
the continuum limit with two different analyses, \ie in a two-step analysis 
\rm{I} or in a one-step analysis \rm{II}~\cite{Takaura:2018vcy, Takaura:2018lpw}, 
but are within errors consistent with the HISQ data at $\beta=7.825$ at the 
similar pion mass at these rather large distances. 
\label{fig:Esea}
}
\end{minipage}
\end{center}
\end{figure}

The nonperturbative lattice calculation of the static energy has a mild 
dependence on the sea quark masses or on the sea quark discretization.  
The former may be visualized most clearly in terms of a ratio of the static 
energy calculated with different sea quark masses and the same bare gauge 
coupling~\cite{Bazavov:2017dsy, Weber:2018bam}, see \Figref{fig:Esea} (left). 
Such a direct comparison shows that the sea quark mass dependence is not 
unambiguously resolved for $r \lesssim 0.55\,r_1$ 
(or $r \lesssim 0.17\,\mr{fm}$).
The latter may be observed at larger distances even by direct comparison 
of such results in the continuum limit or with sufficiently small lattice 
spacing, see \Figref{fig:Esea} (right).
The QCD static energy with the HISQ action or the DWF action and 
a similar pion mass are consistent at distances $r \gtrsim 0.4\,r_1$ 
(or $r \gtrsim 0.13\,\mr{fm}$). 

\subsection{$\als$ from the static energy}
\label{sec:asE}

The calculation of $\als$ from the QCD static energy is straightforward. 
The QCD static energy in any two regularizations differs by an additive 
constant, in which the zeroth order renormalon may be absorbed if necessary. 
In particular, this constant is different for each lattice spacing and has 
to be determined through a fit. 
First, one has to ensure that the lattice data of the QCD static energy are 
at sufficiently small distances that the perturbative expansion shows 
apparent convergence, \ie that the residues per degree of freedom in a fit 
of the lattice data with the weak-coupling result (and with the necessary 
additive constants) become smaller as the perturbative order is increased. 
It has been demonstrated that this condition is satisfied for 
$r \lesssim 0.5\,r_1$~\cite{Bazavov:2014soa}. 
Restriction to such distances automatically implies that the sea quark mass 
effects can be neglected. 
Second, one has to ensure that the lattice data are on sufficiently fine 
lattices that the discretization artifacts can be neglected (\ie after the 
{tree-level correction} or after the {nonperturbative correction}). 
After meeting these two conditions, one may simply compare the lattice 
data with $\Nf$ sea quark flavors to the weak-coupling results with $\Nf$ 
massless quark flavors with a fit of $1+N_\mr{lat}$ parameters, where 
$N_\mr{lat}$ is the number of lattice spacings used as the data set in the 
fit. 
The result of this comparison does not show a strong dependence on the 
distance window used in the fit while $ r \lesssim 0.45\,r_1$ (or 
$r \lesssim 0.13\,\mr{fm}$). 
The analysis of the static energy in the (2+1)-flavor QCD lattice 
simulations with the HISQ action and 
up to six lattice spacings $ 0.08\,r_1 \le a \le 0.20\,r_1$ 
(or $ 0.025\,\mr{fm} \le a \le 0.06\,\mr{fm}$)~\cite{Bazavov:2019qoo} 
yields for $0.076\,r_1 \le r \le 0.24\,r_1$ 
(or $ 0.0237\,\mr{fm} \le r \le 0.0734\,\mr{fm}$) 
\al{\label{eq:as t=0}
 &\phantom{\delta}\als(M_Z) = 0.11660^{+0.00110}_{-0.00056},& 
 &\delta \als(M_Z) = 
 (41)^\mr{stat} (21)^\mr{lat} (10)^{r_1} 
 (^{+95}_{-13})^\mr{soft} (28)^\mr{us}.
} 
The uncertainty due to the residual lattice spacing dependence has been 
estimated from two analyses using the {tree-level corrected} or the 
{nonperturbatively corrected} lattice data with the HISQ action and 
a fit range delimited at small distances by $r/a \ge \sqrt{8}$. 
The uncertainty due to the lattice scale dependence has been included 
at the end, as it is largely independent of the lattice spacing. 
The scale uncertainty includes the uncertainty of the experimental input, 
$f_\pi$, which contributes to the uncertainty of the lattice scale $r_1$. 
Lastly, the perturbative uncertainty has been determined as described 
in \Secref{sec:Eperterr}. 
{Soft scale} variation by the factor $2$ requires a stronger 
restriction than the apparent convergence of the perturbative expansion, 
namely, $r\lesssim ({1}/{3})\,r_1$ (or $r \lesssim 0.1\,\mr{fm}$). 
The uncertainty estimate due to the {soft scale} variation is highly 
asymmetric due to the significant variation of the QCD coupling at such 
low-energy scales, and is the dominant uncertainty in the calculation. 
In order to estimate the influence of the {ultrasoft term} the 
difference between the results obtained via \Eqsref{eq:F3L} 
or~\eqref{eq:FN3LL} has been considered as a symmetric error.  
In fact all of the individual estimates of the perturbative 
uncertainty are substantially reduced when the comparison is restricted 
to a smaller maximal distance $r$ as naively expected.  
\vskip1ex
Two different analyses of the static energy in the (2+1)-flavor QCD 
lattice simulations with the DWF action and 
the three lattice spacings $ 0.14\,r_1 \le a \le 0.26\,\mr{fm}$ 
(or $ 0.044\,\mr{fm} \le a \le 0.080\,\mr{fm}$) 
have been performed~\cite{Takaura:2018vcy, Takaura:2018lpw}.
In the two-step analysis \rm{I}, the continuum limit is obtained by first
interpolating the data on the two finer lattices to the distances, where 
the data on the coarsest lattice are available. 
Second, the (mock) data are extrapolated to the continuum limit at each 
distance $r/a$ on the coarse lattice separately. 
Finally, this continuum limit is compared to the continuum OPE in the 
distance window $0.8\,r_1 \le r \le 1.9\,r_1$ 
(or $0.24\,\mr{fm} \le r \le 0.6\,\mr{fm}$), which yields 
\al{\label{eq:as tsa}
 &\phantom{\delta}\als(M_Z) = 0.1166^{+0.0021}_{-0.0020},& 
 &\delta \als(M_Z) = 
 (^{+0.0010}_{-0.0011})^\mr{stat}(^{+0.0018}_{-0.0017})^\mr{syst}.
} 
In the one-step analysis \rm{II}, the continuum limit is obtained by 
simultaneously fitting the data on all three lattices with the continuum 
OPE result and a parametrization of the discretization artifacts. 
This analysis is performed in the distance window  $0.14\,r_1 \le r \le 1.15\,r_1$ 
(or $0.044\,\mr{fm} \le r \le 0.36\,\mr{fm}$), which yields 
\al{\label{eq:as osa}
 &\phantom{\delta}\als(M_Z) = 0.1179^{+0.0015}_{-0.0014},& 
 &\delta \als(M_Z) =(0.0007)^\mr{stat}(^{+0.0014}_{-0.0012})^\mr{syst}. 
} 
For further details of these systematic error budgets in both analyses 
see~\cite{Takaura:2018vcy, Takaura:2018lpw}. 
The estimate of the contribution from the scale uncertainty, and hence 
from the experimental input, $f_\pi$, is the same as for the HISQ calculation. 
Note that the authors consider the one-step analysis \rm{II} as superior 
to the two-step analysis \rm{I} due to the smaller uncertainties and quote 
\Eqref{eq:as osa} as their final result~\cite{Takaura:2018vcy, Takaura:2018lpw}.

\begin{figure}[tb]
\begin{center}
\begin{minipage}[t]{16.5 cm}
\center
\includegraphics[scale=0.8]{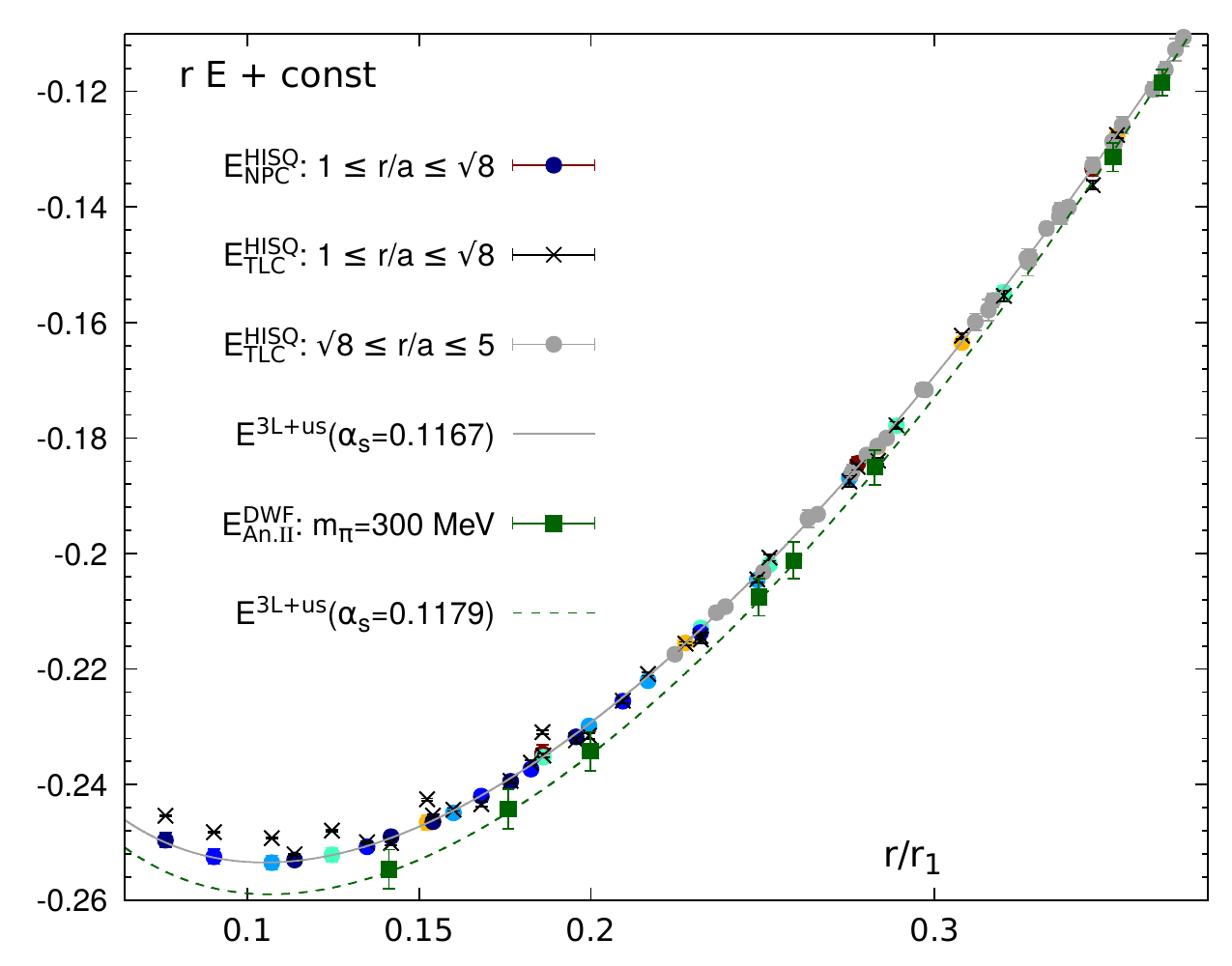}
\end{minipage}
\begin{minipage}[t]{16.5 cm}
\caption{
The nonperturbative lattice and the perturbative continuum results for 
the static energy multiplied by the distance, $r E(r)$. 
The HISQ data~\cite{Bazavov:2019qoo} are {nonperturbatively corrected} 
(NPC, colored bullets) or {tree-level corrected} (TLC, black crosses 
and gray bullets). 
The color indicates the lattice spacing in units of the $r_1$ scale, $a/r_1$. 
The DWF data~\cite{Takaura:2018vcy, Takaura:2018lpw} 
are from a one-step analysis \rm{II} that mixes the continuum extrapolation 
with the fit to the OPE result at $\mr{N^3LO}$, \Eqref{eq:VsRF}, using a 
parametrization of discretization artifacts (green squares). 
The lines represent the 
{three-loop result with resummed leading ultrasoft logarithms}, 
\Eqref{eq:FN3LL}, corresponding to $\als(M_Z,\Nf=5)=0.1167$ (gray, solid) 
or $\als(M_Z,\Nf=5)=0.1179$ (green, dashed). 
The former uses the central value $\als(M_Z,\Nf=5)=0.1167$ of the analysis 
of the (TLC or NPC) HISQ data with $r/a \ge \sqrt{8}$ (gray bullets), the 
latter uses the central value $\als(M_Z,\Nf=5)=0.1179$ of the OPE-based 
one-step analysis \rm{II} of the DWF data~\cite{Takaura:2018vcy, Takaura:2018lpw}. 
The NPC HISQ data with $r/a < \sqrt{8}$ are well-aligned with the fit 
excluding these data, while the TLC HISQ data with $r/a < \sqrt{8}$ cannot 
be consistently described by a continuum result for any value of 
$\als(M_Z,\Nf=5)$.
\label{fig:asE}
}
\end{minipage}
\end{center}
\end{figure}

We show the comparison between the lattice and weak-coupling 
results in \Figref{fig:asE}. 
At small distances the continuum results obtained in the one-step analysis 
\rm{II} of the spatially smeared Wilson loops with the DWF action are 
systematically below the data from the Wilson line correlation function in 
Coulomb gauge with the HISQ action ({nonperturbatively corrected} at 
very small distances $r/a <\sqrt{8}$ or only {tree-level corrected} at 
larger distances $r/a \ge \sqrt{8}$).

\subsection{$\als$ from the singlet free energy}

The QCD singlet free energy is a thermal observable with many properties 
similar to those of the QCD static energy at zero temperature. 
It is defined in terms of the thermal expectation value of the Wilson line 
correlation function at $\tau=1/T$ in a suitable gauge, \ie in the Coulomb gauge, 
or in terms of the thermal expectation value of the cyclic Wilson loop with 
spatially smeared spatial Wilson lines,
\al{\label{eq:FS}
 \FS(r,T) &= -T \Braket { \ln 
 e^{\ri g \int_{0}^{1/T} d{\tau} A_0(\bm 0,{\tau}) }
 e^{-\ri g \int_{0}^{1/T} d{\tau} A_0(\bm r,{\tau}) } }_T, \\
 F_W(r,T) &= -T \ln 
 \Braket { e^{ \ri g \oint_{\bm r,1/T} dz^\mu A_\mu } }_T.
}
However, in contrast to the case of the QCD static energy at zero 
temperature these two quantities $\FS$ and $F_W$ are distinguished 
by their distinct and temperature-dependent UV structures. 
A particular advantage of the QCD lattice calculation at finite 
temperature is that it resolves the IR problem of QCD at 
zero temperature in an elegant way. 
Namely, at high temperatures $T \gtrsim T_c$ the chiral symmetry 
is not spontaneously broken, and there are no associated 
pseudo-Goldstone bosons at the pion scale. 
Those would cause severe finite volume effects in zero temperature 
lattice simulations by propagating across the periodic lattice boundaries. 
At finite temperature, however, the smallest scale is the screening mass 
in the scalar channel, which scales as the temperature 
times the gauge coupling. 
Thus, a volume that is constant in units of the temperature as 
$TV^{1/3} \gtrsim 4$ is generally sufficient for avoiding the large 
finite volume effects. 
\vskip1ex
In particular, $\FS(r,T)$ coincides up to the corrections at the 
order $g^4$ with the zero temperature singlet potential $\Vs(r,\mu_{us})$, 
\al{\label{eq:FSmE}  
 \FS(r,T) = \Vs(r,\mu_{us}) + \delta \FS(r,T,\mu_{us}),
}
and has the same constant renormalon contribution. 
Moreover, $\FS(r,T)$ and $E(r)$ share the same discretization artifacts 
at the tree-level. 
Hence, the renormalon-free difference $E(r)-\FS(r,T)$ is particularly 
convenient for comparing the weak-coupling result and the 
nonperturbative lattice calculation. 
The details of the finite temperature effects  
$\delta \FS(r,T,\mu{us})$, however, depend on the hierarchy 
between the zero temperature and thermal scales. 
One might consider the cases 
$1/r \gg \als/r \gg T \gg m_D \sim g T$ or 
$1/r \gg T \gg m_D \sim g T \gg \als/r $. 
In the former hierarchy, which applies to small distances, the 
{ultrasoft scale} is the same as in the vacuum 
$\mu_{us} = (\Nc\als(1/r))/(2r)$, and  
\al{
 \delta \FS(r,T,\mu_{us})
 =\delta E_{us}(r,\mu_{us}) + 
 \Delta \FS(r,T). 
}
Thus, the thermal effects are expected to be suppressed 
since $T \ll \mu_{us}$. 
In the latter hierarchy, which applies to somewhat larger distances, 
$\FS$ has been calculated up to the order $g^5(T)$ in the multipole 
expansion~\cite{Berwein:2017thy}. 
In this hierarchy $\delta \FS(r,T,\mu_{us})/T$ is a dimensionless 
polynomial in $rT$ that  receives the contributions at order $g^4(T)$ 
from the nonstatic Matsubara modes of the gluons and from the quarks, 
and at order $g^5(T)$ from the static Matsubara modes of the gluons. 
In particular, these contributions to $\delta \FS(r,T,\mu_{us})$ become 
smaller for smaller distances at both orders $g^4(T)$ or $g^5(T)$. 
However, the order $g^6$ contribution to $\delta \FS(r,T,\mu_{us})$ is 
not known for this hierarchy. 
For this reason one cannot be certain that the cancellation of the 
{ultrasoft scale} between $\delta \FS(r,T,\mu_{us})$ and 
$\Vs(r,\mu_{us})$ takes place as in the zero temperature case. 
In particular, one has to expect that there is a temperature-dependent 
contribution $E(r) -\FS(r,T) \sim \als^3(T) \ln(rT)/r$.
Moreover, there is an $r$-independent contribution at the order 
$g^6$ that can be understood as the matching coefficient between 
the NRQCD and the pNRQCD~\cite{Berwein:2017thy}. 
\vskip1ex
In the second hierarchy $1/r \gg T \gg m_D \sim g T \gg \als/r $ 
the two known contributions at the different orders $g^4(T)$ and $g^5(T)$ 
have opposite signs and -- in the case of the phenomenologically 
interesting temperatures $300\,\mr{MeV} \lesssim T \lesssim 10\,\mr{GeV}$ 
that are accessible in the state of the art (2+1)-flavor QCD lattice 
simulations~\cite{Bazavov:2018wmo} -- a similar magnitude and a similar 
$r$ dependence. 
As a consequence, the absolute size of $\delta \FS(r,T,\mu_{us})$ is 
very small due to this approximate cancellation. 
Whereas the QCD lattice simulations do not exhibit a temperature 
dependence that is quite consistent\footnote{
This may be caused by the strong discretization effects in this 
difference that become visible only after the dominant contributions 
have canceled between $E(r)$ and $\FS(r,T)$, see~\cite{Bazavov:2018wmo} 
for a detailed discussion of the discretization effects.}
with this expression at any fixed lattice scale, the general features 
are quite similar for distances up to $r \lesssim 0.3/T$. 
In particular, the approximate cancellation between the contributions 
from the nonstatic and from the static Matsubara modes appears to be 
realized nonperturbatively. 
At even shorter distances $r/a \lesssim \sqrt{6}$ the finite temperature 
effects are not even resolved at the level of the statistical errors. 
Hence, after restriction to such small distances a direct comparison to 
the zero temperature result for the static energy is feasible and appropriate. 
\vskip1ex
After meeting the same two conditions as at zero temperature, and restricting 
to small $r \ll 0.3/T$, one may simply compare the finite temperature lattice 
data with $\Nf$ sea quark flavors to the weak-coupling results at zero 
temperature with $\Nf$ massless quark flavors in a fit of $1+N_\mr{lat}$ 
parameters, where $N_\mr{lat}$ is the number of lattice spacings used as the 
data set in the fit. 
The result of this comparison does not show a strong dependence on the 
distance window used in the fit while $ r \lesssim 0.45\,r_1$ (or 
$r \lesssim 0.13\,\mr{fm}$), and is always perfectly consistent with the 
zero temperature calculation.
The analysis of the singlet free energy in the (2+1)-flavor QCD lattice 
simulations with the HISQ action and 
up to fifteen lattice spacings $ 0.027\,r_1 \le a \le 0.20\,r_1$ 
(or $ 0.0085\,\mr{fm} \le a \le 0.06\,\mr{fm}$)~\cite{Bazavov:2019qoo} 
yields for $0.026\,r_1 \le r \le 0.1\,r_1$ 
(or $ 0.0081\,\mr{fm} \le r \le 0.030\,\mr{fm}$) 

\al{\label{eq:as t>0}
 &\phantom{\delta}\als(M_Z) =0.11638^{+0.00095}_{-0.00087}, 
 &\delta \als(M_Z) = 
 (80)^{\rm stat} (21)^{\rm lat} (17)^{T>0} (10)^{r_1} 
 (^{+40}_{-06})^{\rm soft} (15)^{\rm us}. 
} 
In order to escape a possible contamination by finite temperature effects 
the analysis has been performed with $r/a \le 2$.%
\footnote{A similar analysis with $r/a \le 3$ or $r/a \le \sqrt{12}$ produces 
a consistent result with smaller statistical uncertainties.} 
The uncertainty due to the finite temperature calculation has been 
estimated by comparing the results obtained with $\Nt=12$, or $\Nt=16$ 
with each other, or with the zero temperature calculation for the same 
upper limit of the distance window both in terms of $r/a$ and $r/r_1$. 
The scale uncertainty includes the uncertainty of the experimental input, 
$f_\pi$, which contributes to the uncertainty of the lattice scale $r_1$. 
All of the individual estimates of the perturbative uncertainty are 
dramatically reduced when the comparison is restricted to the smallest 
maximal distance $r$, \cf\Eqref{eq:as t=0}. 

\subsection{Summary and pre-averaging}

\begin{figure}[tb]
\begin{center}
\begin{minipage}[t]{16.5 cm}
\center
\includegraphics[scale=1.8]{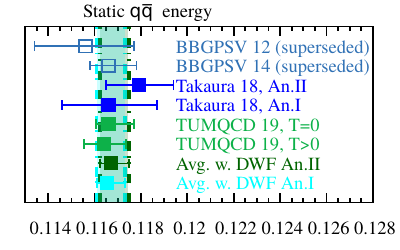}
\end{minipage}
\begin{minipage}[t]{16.5 cm}
\caption{
The results for $\als(M_Z,\Nf=5)$ obtained from the static $\qbq$ 
energy in (2+1)-flavor QCD, \ie the DWF result with a 
two-step analysis \rm{I} and a one-step analysis~\rm{II},
Takaura~18~\cite{Takaura:2018vcy, Takaura:2018lpw}, 
and the HISQ result using $T=0$ and $T>0$ ensembles,
TUMQCD~19~\cite{Bazavov:2019qoo}.
The superseded results BBGPSV~12~\cite{Bazavov:2012ka} and 
BBGPSV~14~\cite{Bazavov:2014soa} (both are superseded by 
TUMQCD~19~\cite{Bazavov:2019qoo}) are shown, too.
\label{fig:asEav}
}
\end{minipage}
\end{center}
\end{figure}

At this point we proceed to the summary and pre-averaging the present 
status of the determination of the strong coupling constant obtained 
from the QCD static energy in (2+1)-flavor QCD lattice simulations. 
The older calculations~\cite{Bazavov:2012ka, Bazavov:2014soa} with 
the HISQ action have been superseded by the most recent 
calculation~\cite{Bazavov:2019qoo} that is reproduced in~\Eqref{eq:as t=0}. 
The comparison with the QCD singlet free energy~\cite{Bazavov:2019qoo} is 
a statistically and systematically independent result that is reproduced 
in~\Eqref{eq:as t>0}. 
Finally, the calculations~\cite{Takaura:2018vcy, Takaura:2018lpw} with 
the DWF action are not independent. 
Hence, we calculate the averages for $\lMSbt$ and $\als(M_Z,\Nf=5)$ 
using either the DWF result from the one-step analysis \rm{II}, 
\Eqref{eq:as osa}, or from the 
two-step analysis \rm{I},~\Eqref{eq:as tsa}. 
\vskip1ex
The lattice averages are obtained by propagating the asymmetric 
errors of the individual results with the bootstrap method, and then 
calculating the central values and the asymmetric errors from the 
cumulative distribution function.
The average of the strong coupling constant at the Z pole reads 
\al{
 \label{eq:as EavII}
 &\als^\mr{stat}(M_Z) =  0.11671^{+0.00076}_{-0.00047},& 
 &\cdf = 0.914/2,& 
 &\text{with DWF~An.~\rm{II}, \Eqref{eq:as osa}}, \\
 \label{eq:as EavI}
 &\als^\mr{stat}(M_Z) =  0.11654^{+0.00076}_{-0.00045},& 
 &\cdf = 0.040/2,& 
 &\text{with DWF~An.~\rm{I}, \Eqref{eq:as tsa}}. 
} 
Treating the uncertainties instead as symmetric (the average of 
the upper or lower error) one would obtain 
$\als(M_Z, \Nf=5) =0.11696(75)$ at $\cdf = 0.580/2$, or 
$\als(M_Z, \Nf=5) =0.11653(78)$ at $\cdf = 0.016/2$ for the two averages, 
respectively. 
The variation between the different possible averages and individual results 
is covered in all cases by the uncertainties. 
However, the central value of the DWF result from the one-step analysis 
\rm{II}~\cite{Takaura:2018vcy, Takaura:2018lpw} is not fully covered by 
the propagated uncertainty in either way of taking the average, 
see \Figref{fig:asEav}. 

\section{Moments of quarkonium correlators}
\label{sec:moments}

Quarkonium correlation functions or the corresponding quarkonium spectral 
functions are observables that are amenable to direct experimental studies 
or to calculation in QCD with perturbative or nonperturbative methods. 
On the one hand, the vector channel is the most favorable for the 
experimental study, since it is known to the best accuracy from the $R$-ratio 
$R=\tfrac{\sigma^{(0)}[e^+e^- \to \text{hadrons}]}
{\sigma^{(0)}[e^+e^- \to \mu^+\mu^-]}$. 
On the other hand, the pseudoscalar channel is the most favorable for a
lattice study, since it offers the best signal-to-noise ratio among the 
lattice observables, and has the least contamination by the discretization 
artifacts. 
In the following we focus exclusively on the pseudoscalar channel. 

\subsection{General definitions and properties}

The renormalization group invariant quarkonium correlation function in the 
pseudoscalar channel is defined using the lattice discretization as
\al{\label{eq:def cfun lat}
 &G(t,a,V,m_h) = 
 \frac{a^3}{V}\sum\limits_{\bm n \in \tfrac{V}{a^3}} \left({am_{h0}}\right)^2 
 \Braket{ j_5\left(\bm n,t\right) j_5 (\bm 0,0) },&
 & j_5\left(\bm n,t\right) = 
 \bar\psi\left(\bm n,t\right) \ga_5 \psi\left(\bm n,t\right).
}
Here, $t=\tau/a$ is the Euclidean time in lattice units, $m_h$ is the mass 
of the heavy quark, and $am_{h0}$ is its bare mass in lattice units. 
Then the $n$th time moment of this quarkonium correlation function is 
defined for the infinite time direction as 
\al{
 &G_n(a,V,m_h) = \sum\limits_{t=0}^{\infty} t^n G(t,a,V,m_h).
}
Due to the periodic boundary condition in the time direction that is used in 
most QCD lattice calculations, which implies a backward propagating 
contribution, this time moment has to be generalized in the lattice approach as 
\al{\label{eq:def mom lat}
 &G_n(a,V,m_h) = \sum\limits_{t=0}^{\Nt/2} t^n 
 \left\{ G(t,a,V,m_h) + G(\Nt-t,a,V,m_h) \right\}. 
}
Due to the symmetry of the correlation function, odd moments vanish exactly. 
The time moments $G_n(a,V,m_h)$ are finite for $n \ge 4$, since the 
correlation function defined in \Eqref{eq:def cfun lat} diverges at 
small $t$ only as $t^{-4}$. 
Obviously the larger values of the heavy quark mass $m_h$ result in
larger discretization artifacts $(am_h)^n$ for the time moments 
$G_n(a,V,m_h)$ with the same lattice spacing. 
Moreover, it is clear from \Eqref{eq:def mom lat} that, on the one hand, the 
lower time moments have a stronger sensitivity to the discretization 
artifacts caused by the larger influence of the lattice correlation function 
at small times $\tau \sim a$, whereas, on the other hand, the higher time 
moments have a stronger sensitivity to the finite time direction (as the 
missing contribution with $t > \Nt/2$ would be given a larger weight). 
\vskip1ex
The time moments of the quarkonium correlation function can be calculated in 
the weak-coupling approach using the $\MS$ scheme, 
\al{\label{eq:mom pert}
 \left. G_n(a,V,m_h)\right\vert_{\nu,\nu_{m}} = 
 \frac{g_n\left[a,V; \als(\nu),\frac{\nu}{m_h(\nu_{m})}\right]}
 {[am_h(\nu_{m})]^{n-4}},
}
where $m_h(\nu_{m})$ is the $\MS$ heavy quark mass at the scale $\nu_{m}$, 
and $\als(\nu)$ is the $\MS$ strong coupling constant at the scale $\nu$. 
In principle the renormalization scale $\nu$ of the strong coupling constant 
could differ~\cite{Dehnadi:2015fra} from the renormalization scale $\nu_{m}$ 
of the heavy quark mass, although most studies assume $\nu=\nu_{m}$. 
The coefficients $g_n\left[a,V; \als(\nu),\frac{\nu}{m_h(\nu_{m})}\right]$ 
are known for the continuous space-time $a \to 0$ and in the infinite volume 
limit $V \to \infty$ at the four-loop order, \ie 
to the order $\als^3$~\cite{Sturm:2008eb, Kiyo:2009gb, Maier:2009fz}. 
\vskip1ex
It has been pointed out~\cite{Allison:2008xk} that it is favorable to consider 
the reduced moments $R_n$ for the calculation in the lattice approach. 
The reduced moments $R_n$ are ratios of the moments in QCD and in the 
free field theory, 
\al{\label{eq:def Rn}
 R_n(a,V,m_h) = 
 \left\{\begin{array}{cc} 
 \left(\frac{G_4^\mr{QCD}(a,V,m_h)}{G_4^{(0)}(a,V,m_h)}\right)
 \phantom{^\frac{1}{n-4}} & (n=4) \\
 \left(\frac{G_n^\mr{QCD}(a,V,m_h)}{G_n^{(0)}(a,V,m_h)}\right)^\frac{1}{n-4} & (n>4) 
 \end{array}\right. .
}
There are various cancellations between the systematic effects in these 
ratios of the lattice moments. 
These cancellations are particularly relevant with regard to the effects of 
the finite lattice spacing $a$, of the heavy quark mass $m_h$, and of the 
periodic time direction, $a\Nt$,  and to some extent, with regard to the 
finite volume $V$, too. 
In particular, the tree-level contribution to the discretization artifacts, 
$\als^0 a^n$, cancels exactly in the reduced moments to all orders $n$. 
Moreover, the uncertainties of the lattice time moments in QCD and in the 
free field theory due to the 
error of the numerical tuning of the heavy quark mass are subject to a 
strong compensation in the reduced moments $R_n$. 
\vskip1ex
Furthermore, the contribution for $t > \Nt/2$ is missing 
both in the numerator and in the denominator. 
It is possible to account for this systematic effect in the moments by
replacing the correlator for $t$ by $\cosh[am_0 (t-\Nt/2)]$ with 
$m_0$ being the ground state quarkonium mass. 
Then one can consider large enough $\Nt$ such that the systematic effects 
are negligibly small. 
In the free theory calculations it is easy to do calculations at large 
enough $\Nt$, where the effects of finite temporal extent can be neglected.
\vskip1ex
Finally, the finite volume effects may also be subject to a partial 
compensation. 
However, this last effect is expected to be less pronounced than any of 
the other compensations. 
A general parametrization of the finite volume error is
\al{
 \frac{R_n(a,\infty,m_h) - R_n(a,V,m_h)}{R_n(a,V,m_h)} = 
 \left[\left(\frac{\delta_V G_n^\mr{QCD}(a,V,m_h)}{G_n^\mr{QCD}(a,V,m_h)}\right)
 -\left(\frac{\delta_V G_n^{(0)}(a,V,m_h)}{G_n^{(0)}(a,V,m_h)}\right) \right]
 \times \left\{\begin{array}{cc} 
 1 & (n=4) \\
 \frac{1}{n-4} & (n>4) 
 \end{array}\right.,
}
but simplifies considerably under the reasonable assumption that the free 
field theory result is much more sensitive to the finite volume effects, 
\ie that the first term in square brackets can be neglected. 
This assumption can be justified from the fact that, on the one hand, the 
QCD result is dominated by the rather volume insensitive bound state 
contributions below the open heavy flavor threshold. 
On the other hand, the free field theory result is dominated by the volume 
sensitive scattering states of a pair of fictitious heavy fermions in a 
finite box. 
With a free field theory calculation using multiple sizes of the box it is 
straightforward to estimate $\delta_V G_4^{(0)}(a,V,m_h)$. 
\vskip1ex
In the weak-coupling approach, the reduced moments in the continuum at 
infinite volume are given as 
\al{\label{eq:Rn pert a}
 \left.R_n(0,\infty,m_h)\right\vert_{\nu,\nu_{m}} & \equiv 
 r_n \left[ \frac{\nu}{m_h(\nu_m)},\als(\nu) \right]
 \left\{\begin{array}{cc} 
 1 & (n=4) \\
 \left(\frac{m_{h0}}{m_h(\nu_{m})}\right)
 & (n>4) 
 \end{array}\right. 
 + ~\text{nonperturbative}, 
 \\
 \label{eq:Rn pert b}
 r_n \left[ \frac{\nu}{m_h(\nu_m)},\als(\nu) \right] &= 1 +
 \sum\limits_{j=1}^{3} r_{nj} 
 ~\left(\frac{\nu}{m_h(\nu_{m})}\right) 
 ~\left(\frac{\als(\nu)}{\pi}\right)^j. 
}
For the choice of scales $\nu=\nu_{m}=m_h$ the coefficients 
$r_{nj} \left(\tfrac{\nu}{m_h(\nu_{m})}\right)$ simplify to the 
mass-independent constants $r_{nj}$ that are reproduced in \Tabref{tab:rnj}. 
These constants are of order one without any evident pattern. 
The quarkonium correlation functions also receive nonperturbative 
contributions through the coupling between the scalar density and the 
QCD condensates. 
The leading contribution among these is due to the gluon 
condensate~\cite{Broadhurst:1994qj}, namely, 
and the reduced moments can be written as
\al{
 &
 \left. R_n(0,\infty,m_h)\right\vert_{\nu,\nu_{m}} =
 \nnnl&
 \left( 1+ \sum\limits_{j=1}^{3} r_{nj} 
 ~\left(\frac{\nu}{m_h(\nu_{m})}\right) 
 ~\left(\frac{\als(\nu)}{\pi}\right)^j
 + \frac{1}{m_h^4(\nu_{m})} \frac{11}{4}\Braket{\frac{\als}{\pi} G^2}
 \right)
 \times
  \left\{\begin{array}{cc} 
 1 & (n=4) \\
 \left(\frac{m_{h0}}{m_h(\nu_{m})}\right)
 & (n>4) 
 \end{array}\right.,
 \label{eq:Rn pert}
} 
where the value of the gluon condensate is known from the $\tau$ 
decays~\cite{Geshkenbein:2001mn},
\al{\label{eq:condensate}
 \Braket{\tfrac{\als}{\pi}\, G^2} = -0.006(12)\,{\rm GeV}^4.
}

\begin{table}
\begin{center}
\begin{minipage}[t]{16.5 cm}
\caption{
The constant coefficients $r_{nj}$ of the perturbative expansion of the 
reduced moments $R_n$ for the continuous space-time and in the infinite 
volume limit with the scales $\nu=\nu_{m}=m_h$. 
\label{tab:rnj}
}
\end{minipage}
\begin{tabular}{|c|lll|}
\hline
n & $r_{n1}$ & $r_{n2}$ &  $r_{n3}$  \\
\hline
4  &  2.3333 & -0.5690  &  1.8325    \\
6  &  1.9352 &  4.7048  & -1.6350    \\
8  &  0.9940 &  3.4012  &  1.9655    \\
10 &  0.5847 &  2.6607  &  3.8387    \\
\hline
\end{tabular}
\end{center}
\end{table}
\vskip1ex
Lastly, from the perspective of the use of data from the lattice calculation 
or from the experimental observation, it is particularly attractive 
to consider the ratios of the reduced moments, 
\al{\label{eq:ratio Rmn}
 R_{m,n}(a,V,m_h) = 
 \frac{R_{m}(a,V,m_h)}{R_{n}(a,V,m_h)}
 \quad\text{and}\quad
 r_{m,n} \left[ \frac{\nu}{m_h(\nu_m)},\als(\nu) \right] = 
 \frac{r_m \left[ \frac{\nu}{m_h(\nu_m)},\als(\nu) \right]}
 {r_n\, \left[ \frac{\nu}{m_h(\nu_m)},\als(\nu) \right]}
 ,
}
where the nonperturbative contributions have been omitted.
The obvious reason is that the fluctuations of the gauge ensembles that 
estimate the QCD path integral modify the underlying quarkonium correlation 
function, and thus also the moments. 
In the ratio of two (reduced) moments of one and the same correlation function, 
the cancellation of some of the statistical fluctuations is natural. 
Moreover, one also expects a partial cancellation of the uncertainties of 
the reduced lattice moments due to the error of the numerical tuning of 
the heavy quark mass in these ratios. 
Lastly, the similar compensation may happen as well for the discretization 
artifacts, for the effects of the periodic time direction, and for 
the finite volume effects, although these compensations can be effective only 
to a lesser extent. 
In the weak-coupling calculation one might consider expanding the numerator 
and the denominator separately in powers of the strong coupling constant 
$\als(\nu)$, or expanding the full ratio of the reduced moments as a total. 
For a fit the difference between both approaches is numerically irrelevant in 
practice.

\subsection{Lattice setup and continuum limit}

The time moments of the heavy quarkonium correlation functions in 
\Eqref{eq:def mom lat} have been calculated using the lattice approach in 
the (2+1)-flavor QCD with the HISQ action~\cite{Follana:2006rc} for 
the valence quarks by the HPQCD collaboration, HPQCD\,08~\cite{Allison:2008xk}, 
or HPQCD\,10~\cite{McNeile:2010ji}, in a partially quenched setting on 
the MILC gauge ensembles~\cite{Aubin:2004wf, Bazavov:2009bb} 
with the asqtad action~\cite{Orginos:1999cr}. 
The time moments have also been calculated on a much wider range of 
gauge ensembles generated for the calculation of the (2+1)-flavor QCD 
equation of state~\cite{Bazavov:2014pvz, Bazavov:2017dsy} with the HISQ 
action, in particular, using unprecedentedly 
fine lattice spacings, Maezawa and Petreczky (MP\,16)~\cite{Maezawa:2016vgv}, 
or Petreczky and Weber (PW\,19)~\cite{Petreczky:2019ozv}. 
Moreover, the time moments have also been calculated by the 
JLQCD collaboration, JLQCD\,16~\cite{Nakayama:2016atf}, in the (2+1)-flavor 
QCD with the domain-wall fermion (DWF) action~\cite{Kaplan:1992bt} for 
the valence quarks on DWF gauge ensembles with three steps of stout 
smearing~\cite{Morningstar:2003gk}. 
Lastly, the time moments have also been calculated in the (2+1+1)-flavor 
QCD by the HPQCD collaboration, HPQCD\,14~\cite{Chakraborty:2014aca}, using 
the HISQ action for the valence quarks on 
the MILC gauge ensembles~\cite{Bazavov:2012xda} with the HISQ action, but 
no continuum extrapolated results have been given in terms of the moments.
\vskip1ex
In the following we focus on the most recent calculation in 
the (2+1)-flavor QCD lattice simulations with the HISQ action. 
Since the underlying gauge ensembles have been generated for the study of the 
(2+1)-flavor QCD equation of state~\cite{Bazavov:2014pvz, Bazavov:2017dsy}, 
they suffer from the rather small spatial volumes and rather short Euclidean 
time directions. 
These two restrictions are partly responsible for the fact that the finite 
size effects tend to be the dominant systematic uncertainties for the finer 
lattices, whereas the mis-tuning of the valence heavy quark masses tends to 
be the dominant systematic uncertainty for the coarser lattices. 
The experimental errors of the pseudoscalar or spin-averaged meson masses 
are statistically insignificant in this calculation.
Moreover, since the bulk properties in the QCD thermodynamics have only a 
relatively mild dependence on the sea quark masses, the light sea quark masses 
are not fixed at the physical point, but at $m_l=m_s/20$ or $m_l=m_s/5$. 
The strange-quark mass has been tuned using 
the fictitious $\eta_{s\bar s}$ meson. 
These choices of the parameters ($m_l/m_s$) correspond to a pion mass of 
$160\,\mr{MeV}$ or $320\,\mr{MeV}$ in the continuum limit, 
respectively.
The kaon mass of the former corresponds to $504\,\mr{MeV}$ in the continuum. 
The valence heavy quark masses were tuned with the spin-average of the 
pseudoscalar and vector channels for the $m_l=m_s/20$ ensembles, and with 
just the pseudoscalar mass for the very fine $m_l=m_s/5$ ensembles. 
In the HISQ calculations with the exception of~\cite{Maezawa:2016vgv} the 
random color wall sources have been utilized, \ie gaussian $Z(2)$ noise 
distributed over an entire time slice. 
The expectation values from the random color wall sources reproduce the 
expectation values from the more simple point sources, but achieve a
reduction of the statistical errors by more than one order of magnitude. 
The quarkonium correlation functions have been calculated on ensembles with 
multiple light sea quark masses and/or multiple strange sea quark masses. 
Effects due to the different sea quark masses are generally at the same 
level as the statistical or systematic errors (due to finite volume or 
mis-tuning of the valence quark mass), \ie they do not cause a statistically 
significant trend, and thus can be neglected.
\vskip1ex

\begin{figure}[tb]
\begin{center}
\begin{minipage}[t]{16.5 cm}
\begin{minipage}[t]{8 cm}
\includegraphics[scale=0.7]{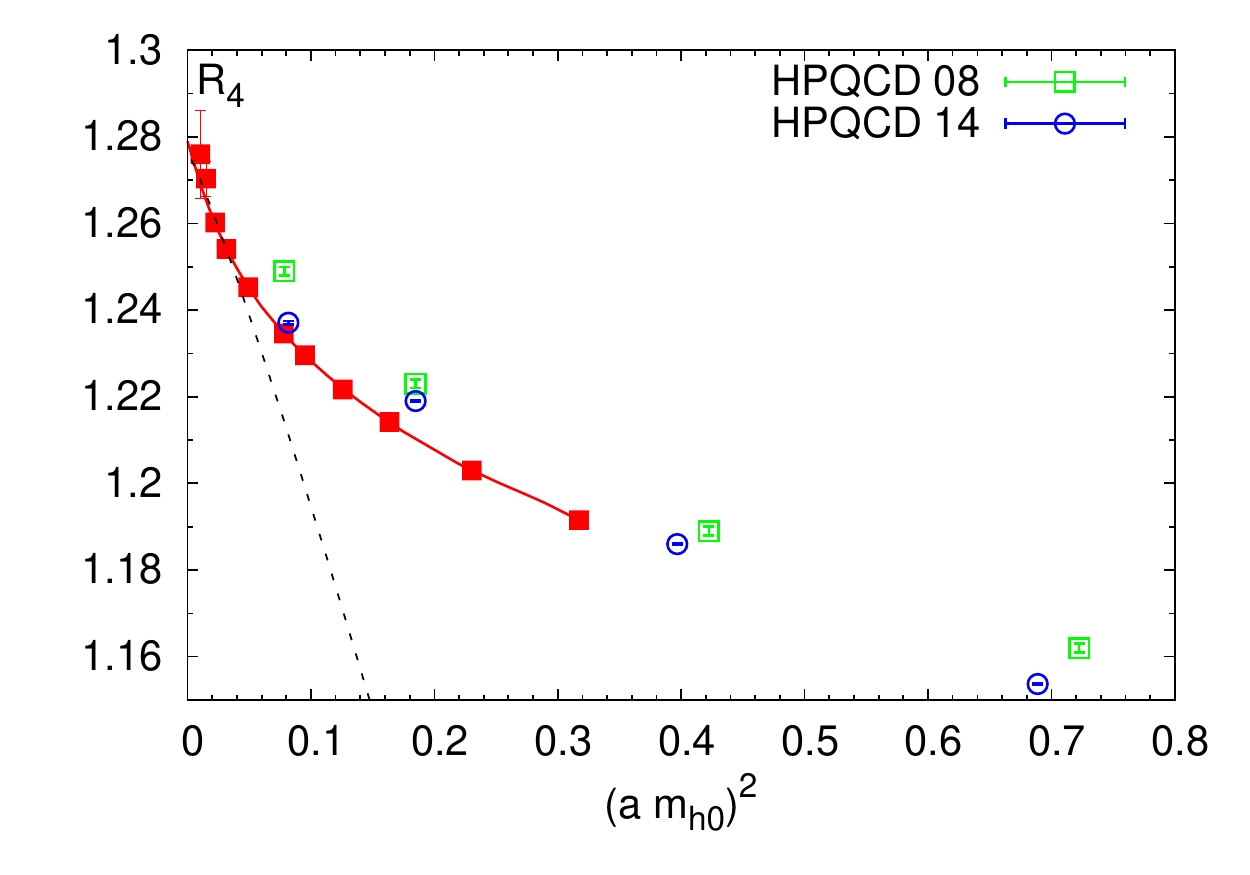}
\end{minipage}
\begin{minipage}[t]{8 cm}
\includegraphics[scale=0.7]{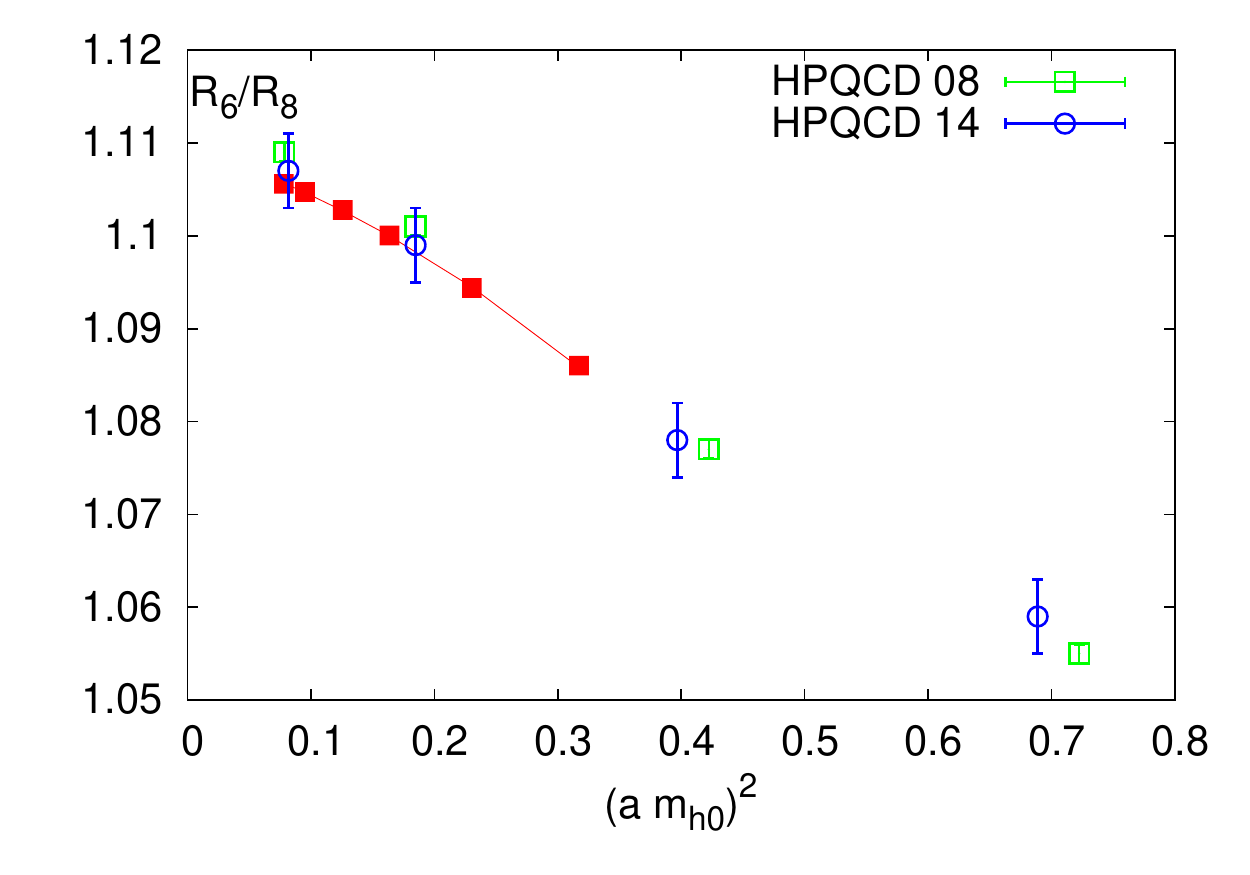}
\end{minipage}
\end{minipage}
\begin{minipage}[t]{16.5 cm}
\caption{
The lattice spacing dependence of the reduced moment $R_4$, or of the 
ratio $R_{6,8}$ with the HISQ action at the valence charm quark mass. 
The red filled squares correspond to the most recent valence HISQ result 
on the (2+1)-flavor HISQ ensembles, PW19~\cite{Petreczky:2019ozv}, and 
clearly resolve the logarithmic lattice spacing dependence. 
The green open squares correspond to the first valence HISQ result 
on the (2+1)-flavor asqtad ensembles, HPQCD08~\cite{Allison:2008xk}, while 
the blue open circles correspond to the most recent valence HISQ result 
on the (2+1+1)-flavor HISQ ensembles, HPQCD14~\cite{Chakraborty:2014aca}. 
The most simple fit of $R_4$ using only $\als^\mr{lat}~(am_h)^{2}$ (black 
dashed line) is only feasible upon restricting the coarsest lattice spacing 
to $a \lesssim 0.04\,\mr{fm}$. 
\label{fig:Rmc-ext}
}
\end{minipage}
\end{center}
\end{figure}

In all of these calculations, the tree-level or the one-loop improved 
Symanzik gauge action has been used. 
Improved lattice fermions (asqtad, HISQ, or DWF) guarantee 
that there are no odd powers of the lattice spacing $a$ permitted for the 
discretization artifacts. 
Moreover, with the exception of the contributions from the condensates, 
which are from the scale $\lQ$, or from the even lower and less important 
scales of the sea quark masses, the relevant scale in the problem is 
given by the (bare) heavy valence quark mass $m_{h0}$. 
As the contribution from the condensates has 200\% uncertainty in the 
continuum limit, \cf\Eqref{eq:condensate}, and is suppressed by four powers 
of the heavy valence quark mass, $1/m_h^4$, \cf\Eqref{eq:Rn pert}, the 
associated discretization errors are completely negligible and cannot be 
resolved in the analysis. 
Hence, in the infinite volume limit the most general fit form for the 
discretization artifacts is given by 
\al{\label{eq:Rn ext}
 R_n(a,\infty,m_h) - R_n(0,\infty,m_h)= 
 \sum\limits_{n=1}^{N} \sum\limits_{j=1}^{J} 
 c_{nj}[\ln(a m_{0h})]~(\als^\mr{lat})^j~(am_{h0})^{2n}, 
}
where the coefficients $c_{nj}$ could depend logarithmically on the 
heavy valence quark mass $a m_{h0}$. 
Here, the tadpole-improved gauge coupling $\als^\mr{lat}=\als^\mr{bare}/u_0^4$ 
is a function of the bare lattice gauge coupling and of the 
``average'' link variable $u_0$\footnote{
Here, $u_0$ in the tadpole improvement is defined using 
the expectation value of the plaquette.}.
%
%
The tadpole-improved gauge coupling $\als^\mr{lat}$ parametrizes the 
logarithmic dependence on the lattice spacing; see \Figref{fig:Rmc-ext}. 
For lattice spacings coarser than $a \lesssim 0.04\,\mr{fm}$ some higher 
order terms in $(\als^\mr{lat})^j$ or $(am_{h0})^{2n}$ have to be included in 
a fit, \ie see ~\cite{Petreczky:2019ozv} for a systematic discussion of 
the continuum extrapolation with up to eleven different lattice spacings 
in the range $0.025\,\mr{fm} \lesssim a \lesssim 0.109\,\mr{fm}$. 
If not enough data at fine enough lattice spacings or at different lattice 
spacings are available for such a fit, Bayesian techniques have to be 
employed to include additional constrained fit parameters, 
\ie see~\cite{Allison:2008xk, McNeile:2010ji, Chakraborty:2014aca}, where 
only four lattice spacings in the 
range $0.059\,\mr{fm} \lesssim a \lesssim 0.154\,\mr{fm}$ were available 
and high powers $(a m_{h0})^{2n}$ of the heavy valence quark mass were used. 
These more sophisticated analyses indicate the upward curvature for $R_4$, 
and the downward curvature for the ratios $R_6/R_8$, or $R_8/R_{10}$ in the 
approach to the continuum limit for all heavy valence quark masses;
see \Figref{fig:Rmc-ext}. 
%
%
However, with the larger statistical uncertainties due to the use of point 
sources such subtle effects could not be resolved in some of the presented 
analyses~\cite{Maezawa:2016vgv, Nakayama:2016atf}, which used the simplest 
linear fits without even the factor $(\als^\mr{lat})^1$. 
For these reasons, these analyses failed to identify the logarithmic 
lattice spacing dependence. 
In particular, in the JLQCD analysis three lattice spacings 
$a=0.080,~0.055$ and $0.044\,\mr{fm}$ were used and simple $(a m_{h0})^2$ 
extrapolations were performed.
The $\cdf$ of such extrapolations turned out to be large. 
For example $\cdf=2.1,~4.1$ and $5.1$ for $R_6/m_{h0}$, $R_8/m_{h0}$ and 
$R_{10}/m_{h0}$, respectively.
On the other hand, for the higher moments the overall uncertainty is 
dominated by the error of the lattice scale. 
For this reason, the most simple fit form $\als^\mr{lat}(am_{h0})^2$ is 
often sufficient, \ie including any higher order terms does not have 
a statistically significant impact on the continuum limit. 
\vskip1ex
The continuum extrapolated results of the most recent valence 
HISQ results~\cite{Petreczky:2019ozv} on the (2+1)-flavor HISQ ensembles 
are reproduced in \Tabref{tab:Rconthisq}. 
Given that no statistically significant sea quark effects are observed, 
these results can be considered to correspond to physical quark masses.

\begin{table}
\begin{center}
\begin{minipage}[t]{16.5 cm}
\caption{
The continuum results for $R_4$, $R_6/R_8$, or $R_8/R_{10}$, or 
$R_n/m_{h0}$, $n \ge 6$ at the different valence heavy quark masses, 
$m_h$, in (2+1)-flavor QCD with the HISQ action, PW19~\cite{Petreczky:2019ozv}. 
The continuum extrapolation for  $m_h=4m_c$ or $m_h=m_b$ was not possible 
due the lack of a sufficiently large number of fine lattice spacings. 
Note that the error of the $R_n/m_{h0}$ is dominated by the error of 
lattice scale in units of the lattice spacing, $r_1/a$, 
which includes the error of the experimental input, $f_\pi$, 
that was used for setting the lattice scale.
\label{tab:Rconthisq}}
\end{minipage}
\begin{tabular}{|c|lll|lll|}
\hline
$m_h$    &    $R_4$     &   $R_6/R_8$  &   $R_8/R_{10}$  & $R_6/m_{h0}$ & $R_8/m_{h0}$ & $R_{10}/m_{h0}$ \\
\hline
$1.0m_c$ &    1.279(4)  &   1.1092(6)  &     1.0485(8)   & 1.0195(20)   & 0.9174(20)   & 0.8787(50) \\
$1.5m_c$ &    1.228(2)  &   1.0895(11) &     1.0403(10)  & 0.7203(35)   & 0.6586(16)   & 0.6324(13) \\
$2.0m_c$ &    1.194(2)  &   1.0791(7)  &     1.0353(5)   & 0.5584(35)   & 0.5156(17)   & 0.4972(17) \\
$3.0m_c$ &    1.158(6)  &   1.0693(10) &     1.0302(5)   & 0.3916(23)   & 0.3647(19)   & 0.3527(20) \\
$4.0m_c$ &  &  &  & 0.3055(23)   & 0.2859(12)   & 0.2771(23) \\
$m_b$    &  &  &  & 0.2733(17)   & 0.2567(17)   & 0.2499(16) \\
\hline
\end{tabular}
\end{center}
\end{table}

\section{Comparison of the reduced moments from various groups}

\begin{figure}[tb]
\begin{center}
\begin{minipage}[t]{16.5 cm}
\centering
\includegraphics[width=5.3cm]{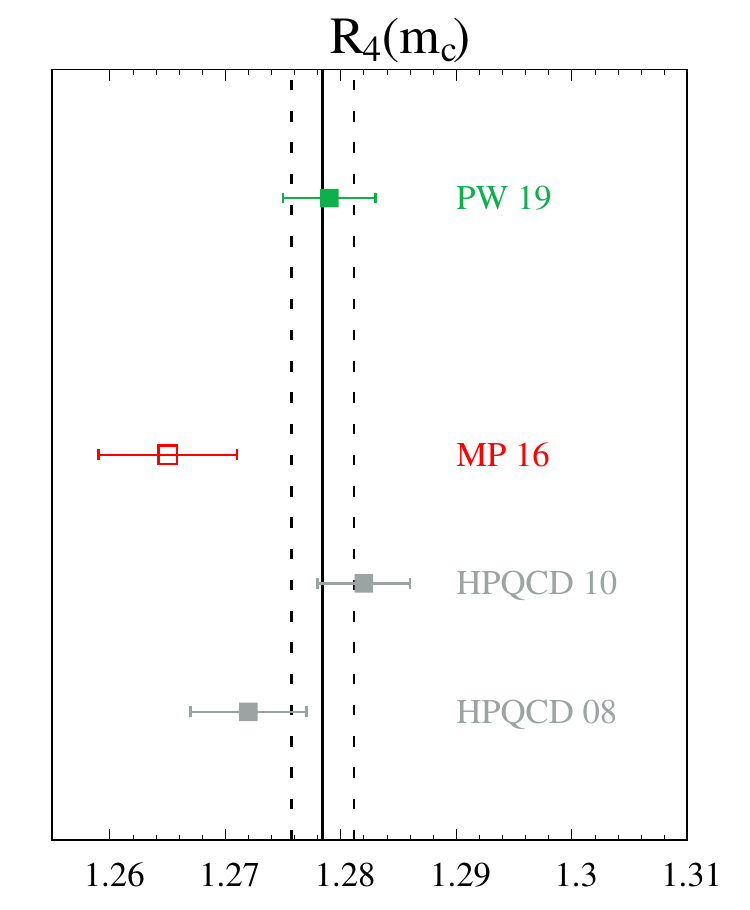}
\includegraphics[width=5.3cm]{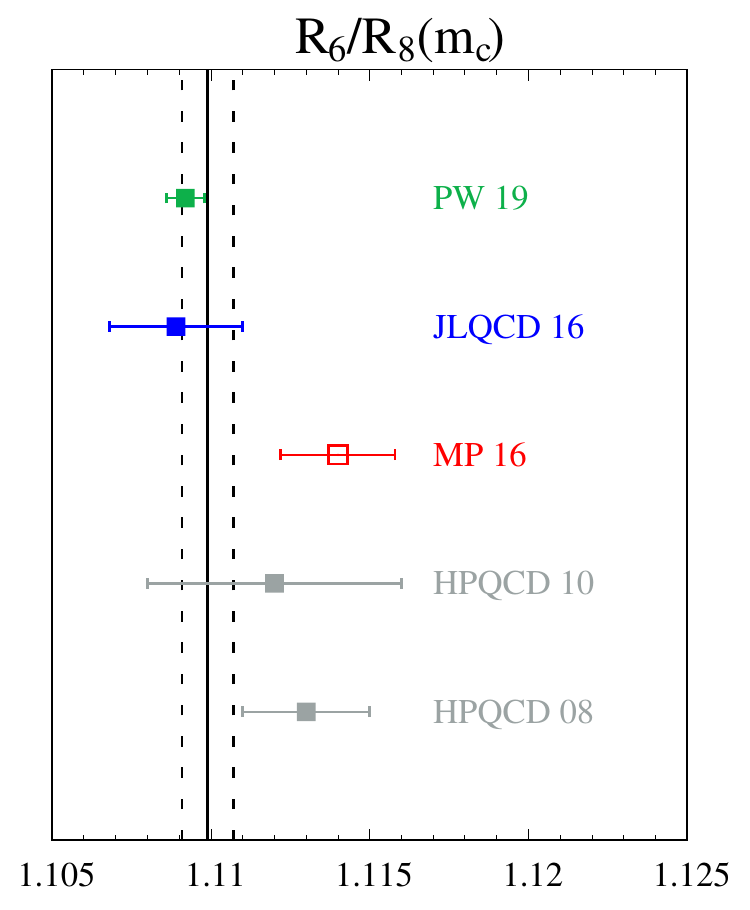}
\includegraphics[width=5.3cm]{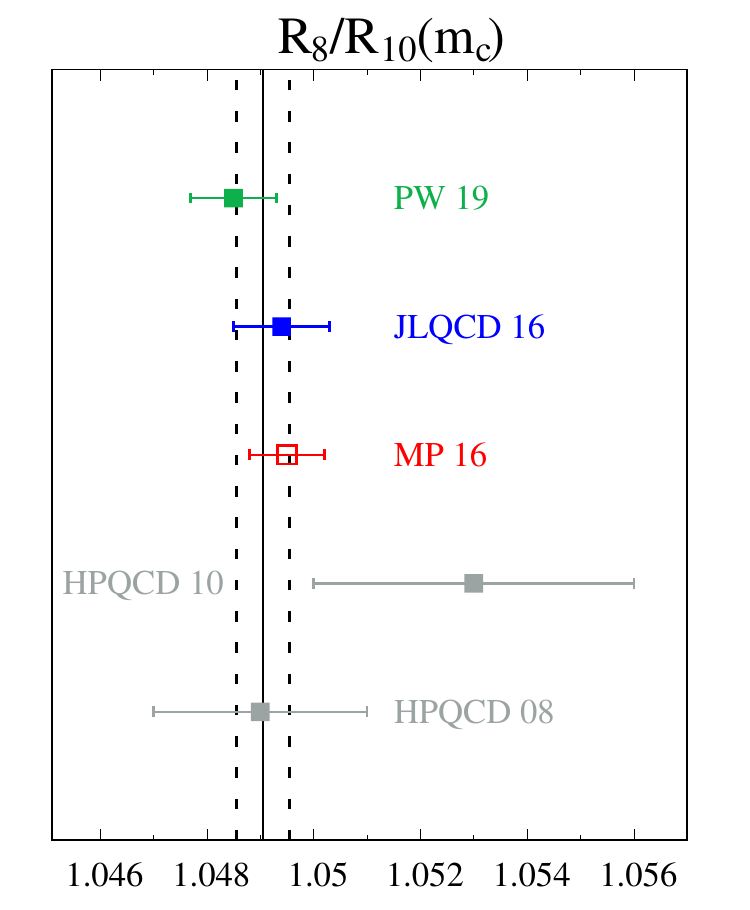}
\end{minipage}
\begin{minipage}[t]{16.5 cm}
\caption{Comparison of different lattice results for $R_4$ (left), 
$R_6/R_8$ (middle), and $R_{8}/R_{10}$ (right). 
Shown are the lattice results from HPQCD 08~\cite{Allison:2008xk},
HPQCD 10~\cite{McNeile:2010ji}, JLQCD 16~\cite{Nakayama:2016atf}, 
Maezawa and Petreczky (MP 16)~\cite{Maezawa:2016vgv}, 
Petreczky and Weber (PW 19)~\cite{Petreczky:2019ozv}. 
The vertical solid and dashed lines show the weighted average and the 
corresponding uncertainty. 
The data corresponding to the open symbols do not enter the averages 
since they are superseded by PW 19.
\label{fig:compR4}
}
\end{minipage}
\end{center}
\end{figure}

\begin{figure}[tb]
\begin{center}
\begin{minipage}[t]{16.5 cm}
\centering
\includegraphics[width=5.3cm]{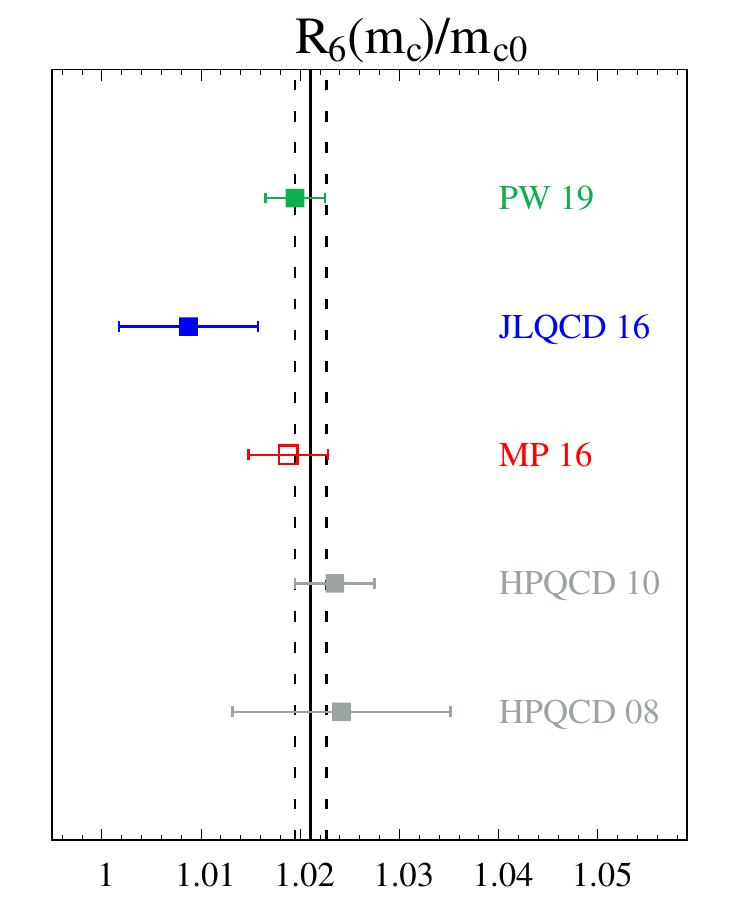}
\includegraphics[width=5.3cm]{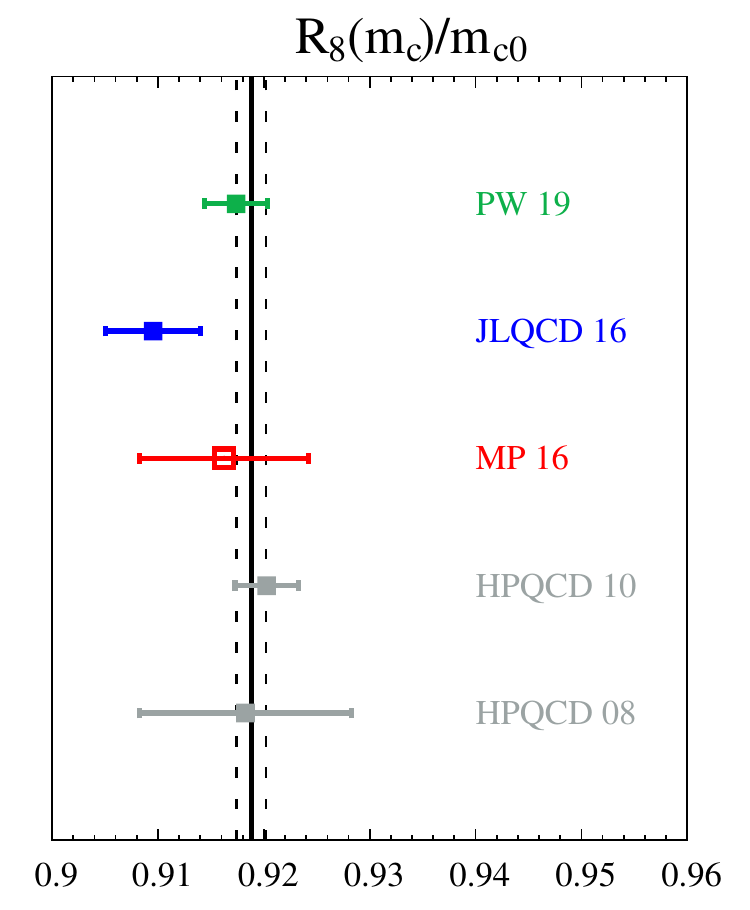}
\includegraphics[width=5.3cm]{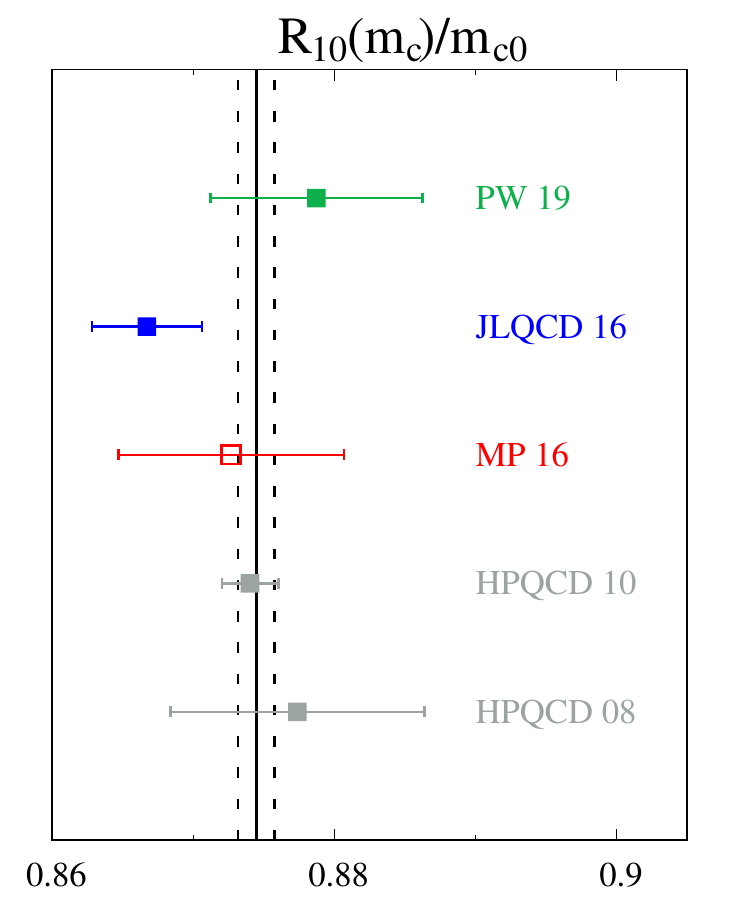}
\end{minipage}
\begin{minipage}[t]{16.5 cm}
\caption{Comparison of different lattice results for $R_6/m_{h0}$ (left), 
$R_8/m_{h0}$ (middle), and $R_{10}/m_{h0}$ (right). 
Shown are the lattice results form HPQCD 08~\cite{Allison:2008xk},
HPQCD 10~\cite{McNeile:2010ji}, JLQCD 16~\cite{Nakayama:2016atf}, 
Maezawa and Petreczky (MP 16)~\cite{Maezawa:2016vgv},
Petreczky and Weber (PW 19)~\cite{Petreczky:2019ozv}. 
The vertical solid and dashed lines show the weighted average and 
the corresponding uncertainty. 
The data corresponding to the open symbols do not enter the averages 
since they are superseded by PW 19.
\label{fig:compR6}
}
\end{minipage}
\end{center}
\end{figure}

In the previous section we discussed the calculations of the moments 
of quarkonium correlators and demonstrated the main features of such 
calculations using the HISQ action with $N_f=2+1$.
We showed that the continuum extrapolation of the lowest moment or the 
ratio of the moments is quite challenging. 
Since the reliable continuum results for the moments are pre-requisite 
for an accurate determination of heavy quark masses and the strong coupling 
constant, it is important to compare different lattice calculations at the 
level of the continuum results for the moments. 
Comparison with perturbation theory and estimates of the corresponding 
systematic errors introduces another level of complications.
In \Figref{fig:compR4} we show the continuum extrapolated results for 
$R_4$, $R_6/R_8$ and $R_8/R_{10}$ from different groups. 
We calculated the weighted average of different lattice results excluding 
MP\,16~\cite{Maezawa:2016vgv}, since it was superseded by 
PW\,19~\cite{Petreczky:2019ozv}. 
The averaging procedure gives $\cdf=1.24$ ($R_4$), $2.26$  ($R_6/R_8$)
and $0.79$ ($R_8/R_{10}$). 
There is some scattering in the data, which together with large $\cdf$ 
may indicate that some errors are underestimated.
In \Figref{fig:compR6} we show a comparison of different lattice results for 
$R_6/m_{h0}$, $R_8/m_{h0}$, and $R_{10}/m_{h0}$.
We also performed averaging of the different lattice results, which 
resulted in $\cdf=2.61,~2.93$ and $2.73$ for $R_6$, $R_8$ and $R_{10}$, 
respectively.
The JLQCD~16 value is lower than the average in all cases. 
This may indicate that there is a problem with the continuum extrapolation 
in the JLQCD~16 analysis, which would not be too surprising, since 
the $a^2$ extrapolation did not work well. 
If we omit the JLQCD~16 results from the average we obtain 
$\cdf=0.81$ ($R_6$), $0.52$ ($R_8$) and $0.53$ ($R_{10}$), which appears 
to be much more reasonable.
Therefore, we will quote the averages for $R_6/m_{h0}$, $R_8/m_{h0}$, and 
$R_{10}/m_{h0}$ excluding the JLQCD~16 result. 
The average continuum values for the moments and their ratios are summarized 
in \Tabref{tab:sumR}.
These can be used to extract the strong coupling constant and heavy quark 
masses in an independent way by other groups.

\begin{table}
\begin{center}
\begin{minipage}[t]{16.5 cm}
\caption{
Lattice averages for $R_4$, $R_6/R_8$, or $R_8/R_{10}$, or 
$R_n/m_{h0}$, $n \ge 6$ at the charm quark mass. 
The errors of the experimental inputs, \ie the scale setting error originating 
in the different experimental scales (here: $f_\pi$ or $m_\Omega$), and the 
mass min-tuning error originating the experimental meson masses, are 
sub-leading among the error sources of the averages.
\label{tab:sumR}}
\end{minipage}
\begin{tabular}{|c|c|c|c|c|c|}
\hline
$R_4$      &  $R_6/R_8$   & $R_8/R_{10}$  & $R_6/m_{h0}$ &  $R_8/m_{h0}$ & $R_{10}/m_{h0}$ \\
\hline
1.2784(27) &  1.10990(81) & 1.04905(50)    & 1.0211(16) &  0.9188(14) & 0.8744(13) \\ 
\hline
\end{tabular}
\end{center}
\end{table}

\section{Heavy quark masses and $\als$ from the moments}

Given the time moments of the quarkonium correlation function and their 
ratios in the continuum limit, one may proceed with the extraction of the 
heavy quark masses and the strong coupling constant in the $\MS$ scheme 
at the next-to-next-to-next-to-leading order. 
\vskip1ex
First of all, one may choose the common renormalization scale 
$\nu=\nu_{m}=m_h$ for the strong coupling and the quark masses 
in \Eqref{eq:Rn pert}, leading to the simplification 
$r_n[\nu/m_h(\nu_m),\als(nu)] \to r_n[\als(m_h)]$ as in 
PW~19~\cite{Petreczky:2019ozv}. 
Then, using the continuum extrapolated HISQ results for $R_4$, $R_6/R_8$, 
or $R_8/R_{10}$ that are given in \Tabref{tab:Rconthisq} in a first step, 
one may solve for the strong coupling constant $\als(m_h,\Nf=3)$ 
in $\Nf=3$ QCD. 
In a second step, one may use this $\als(m_h,\Nf=3)$ and fit 
the higher moments $R_n/m_{h0}$ for $n>4$, which are given in \Tabref{tab:Rconthisq}, 
too. 
These higher moments satisfy the scheme independent relation 
$m_h^{(a)}/m_h^{(b)} = R_n^{(a)}/R_n^{(b)}$ for any schemes $a$ and $b$, 
and thus 
\al{
 {m_h(m_h)} = \frac{r_n[\als(m_h)]}{R_n}~{m_{h0}}.
}
Hence, one can directly calculate the heavy quark mass $m_h(m_h)$ in the 
$\MS$ scheme. 

\begin{figure}[tb]
\begin{center}
\begin{minipage}[t]{16.5 cm}
\centering
\includegraphics[scale=2.0]{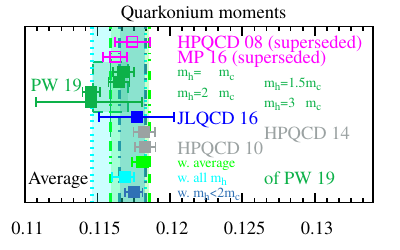}
\end{minipage}
\begin{minipage}[t]{16.5 cm}
\caption{
The results for $\als(M_Z)$ obtained from the time moments of 
quarkonium correlation functions in (2+1)-flavor QCD, \ie 
HPQCD~10~\cite{McNeile:2010ji}), JLQCD~16~\cite{Nakayama:2016atf}, 
Petreczky and Weber (PW~19)~\cite{Petreczky:2019ozv}, 
or in (2+1+1)-flavor QCD from HPQCD~14~\cite{Chakraborty:2014aca}.
The superseded results HPQCD~08~\cite{Allison:2008xk} 
(superseded by HPQCD~10~\cite{McNeile:2010ji}), 
and Maezawa and Petreczky (MP~16)~\cite{Maezawa:2016vgv} (superseded 
by PW~19~\cite{Petreczky:2019ozv}) are shown, too. 
\label{fig:asmom}
}
\end{minipage}
\end{center}
\end{figure}

\vskip1ex
For the first step one could proceed with multiple levels of combined or 
separate fits. 
On the one hand, one might fit either multiple results among the $R_4$, 
$R_6/R_8$, and $R_8/R_{10}$ at multiple valence heavy quark masses in a 
simultaneous fit, which, however has to include (reasonable) assumptions 
about the running of the strong coupling. 
On the other hand, one might use a combined fit of $R_4$, $R_6/R_8$, and 
$R_8/R_{10}$ at each heavy quark mass, or one could even consider these 
results in separate fits. 
The latter provides the advantage of a consistency check of the three 
individual continuum extrapolations, which is the approach used in 
PW~19~\cite{Petreczky:2019ozv}. 
In order to estimate the perturbative truncation error, one may consider 
including a generic higher order term $r_{n4} (\als(m_h)/\pi)^4$ in 
the analysis. 
In order to be conservative its coefficient has been varied between 
$-5\,r_{n3} \le r_{n4} \le +5\,r_{n3}$ in PW~19~\cite{Petreczky:2019ozv}. 
The $r_{nj}$ can be considered as independent for $R_4$, $R_6/R_8$, or 
$R_8/R_{10}$. 
Since the strong coupling constant $\als(m_h)$ becomes much smaller for the 
higher scales, appropriate for higher masses, the perturbative error is 
systematically reduced for the larger heavy quark masses. 
The nonperturbative contribution from the gluon condensate is just added 
as a constant shift, and varied within its error. 
Since it is weighted with the fourth power of the inverse heavy quark mass, 
this uncertainty quickly becomes numerically insignificant for the larger 
heavy quark masses. 
We reproduce the results for $\als(m_h,\Nf=3)$ in the $\MS$ scheme 
from the different possible fits in \Tabref{tab:Rnas}.
\begin{table}
\begin{center}
\begin{minipage}[t]{16.5 cm}
\caption{
The values of $\als(m_h,\Nf=3)$ for different heavy quark masses, $m_h$, 
extracted from $R_4$, $R_6/R_8$, and $R_8/R_{10}$. 
The first, second, and third errors correspond to the lattice error, the 
perturbative truncation error, and the error due to the gluon condensate.
The error of the experimental inputs, \ie the scale setting error originating 
in the experimental error of $f_\pi$, and the mass min-tuning error 
originating the experimental meson masses are subleading contributions to 
the lattice error.
In the fifth column the averaged value of $\als(m_h,\Nf=3)$ is shown (see text).
The last column gives the value of $\lMSbt$ in MeV.
\label{tab:Rnas}}
\end{minipage}
\begin{tabular}{|c|lllll|}
\hline
$\frac{m_h}{m_c}$    & $R_4$              & $R_6/R_8$           &  $R_8/R_{10}$        & av.         & $\Lambda_{\overline{MS}}^{n_f=3}$ MeV \\
\hline
$1.0$ & 0.3815(55)(30)(22) & 0.3837(25)(180)(40) &  0.3550(63)(140)(88) & 0.3782(65)  & 314(10)  \\
$1.5$ & 0.3119(28)(4)(4)   & 0.3073(42)(63)(7)   &  0.2954(75)(60)(17)  & 0.3099(48)  & 310(10) \\
$2.0$ & 0.2651(28)(7)(1)   & 0.2689(26)(35)(2)   &  0.2587(37)(34)(6)   & 0.2648(29)  & 284(8)  \\
$3.0$ & 0.2155(83)(3)(1)   & 0.2338(35)(19)(1)   &  0.2215(367)(17)(1)  & 0.2303(150) & 284(48) \\
\hline
\end{tabular}
\end{center}
\end{table}
On the one hand, the $\als(m_h,\Nf=3)$ value for $m_h=m_c$ or 
$m_h=1.5\,m_c$ from the (2+1)-flavor HISQ result for $R_8/R_{10}$ is slightly 
lower than the corresponding values from $R_4$ or $R_6/R_8$. 
A possible reason might be that the (2+1)-flavor HISQ result for $R_{10}$ is 
slightly higher, than the associated (2+1)-flavor world average, whereas the 
(2+1)-flavor HISQ results for $R_{6}/R_{8}$, $R_{6}/m_{h0}$, and $R_8/m_{h0}$ 
are much closer to their respective averages; see \Figref{fig:compR6}. 
On the other hand, for $m_h=2\,m_c$ the agreement between the three values 
from $R_4$, $R_6/R_8$, or $R_8/R_{10}$ is better. 
Lastly, the $m_h=3\,m_c$ result has quite a large lattice error, which 
appears to be slightly lower $\als(m_h,\Nf=3)$ value from $R_4$. 
The averages in \Tabref{tab:Rnas} can be obtained as weighted averages, since 
the dominant uncertainties, \ie the perturbative truncation errors, can be 
considered as sufficiently independent for the different quantities. 
\vskip1ex
In the second step, one may use the thus obtained $\als(m_h,\Nf=3)$ to 
calculate values of $m_h(m_h)$ for $m_h \le 3\,m_c$ from the different 
reduced moments $R_6/m_{h0}$, $R_8/m_{h0}$, or $R_{10}/m_{h0}$, 
which are generally in very good agreement. 
This is not surprising, since their approach to the continuum limit is quite 
simple. 
We reproduce the results for $m_h(m_h)$ in the $\MS$ scheme 
from the different possible fits in \Tabref{tab:Rnmh}.
\begin{table}
\begin{center}
\begin{minipage}[t]{16.5 cm}
\caption{
The heavy quark masses $m_h(m_h)$ in the $\MS$ scheme in units of GeV 
for the different values of $m_h$.
The first, second, third, and fourth errors correspond to the errors of the 
lattice result, the perturbative truncation error, the error due to the 
gluon condensate, and the error from $\als(m_h,\Nf=3)$, respectively.
The last column shows the average of the masses determined from $R_6/m_{h0}$, 
$R_8/m_{h0}$, and $R_{10}/m_{h0}$.  
This error budget does not include the overall error of the lattice scale, 
$r_1$. 
\label{tab:Rnmh}}
\end{minipage}
\begin{minipage}[t]{16.5 cm}
\begin{tabular}{|l|cccc|}
\hline
$\frac{m_h}{m_c}$     &   $m_h$ from $R_6/m_{h0}$~[GeV]  &       $m_h$ from $R_8/m_{h0}$~[GeV]  &         $m_h$ from $R_{10}/m_{h0}$~[GeV] &   average $m_h$~[GeV]  \\ 
\hline
$1.0$  &  1.2740(25)(17)(11)(61)   &  1.2783(28)(23)(00)(43)   &  1.2700(72)(46)(13)(33)   & 1.2741(42)(29)(8)(46)  \\
$1.5$  &  1.7147(83)(11)(03)(60)   &  1.7204(42)(14)(00)(40)   &  1.7192(35)(29)(04)(30)   & 1.7181(53)(18)(2)(43)  \\
$2.0$  &  2.1412(134)(07)(01)(44)  &  2.1512(71)(10)(00)(29)   &  2.1531(74)(19)(02)(21)   & 2.1481(93)(12)(1)(31)  \\
$3.0$  &  2.9788(175)(06)(00)(319) &  2.9940(156)(08)(00)(201) &  3.0016(170)(16)(00)(143) & 2.9915(167)(10)(0)(220) \\
$4.0$  &  3.7770(284)(06)(00)(109) &  3.7934(159)(08)(00)(68)  &  3.8025(152)(15)(00)(47)  & 3.7910(198)(10)(0)(75)  \\
$\frac{m_b}{m_c}$     &  4.1888(260)(05)(00)(111) &  4.2045(280)(07)(00)(69)  &  4.2023(270)(14)(00)(47)  & 4.1985(270)(9)(0)(76) \\
\hline
\end{tabular}
\end{minipage}
\end{center}
\end{table}
The averages in \Tabref{tab:Rnmh} can be obtained straightforwardly, since 
the agreement is generally better than the errors would suggest. 
\vskip1ex
Combining the results for $\als(m_h)$ and $m_h(m_h)$ at each heavy 
quark mass, one may use the perturbative running to obtain the values of the 
QCD Lambda parameter $\lMSbt$.  
A small systematic error estimate due the difference between the implicit or 
explicit schemes\footnote{The implicit/explicit schemes 
are defined in terms of \mbox{Eqs.~(5) or~(4)} in~\cite{Chetyrkin:2000yt}.
}
has been included. 
Together with the errors of the numerical results for $\als(m_h,\Nf=3)$ and 
$m_h(m_h)$ all errors are added in quadrature to determine the uncertainty 
of $\lMSbt$.  
The central values and errors are reported in the last column of 
\Tabref{tab:Rnas}. 
\vskip1ex
There is some concern that the central value of $\lMSbt$ obtained from the 
data at $m_h=2\,m_c$ is $2.5\sigma$ lower than the central values obtained 
from the lattice data at the smaller heavy quark masses, 
$m_h=m_c$ or $m_h=1.5\,m_c$. 
This could imply that some continuum extrapolations at $m_h=2\,m_c$ 
might suffer from underestimated systematic uncertainties. 
However, given the good consistency between the corresponding results 
in \Tabref{tab:Rnas}, it appears unlikely that multiple among the separate 
continuum extrapolations are flawed in order to produce such a result. 
Moreover, even eliminating the lowest or the lower two results of 
$\als(2\,m_c,\Nf=3)$ does not change the outcome of the analysis in a 
statistically significant manner.
For these reasons, the spread between the different results has to be taken 
as a more conservative estimate for the error than the smaller result from 
the direct error propagation. 
Hence, the (unweighted) average of the results is 
\al{\label{eq:lq mom}
 \lMSbt= 298(16)\,\mr{MeV}. 
}
This result has been used to determine the strong coupling $\als(m_h,\Nf=3)$ 
and the heavy quark masses $m_h(m_h)$ at the higher scales $m_h=4\,m_c$ or 
$m_h=m_b$; see \Tabref{tab:Rnmh}.
\vskip1ex
The determination of the strong coupling constant from the moments is 
strongly affected by the uncertainties in the continuum extrapolation of 
the moments. 
Using the perturbative decoupling of the charm quark at $1.5\,\mr{GeV}$ and 
of the bottom quark at $4.7\,\mr{GeV}$ this average result for $\lMSbt$ can 
be converted to the strong coupling constant at the Z pole, $\nu=M_Z$, 
\al{\label{eq:as mom}
 \als(M_Z,\Nf=5) = 0.1159(12). 
}
In particular, we show the four separate results 
of PW~19~\cite{Petreczky:2019ozv} due to the four values of the QCD Lambda 
parameter obtained with the four different valence heavy quark masses, 
\cf\Tabref{tab:Rnas}, as separate symbols and the average in \Eqref{eq:as mom} 
as a shaded region. 
One might calculate a lattice average using either this average, or the 
individual results obtained with all four different valence heavy quark 
masses, or the results obtained by restricting to the smaller valence heavy 
quark masses $m_h < 2\,m_c$ (the largest valence heavy quark mass 
$m_h=3\,m_c$ has practically no impact on the average). 
For these three averages one obtains 
\al{
 &\als(M_Z) = 0.1180^{+0.0006}_{-0.0006},&& \cdf=\frac{3.35}{3},&
 & \text{ with PW~19 average, \Eqref{eq:as mom}}, 
 \label{eq:as momav1} \\
 &\als(M_Z) = 0.1169^{+0.0006}_{-0.0009},&& \cdf=\frac{22.7}{7},&
 & \text{ with PW~19 all } m_h, 
 \label{eq:as momav2} \\
 &\als(M_Z) = 0.1175^{+0.0006}_{-0.0006},&& \cdf=\frac{5.24}{4},&
 & \text{ with PW~19 } m_h < 2\,m_c, \label{eq:as momav3}  
}
respectively. 
The $\cdf$ is never below $1$, which suggests that the errors in 
at least some of the averaged results are indeed underestimated as 
mentioned earlier on the level of the continuum extrapolated moments. 
The last average in \Eqref{eq:as momav3} excluding the two larger valence 
heavy quark masses of PW~19~\cite{Petreczky:2019ozv} appears to be the most 
consistent with the individual lattice results, and covers 
the central values of the other averages, too. 
We assign the spread between all individual results that enter this average 
as the estimate of the uncertainty and show this as a colored band in 
\Figref{fig:asmom}, namely,
\al{\label{eq:as momav4}  
 &\als^\mr{mom}(M_Z) = 0.1175^{+0.0008}_{-0.0010}&
 & \text{ pre-average from the moments}. 
}

\vskip1ex
On the contrary, the quark masses are in much better agreement between 
the different lattice calculations. 
Here, we reproduce the most recent (2+1)-flavor QCD results with the HISQ 
action, PW~19~\cite{Petreczky:2019ozv}. 
After combining the uncertainty of the lattice scale $r_1$ with the result in 
\Tabref{tab:Rnmh}, we obtain the $\MS$ charm quark mass in the three and four 
flavor theories,
\al{
 m_c(\nu_m=m_c,\Nf=3) &=1.2741(101)\,\mr{GeV}, \\
 m_c(\nu_m=m_c,\Nf=4) &=1.265(10)\,\mr{GeV}, \\
 m_c(\nu_m=3\,\mr{GeV},\Nf=4) &=1.001(16)\,\mr{GeV}. 
}
Finally, we obtain the $\MS$ bottom quark mass in the three and five 
flavor theories,
\al{
 m_b(\nu_m=m_b,\Nf=3) &=4.1985(371)\,\mr{GeV}, \\
 m_b(\nu_m=m_b,\Nf=5) &=4.188(37)\,\mr{GeV}.
}
As previously, the matchings to the four or five flavor theories has 
been performed at $1.5\,\mr{GeV}$ or $4.7\,\mr{GeV}$, respectively. 
\vskip1ex
So far we discussed the determination of $\als$ using reduced moments by 
different lattice groups. 
Very recently the continuum extrapolated lattice results on the reduced 
moments of pseudoscalar correlators have been combined with the reduced 
moments of vector correlation function extracted from the experimental
results on $e^+ e^-$ collisions to obtain $\als$~\cite{Boito:2020lyp} 
(see also \mbox{Ref.}~\cite{Boito:2019pqp} for a related extraction). 
This study used a different methodology to extract the strong coupling 
constant and its uncertainty.
The perturbative error has been estimated by varying the scale $\nu$ in 
$\als(\nu)$ as well as the scale $\nu_{m}$ at which the heavy quark mass 
$m_h(\nu_m)$ is defined contrary to the analysis by the lattice groups. 
This resulted in a perturbative error which is significantly larger.
Since the renormalization scale dependence of the heavy quark masses is 
known to high order it is not clear if one should count the scale dependence 
of the quark masses at fixed order as part of the perturbative error. 
In any case these issues need further studies.

\section{Heavy quark masses from a combined EFT and lattice QCD analysis}
\label{sec:heavylight}
\label{sec:EFT}

Because of confinement, one cannot determine quark masses by isolating them 
from the rest of the world and measuring their rest masses.
For a heavy quark, however, there are physical states, such as a heavy-light 
meson composed of the heavy quark and a light antiquark, with this property 
that the bulk of the meson mass comes from the mass of the heavy quark;
thus, the mass of the heavy quark can be identified at leading order with the 
masses of these physical states.
Investigating the dependence of the mass of such a system on the mass of the 
heavy quark, one can in principle extract the heavy quark mass.
We now discuss how a combination of effective field theories and lattice QCD 
can be used to quantify this dependence and in turn determine the charm- and 
bottom-quark masses.

Heavy quark effective theory (HQET) formula for the mass of a heavy-light 
pseudoscalar meson is~\cite{Falk:1992wt}
\begin{equation}
    M_{H} = m_h + \Lambdabar + \frac{\mu_\pi^2}{2m_h} 
    -  \frac{\mu_G^2(m_h)}{2m_h} + \order\left( m_h^{-2} \right),
   \label{eq:mh_2_MH}
\end{equation}
where $m_h$ is the heavy-quark mass and $\Lambdabar$, $\mu_\pi^2$, and 
$\mu_G^2(m_h)$ are matrix elements of HQET operators with dimension~4 and~5. 
These matrix elements have simple physical meanings.
They correspond, respectively, to the energy of the light quarks and gluons, 
the heavy quark's kinetic energy, and the chromomagnetic energy, which has 
an anomalous dimension and, therefore, has a logarithmic dependence on the 
mass~$m_h$.

A precise determination of the HQET matrix elements is necessary for precise 
determination of a heavy quark mass from \Eqref{eq:mh_2_MH}. 
Lattice QCD makes it possible to compute many quantities including these 
matrix elements.
One approach is to exploit the very \Eqref{eq:mh_2_MH}:
use lattice QCD to compute $M_H$ as a function of~$m_h$
(for many fictitious values of the heavy quark mass) and fit \Eqref{eq:mh_2_MH}
to compute the HQET matrix elements on the right-hand side including higher 
orders in $1/m_h$~\cite{Kronfeld:2000gk}.
When the HQET matrix elements are determined, one can plug them into 
\Eqref{eq:mh_2_MH} along with the physical masses of the $D$ and $B$ systems, 
and extract the masses of the charm and bottom quarks, respectively.

There are several challenges that should be addressed in order to implement 
the above program.
First, as we discuss below, one should give a precise definition for the HQET matrix elements;
this is tied to a precise definition for $m_h$ in Eq.~\eqref{eq:mh_2_MH}.
Second, the simulation masses of quarks on a lattice
must be renormalized in order to be used in expressions such as Eq.~\eqref{eq:mh_2_MH}.
The renormalization can be performed using lattice perturbation theory,
which is in practice very limited and consequently restricts the accuracy,
or using nonperturbative approaches such as the one we explain below,
which is designed to improve the accuracy.
Third, one should include the effects of light quark masses in Eq.~\eqref{eq:mh_2_MH}.
These effects can be incorporated into the analysis using heavy-meson chiral perturbation theory
(HM\chpt)~\cite{Burdman:1992gh,Yan:1992gz,Wise:1992hn},
which is a merger of HQET and \chpt.
Fourth, because the data calculated by lattice simulations 
suffer from lattice artifacts that depend on the employed lattice action and are present at any nonzero lattice spacing,
one cannot simply rely on HQET and \chpt\ to describe how the computed mass of a heavy-light meson on a lattice depends
on the masses of its heavy and light quarks.%
\footnote{Except for electromagnetic effects, we use the term ``quark'' to refer
to both components of a heavy-light meson.}
In many cases the lattice artifacts can be modeled by simple analytic expansions in powers of the lattice spacing.
This might not be a useful practice when light quarks are involved.
In this case, to take the lattice artifacts into account, one can modify HM\chpt\
within the framework of another EFT: the Symanzik effective theory~\cite{Symanzik:1983dc}.

When all these challenges are addressed one obtains an effective formula
to describe the mass of a heavy-light meson in terms of its
quark masses as well as the lattice artifacts related to the employed lattice-QCD action.
The free (mass-independent) parameters of such a formula would consist of several HQET matrix elements,
such as those introduced in Eq.~\eqref{eq:mh_2_MH},
several low energy constants in \chpt,
as well as many parameters that describe the dependence on the lattice spacing $a$,
which disappear at the continuum limit where $a=0$.
Fitting such a formula to lattice-QCD data,
one can extract the free parameters of the function.
Then by setting $a=0$, one passes to the continuum,
where the masses of the charm and bottom quarks can be obtained
at points where, for example, the $D_s$- and $B_s$-meson masses
take their physical values.

In a series of papers~\cite{Kronfeld:2000gk, Komijani:2016jrh, 
Komijani:2017vep, Brambilla:2017mrs, Bazavov:2018omf, Bazavov:2018tvt} 
by members of the Fermilab lattice, MILC, and TUMQCD collaborations,
different aspects of the above program have been discussed and implemented 
leading to a precise determination of the charm- and bottom-quark masses. 
Combining the analysis with a separate determination of ratios of light-quark masses, 
they present one of the most precise determinations of the up-, down-, 
strange-, charm-, and bottom-quark masses~\cite{Bazavov:2018omf}.
In this section we highlight the main features of this program, 
and present the results for the quark masses. 
For details one should consult with Ref.~\cite{Bazavov:2018omf}.

This section is organized as follows.
In Sec.~\ref{sec:EFT:func}, we address the mentioned challenges
and present our formula for describing heavy-light meson masses.
Then, in Sec.~\ref{sec:EFT:latticesetup}, we briefly describe the 
lattice data used in this analysis, including a discussion on tuning 
quark masses and dealing with unequilibrated topological charge.
Finally, in Sec.~\ref{sec:EFT:results}, we summarize the analysis 
and present the quark masses.

\subsection{Relations between heavy-light meson and quark masses}
\label{sec:EFT:func}

In this part, we discuss different aspects of relating heavy-light meson masses to quark masses.
We start with a definition of a quark mass in the continuum that is useful in our analysis.
We then discuss how such a mass can be calculated from the simulation mass of a quark on a lattice.
Equipped with this definition of mass for a heavy quark, we present the EFT description
of the mass of a heavy-light meson simulated on a lattice with staggered quarks.

\subsubsection{Minimal-Renormalon-Subtracted mass}
\label{sec:EFT:func:MRS}

In HQET in general and in Eq.~\eqref{eq:mh_2_MH} in particular,
the meaning of the matrix elements such as $\Lambdabar$
is tied to the definition of the heavy-quark mass $m_h$ used in the expansion.
The natural definition, which was initially assumed when HQET was introduced, is the pole mass of the heavy quark.
The pole mass is infrared finite~\cite{Kronfeld:1998di} and gauge independent~\cite{Kronfeld:1998di,Breckenridge:1994gs} at
every order in perturbation theory, but it is not well-defined when all orders are considered
because at large orders the coefficients grow factorially rendering a divergent series~\cite{Bigi:1994em,Beneke:1994sw}.
This particular pattern of divergence in the pole mass is referred to as a renormalon divergence.
An interpretation of the divergent series is possible via Borel summation up to ambiguities
caused by a series of (renormalon) singularities.
In particular, the leading renormalon introduces an intrinsic ambiguity
of order of QCD Lambda in the Borel sum of the pole mass.
From Eq.~\eqref{eq:mh_2_MH}, one can see that because the meson mass is unambiguous,
this ambiguity must be canceled by another quantity of similar order: $\Lambdabar$.
Similar ambiguities from subleading infrared renormalons must be canceled by
higher-dimension matrix elements such as $\mu_\pi^2$.

Several so-called threshold masses, such as the kinetic mass~\cite{Uraltsev:1996rd,Uraltsev:2001ih},
the renormalon-subtracted (RS) mass~\cite{Pineda:2001zq}, the MSR mass~\cite{Hoang:2008yj},
the 1S mass~\cite{Hoang:1999ye}, and the potential-subtracted mass~\cite{Beneke:1998rk},
have been introduced to handle the renormalon divergence in the pole mass.
These threshold masses are defined by introducing an arbitrary factorization scale
as an infrared cutoff to remove the renormalon from the pole mass.
Instead of the above threshold masses,
here we briefly describe a recently introduced mass, the minimal 
renormalon-subtracted (MRS) mass~\cite{Brambilla:2017mrs},
which is used by the Fermilab lattice, MILC, and TUMQCD collaborations in their analysis.
It is a gauge- and scale-independent mass
with an asymptotic expansion identical to the perturbative pole mass.
Because of the last feature, one can use the MRS mass of a heavy quark for the HQET expansion.
The main advantage of using the MRS mass is that it handles the renormalons
without introducing any factorization scale unlike the other threshold masses listed above.
This simplifies the analysis (with one less scale) and improves the precision of the analysis.

The MRS mass is defined based on an investigation of the large-order behavior
of the pole mass in perturbation theory in Ref.~\cite{Komijani:2017vep}.
Using $r_n$ to denote the coefficients relating the \msb\ mass to the pole mass
and $R_n$ to denote their asymptotic behavior,
the idea behind the MRS mass is to interpret the pure asymptotic part, i.e. $\sum R_n\alpha_s^{n+1}$,
via Borel summation, split the resulting integral into an unambiguous piece
and an ambiguous one, and finally discard the ambiguous piece.
The last step has an interesting meaning in the context of HQET.
As discussed below, it means to sweep the ambiguous piece into
a specific quantity related to the concept of residual mass in HQET~\cite{Falk:1992fm}.
The MRS mass defined in Eq.~(2.24) of Ref.~\cite{Brambilla:2017mrs} is
\begin{equation}
    m_\MRS = \mbar\left(1+\sum_{n=0}^{\infty} \left[r_n-R_n\right] \alpha_s^{n+1}(\mbar) 
        + J_\MRS(\mbar) \right),
    \label{eq:mMRS}
\end{equation}
where $\mbar=m_\msb(m_\msb)$,
and $J_\MRS(\mbar)$, defined in Eqs.~(2.25) and~(2.26) of Ref.~\cite{Brambilla:2017mrs},
is the unambiguous part of the Borel sum of $\sum R_n\alpha_s^{n+1}$.
Note that the $R_n$ depend only on the coefficients of the beta function
up to an overall normalization~\cite{Beneke:1994rs,Komijani:2017vep}.

Although the MRS mass is introduced such that one can in principle handle the effects of all renormalons,
in practice (with four-loop calculations) one needs only to take care of the leading renormalon.
To see that $m_\MRS$ admits a well-behaved perturbative expansion in $\alpha_s$,
even when the possible subleading renormalons are overlooked,
let us briefly discuss the relation between the MRS and $\msb$ masses
in a theory with three massless quarks in the sea ($n_l=3$).
Exploiting four-loop calculations in Ref.~\cite{Marquard:2016dcn}, and
setting the overall normalization of the leading renormalon to $R_0^{(n_l=3)}=0.535$~\cite{Komijani:2017vep},
one obtains~\cite{Brambilla:2017mrs}
\begin{align}
    r_n^{(n_l=3)} &= (0.4244,\   1.0351,\   3.6932,\  17.4358, \ldots),
    \label{eq:EFT:rn3} \\
    R_n^{(n_l=3)} &= (0.5350,\   1.0691,\   3.5966,\  17.4195, \ldots),  
    \label{eq:EFT:Rn3}
\end{align}
for $n=0,1,2,3,\ldots$.
The differences
\begin{equation}
    r_n^{(n_l=3)}-R_n^{(n_l=3)} = (-0.1106,\  -0.0340,\  0.0966,\  0.0162, \ldots)
  \label{eq:EFT:rn-Rn}
\end{equation}
are much smaller than the $r_n$; consequently, the series in powers of $\alpha_s$ in Eq.~\eqref{eq:mMRS}
is a well-behaved series through order $\alpha_s^4$.

There are subtleties in the definition of the MRS mass in presence of massive fermions in the sea.
These subtleties are discussed in detail in Ref.~\cite{Brambilla:2017mrs}.

Let us now discuss the MRS mass in connection to the HQET description
of heavy-light meson masses in Eq.~\eqref{eq:mh_2_MH}.
HQET as an effective theory of QCD can be reformulated in presence
of a residual mass term~\cite{Falk:1992fm}.
With this reformulation, one is allowed to shift the mass of the heavy quark by an amount
of order of QCD Lambda and redefine the HQET matrix elements such as $\Lambdabar$.
In deriving Eq.~\eqref{eq:mMRS},
the ambiguous part of the Borel sum of $\sum R_n\alpha_s^{n+1}$, which itself is of order of QCD Lambda,
is swept from $m_h$ to~$\Lambdabar$ rendering it unambiguous too.
We write $m_{h,\MRS}$ and $\Lambdabar_\MRS$ to denote the unambiguous definitions of $m_h$
and \Lambdabar\ in the MRS scheme. Equation~\eqref{eq:mh_2_MH} then reads
\begin{equation}
    M_{H} = m_{h,\MRS} + \Lambdabar_\MRS + \frac{\mu_\pi^2 - \mu_G^2(m_h)}{2m_{h,\MRS}} + \order\left( m_{h,\MRS}^{-2} \right)
   \label{eq:mhMRS_2_MH}
\end{equation}
in the MRS scheme.
We emphasize that the only difference between Eqs.~\eqref{eq:mh_2_MH} and~\eqref{eq:mhMRS_2_MH}
is that in the later one the definitions of the heavy quark mass and
the matrix element corresponding to the energy of the light quarks and gluons
are fixed to the MRS scheme.
We do not attribute the MRS scheme to the kinetic term $\mu_\pi^2$ nor the chromomagnetic term $\mu_G^2$ in Eq.~\eqref{eq:mhMRS_2_MH}.
The chromomagnetic term is indeed related to the hyperfine splitting in the masses of
the vector and pseudoscalar mesons; therefore, it can be simply estimated from experimental data
up to an uncertainty suppressed by an inverse power of the heavy quark mass.
The kinetic term is expected to have very small intrinsic ambiguity (if not zero)
simply because its corresponding ambiguity in the pole mass is very small.
This can be seen from the fact that in Eq.~\eqref{eq:EFT:rn-Rn}
the coefficients are all small without any sign of divergence through order $\alpha_s^4$.

As an alternative scheme to organize the HQET expansion,
one can consider the so-called principal value (PV) mass~(e.g.,~\cite{Beneke:1994rs,Lee:2003hh,Bali:2003jq}).
A practical and transparent definition for the PV mass is presented in Ref.~\cite{Brambilla:2017mrs} via the MRS mass as 
\begin{align}
   m_\text{PV} &= m_\MRS + c \Lambda_\msb \, , \label{eq:mMRS2mPV}
\end{align}
where the constant $c$ is given in Eq.~(2.30) of Ref.~\cite{Brambilla:2017mrs}.
The scheme conversion is simply a shift by an amount independent of the quark mass.
The shift is indeed small for QCD with three light quarks; $c\Lambda_\msb\approx 0.1$~GeV.
Similarly, we have
\begin{align}
   \Lambdabar_\text{PV}  &= \Lambdabar_\MRS - c \Lambda_\msb  \label{eq:LambdaMRS2PV}
\end{align}
so that Eq.~\eqref{eq:mhMRS_2_MH} remains unchanged with this scheme conversion.
Both the MRS and PV masses can be used in the analysis.
In practice, for QCD with three light quarks,
there is no need to perform a simple shift to the MRS mass to obtain the PV mass.
Moreover, it turns out that the PV mass becomes slightly less precise than the MRS mass
due to uncertainties in the ingredients of $c \Lambda_\msb$ including $\Lambda_\msb$.
As a result this analysis prefers the MRS scheme.

This concludes the derivation and implementation of the MRS scheme in the continuum.
In the context of lattice QCD, one still needs to discuss how to calculate
the MRS mass of a heavy quark from its simulation mass on a lattice.

\subsubsection{Simulation and renormalized masses}
\label{sec:EFT:func:rabbit}

The Fermilab lattice, MILC, and TUMQCD collaborations use a nonperturbative approach to
relate the simulation mass of a heavy quark on a lattice to its MRS mass.
With staggered fermions, they introduce a ``reference quark'' and rewrite $m_{h,\MRS}$ as
\begin{align}
   m_{h,\MRS}  
      &= m_{r,\msb}(\mu) \frac{\mbar_h}{m_{h,\msb}(\mu)} \frac{m_{h,\MRS}}{\mbar_h} \frac{am_h}{am_r}
      \label{eq:rabbit}
\end{align}
with the four factors as follows.
The first factor is the mass of the reference quark in the \msb\ scheme, which can be treated as a fit parameter.
The second is a factor to run the heavy quark mass in the \msb\ scheme from scale $\mu$ to the self-consistent
scale $\mbar_h = m_{h,\msb}(\mbar_h)$
\begin{equation}
    \frac{\mbar_h}{m_{h,\msb}(\mu)} = \frac{C\left(\alpha_\msb(\mbar_h)\right)}{C\left(\alpha_\msb(\mu)\right)},
    \label{eq:EFT:mass-anomalous}
\end{equation}
where with four active flavors~\cite{Baikov:2014qja} 
\begin{equation}
    C(\pi u) = u^{12/25} \left[1 + 1.01413  u + 1.38921 u^2 + 1.09054 u^3  + 5.8304 u^4 + 
        \order(u^5)\right].
    \label{eq:EFT:mass-anomalous_c}
\end{equation}
The coefficient of $u^4$ is obtained from the five-loop results for the beta function~\cite{Baikov:2016tgj}
and the quark-mass anomalous dimension~\cite{Baikov:2014qja}.
The third factor is basically the big parentheses in Eq.~(\ref{eq:mMRS}).
Finally, the fourth factor is the ratio of the simulation masses (in lattice units)
of the heavy quark and the reference quark.

It is worth mentioning that the remnant chiral symmetry of staggered fermions
plays a key role in Eq.~\eqref{eq:rabbit}.
To be more specific, Eq.~\eqref{eq:rabbit} is constructed based on the identity
\begin{equation}
      \frac{m_{h,\msb}(\mu)}{m_{r,\msb}(\mu)} = \frac{am_h}{am_r} + \order(a^2)\, ,
\end{equation}
which holds for staggered fermions.

In Ref.~\cite{Bazavov:2018omf}, the reference mass $am_r$ is set to $0.4$ times the bare mass of the strange quark
obtained from an analysis of light mesons with the so-called $p4s$ method.
Therefore, when the analysis is completed, 
this method yields $m_s$ in addition to the heavy-quark masses $m_c$ and $m_b$.

\subsubsection{Heavy-light mesons in EFTs and construction of a fit function}
\label{sec:EFT:func:construction}

The technique of constructing EFTs is very powerful in tackling problems with
several scales such as problems involving very heavy and/or very light quarks.
HQET, \chpt, and a merger of them HM\chpt\ are among the well-known EFTs
developed to study the low-energy dynamics of QCD.
Our main aim is to use EFTs along with lattice-QCD data to extract quark masses
from heavy-light meson masses.
To this end we could simply use the EFTs that are developed in the continuum to analyze lattice data.
An alternative approach, which we use here, is to develop new EFTs based on symmetries of the lattice action.
This helps us have more control on artifacts of discretization
of the lattice action and be able to extrapolate the lattice results to the continuum in a more
systematic way.

With staggered fermions, the appropriate theory for this analysis is known as heavy-meson,
rooted, all-staggered chiral perturbation theory (HMrAS\chpt)~\cite{Bernard:2013qwa}.
This is a cascade of EFTs describing, for example, how the mass of a heavy-light
meson composed of staggered quarks depends on the simulation quark masses, the lattice spacing,
and the volume of the lattice.
Within the framework of this theory, one-loop corrections to heavy-light meson masses
are calculated in Ref.~\cite{Brambilla:2017mrs}.
We summarize the results in \ref{sec:app:heavylight}.

The one-loop corrections contain effects of flavor, hyperfine, and (staggered) 
taste splittings as well as effects of finite lattice size.
They are needed to describe the next-to-leading order behavior of the lattice data.
They, however, are not enough to fully describe the quark-mass and lattice-spacing dependence.
Therefore, the authors of Ref.~\cite{Bazavov:2018omf} extend the function
containing one-loop corrections by adding
higher-order analytic corrections in powers of light-quark
masses, the lattice spacing, and the heavy quark mass in lattice units,
and also in inverse powers of the heavy-quark mass.
The extended function, which is given in Eq.~(3.26) of Ref.~\cite{Bazavov:2018omf},
contains 67 fit parameters.
Because of lengthy expressions, we do not reproduce them here.
We simply use

\begin{align}
  & M_{H_x}\big(\{m_h, m_x\},\{m'_l,m'_l,m'_s,m'_c\}, \{a,V\}; \{p_1,\cdots,p_{67}\}\big) \label{eq:EFT:M_hx}
\end{align}
to denote this function, which takes as inputs 
the simulation masses of the valence heavy and light quarks (in the first braces),
the simulation masses of four flavors of sea quarks (in the second braces),
and the lattice spacing and volume (in the third braces).
Note that in this analysis the up and down sea-quark masses are
taken equal and denoted by $m'_l$ in Eq.~\eqref{eq:EFT:M_hx}.
The fourth braces in Eq.~\eqref{eq:EFT:M_hx} lists the 67 parameters of $M_{H_x}$.
The main physical parameters correspond to $m_{r,\msb}(\mu)$ in Eq.~\eqref{eq:rabbit},
which is set to 0.4 times the strange quark mass $m_{s,\msb}(2\,\GeV)$,
and the HQET matrix elements in Eq.~\eqref{eq:mhMRS_2_MH}:
\begin{equation}
    p_1 = m_{p4s,\msb}(2\,\GeV), ~p_2 = \Lambdabar_\MRS, ~p_3 = \mu_\pi^2, ~p_4=\mu_G^2(m_b), \cdots ,
    \label{eq:EFT:fit-params}
\end{equation}
where $m_{p4s,\msb} \equiv 0.4\, m_{s,\msb}\,$.

The effective function $M_{H_x}$ can be readily used to perform a combined EFT fit to lattice data of heavy-light meson masses
computed at multiple lattice spacings and various valence- and sea-quark masses.
This determines the fit parameters in Eq.~\eqref{eq:EFT:fit-params},
including the strange quark mass $m_{s,\msb}$,
and enables us to calculate heavy quark masses $m_c$ and $m_b$.
This is discussed further in Sec.~\ref{sec:EFT:results}.

\subsection{Lattice setup and simulations parameters}
\label{sec:EFT:latticesetup}

This analysis uses a data set from 24 ensembles generated by the MILC collaboration~\cite{Bazavov:2010ru,Bazavov:2012xda}
with four flavors of sea quarks using the HISQ action~\cite{Follana:2006rc}.
The data set includes ensembles with six values of lattice spacings ranging from approximately 0.15~fm to 0.03~fm;
ensembles with the light (up-down), strange, and charm sea masses close to their physical values;
and ensembles with the light and strange sea masses heavier or lighter than in nature, respectively.
These ensembles with a wide range of simulation parameters
provide a good control over the continuum extrapolation
and the dependence on sea quark masses.

In order to control the dependence of heavy-light meson masses on their valence-quark masses,
this analysis uses a wide range of masses for valence heavy and light quarks.
On the coarsest ensembles, there are only two different values for the valence
heavy quark: $m_h=m'_c$ and $m_h=0.9m'_c$.
Here $m'_c$ denotes the simulation value of the sea charm-quark mass in each ensemble,
which is slightly different from the physical charm mass $m_c$ because of tuning errors.
On the finest ensembles, there are several data with $0.9m'_c\le m_h\le 5m'_c$.
The analysis uses the data with $am_h<0.9$ to avoid large lattice artifacts.
For every valence heavy quark, several light valence quarks are used with masses $m_l \lesssim m_x \lesssim m_s$.
Altogether 384 data points are used in the main analysis of heavy-light meson masses.

The full description of the lattice setup and simulation parameters
is given in Table~I of Ref.~\cite{Bazavov:2017lyh}.
The procedures for calculating pseudoscalar meson correlators and for finding masses from these correlators are
also described in Refs.~\cite{Bazavov:2017lyh,Bazavov:2014wgs}.
Moreover, the scale setting and tuning the masses of light quarks are performed in Ref.~\cite{Bazavov:2017lyh}
with a two-step procedure that uses the pion decay constant $f_\pi$ to set the overall scale combined
with the so-called $p4s$ method, which yields a simultaneous determination of the lattice spacing $a$
and $m_{p4s}{\,\equiv\,}{0.4\,m_s}$.

In the tuning procedure of the light quark masses, the values of the bare masses corresponding to
the light and strange quarks are obtained from combinations of the physical pion and kaon masses.
Because the gauge-field ensembles omit electromagnetism,
electromagnetic effects should be subtracted (in a specific scheme) from the experimentally measured masses.
In this analysis, the subtraction is performed using the results of Ref.~\cite{Basak:2018yzz},
which presents a lattice computation of the electromagnetic contributions to kaon and pion masses.
As we discuss below, a similar treatment is used for heavy quarks.

We conclude this part by a comment on the topological charge in lattice simulations.
In production of the gauge configurations at very small lattice spacings
a very slow evolution of topological charge has been seen.
This in principle leads to incorrect sampling and causes small finite-volume effects.
In Ref.~\cite{Bernard:2017npd}, the topological-charge evolution in the ensembles
used in this analysis is investigated, and it is discussed how to correct lattice data
for the leading effects of unequilibrated topological charge.
Although the corrections are negligible for masses of heavy-light mesons,
lattice data at the finest ensembles are corrected accordingly,
and uncertainties in the corrections are included in the systematic errors.

\subsection{Quark mass results from EFT fit to lattice data}
\label{sec:EFT:results}

In Sec.~\ref{sec:EFT:func}, we discussed how using a cascade of EFTs we construct
an effective function $M_{H_x}$ with 67 free parameters.
Here, we discuss how we determine the free parameters by fitting $M_{H_x}$ to
the lattice data summarized in Sec.~\ref{sec:EFT:latticesetup}.

As discussed in Sec.~\ref{sec:EFT:latticesetup}, the data set used in this analysis
contains several lattice spacings and various masses for valence and sea quarks.
In particular, a wide range of heavy-quark masses are used from near charm-quark mass to bottom-quark mass.
To avoid large lattice artifacts, the heavy quark masses are restricted to $am_h<0.9$ in the fitting procedure.
The MRS mass of each heavy quark is calculated using Eq.~\eqref{eq:rabbit}.

The calculation of the MRS mass relies on having a precise value for the strong coupling.
This analysis uses
\begin{equation}
    \alpha_\msb(5~\GeV; n_f=4) = 0.2128(25),
    \label{eq:EFT:alpha5GeV}
\end{equation}
which is obtained by HPQCD collaboration~\cite{Chakraborty:2014aca};
this value corresponds to $\alpha_\msb(m_Z; n_f=5)=0.11822(74)$.
The mean value is used in the base fit, and an uncertainty
associated with $\alpha_\msb$ is introduced by varying its value by $1\sigma$.
To run the coupling constant to the scale $\mu$, the QCD beta function at five-loop order accuracy~\cite{Baikov:2016tgj}
is employed and the differential equation is integrated numerically.

The data for the heavy-light meson masses are more precise than the data for scale-setting quantities.
To account for the uncertainties in scale setting quantities,
the so-called penalty trick~\cite{Ball:2009qv} is employed.
With this trick the optimized values for the scale setting quantities are obtained simultaneously in the EFT fit.
Because the base fit uses data at 5 different lattice spacings, 10 additional parameters are required.

Altogether there are 384 lattice data points and 77 parameters in the base fit:
67 parameters in the effective fit function and 10 parameters for optimized values of scale-setting quantities.
The analysis is performed using a constrained fitting procedure \cite{Lepage:2001ym} with prior distributions
of parameters set mainly according to expectations from EFTs and in several cases from external considerations.
(See Ref.~\cite{Bazavov:2018omf} for details.)
The fit returns a correlated $\chi^2_{\text{data}}/\text{dof}=320/307$, giving a $p$~value of $p=0.3$.

\begin{figure}
    \includegraphics[width=0.48\textwidth]{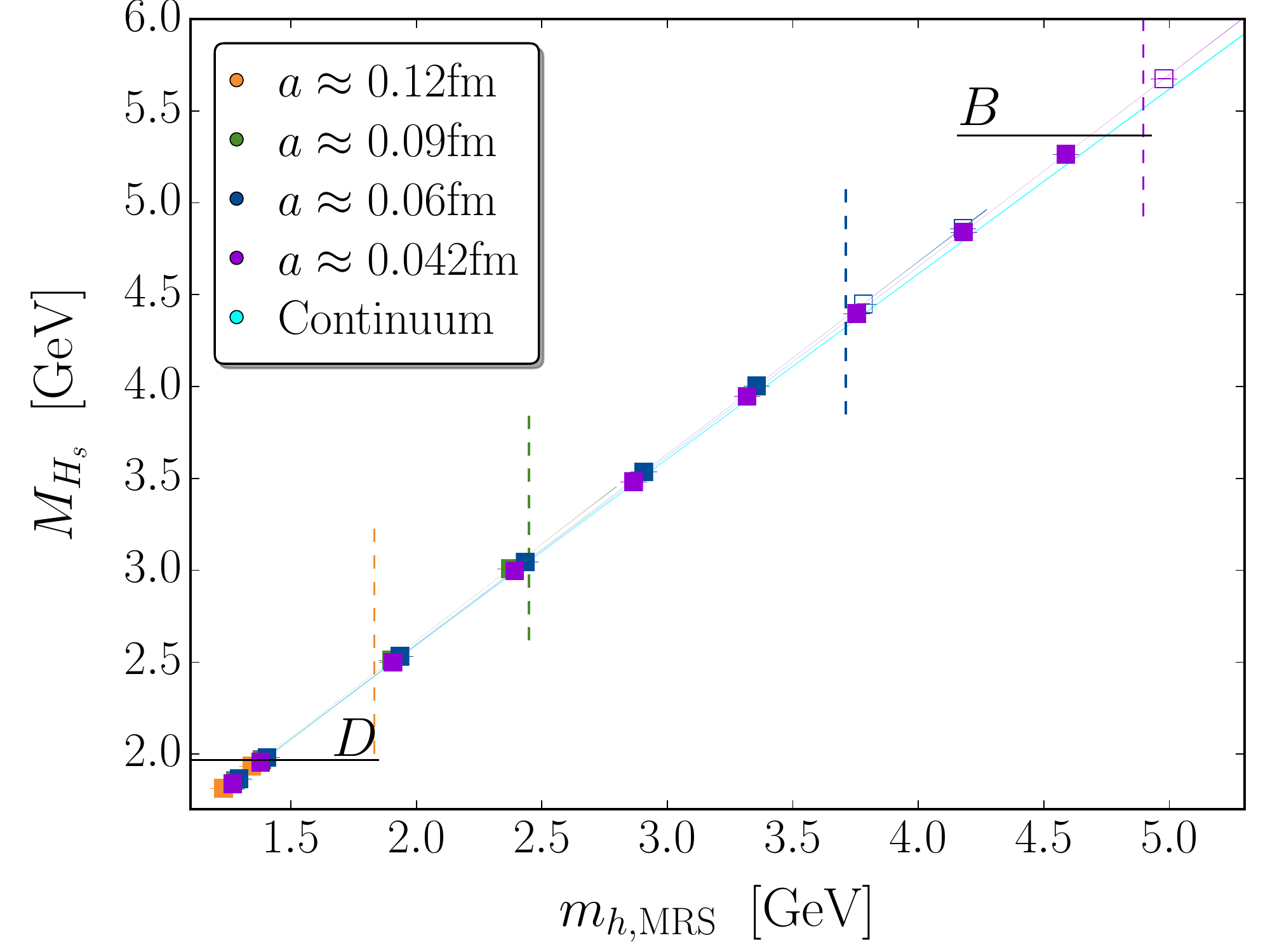} \hfill
    \includegraphics[width=0.48\textwidth]{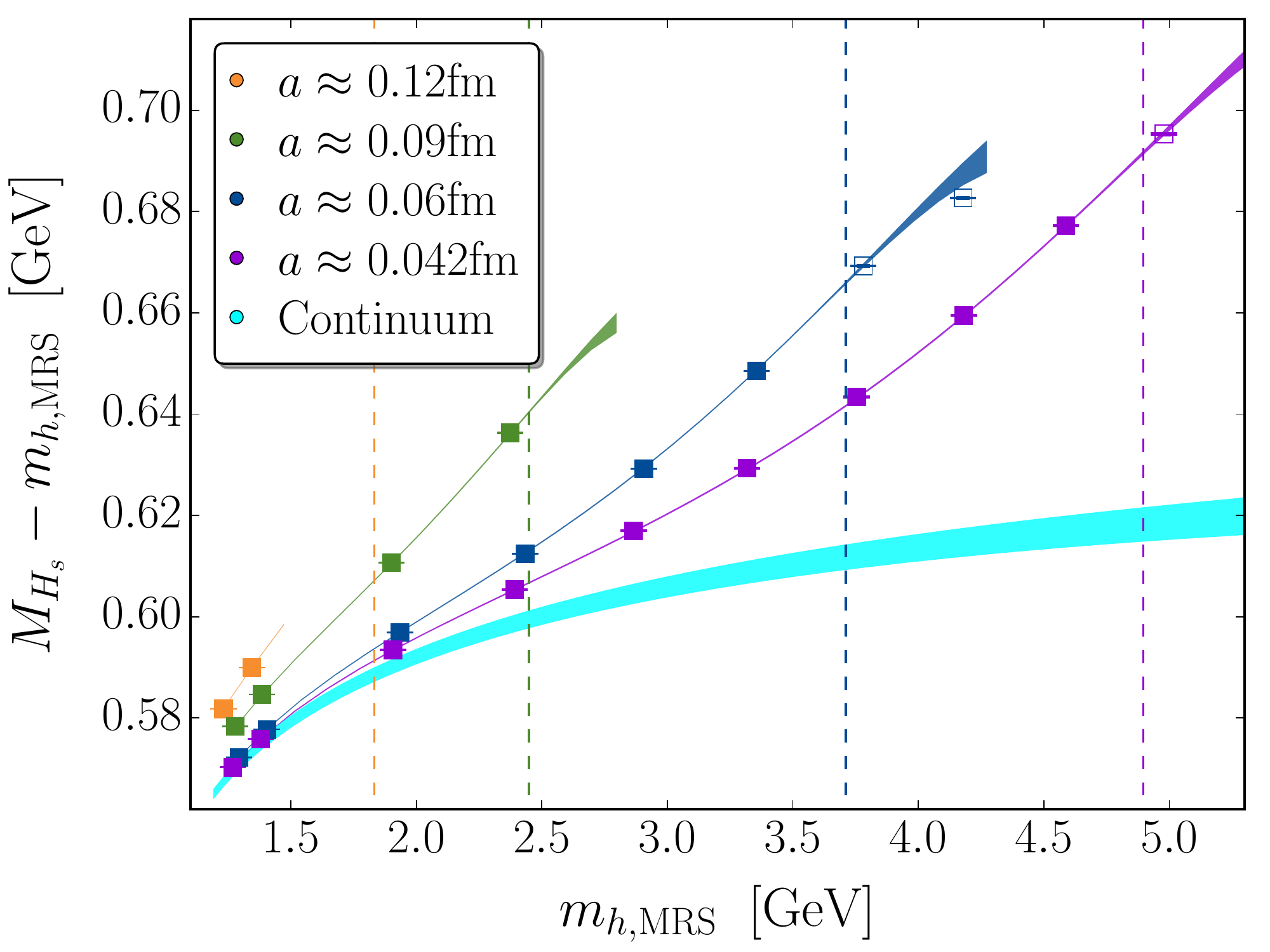}
    \caption{A snapshot of the base fit and the lattice data for heavy-strange meson masses.
	Only ensembles with physical light sea mass are shown.
        Left:~heavy-strange meson mass vs.\ heavy-quark MRS mass.
        Right:~difference of the heavy-strange meson mass and the heavy-quark MRS mass vs.\ heavy-quark MRS mass.
        The dashed vertical lines indicate the cut $am_h = 0.9$ for each lattice spacing,
        and data points with open symbols to the right of them are not included in the fit.
        Here $m_{h,\MRS}$ is the continuum limit of the MRS mass of the heavy quark~$h$,
        and its error bar is suppressed for clarity. From Ref.~\cite{Bazavov:2018omf}.}
    \label{fig:mass-vs-mMRS}
\end{figure}
Figure~\ref{fig:mass-vs-mMRS} illustrates a snapshot of the fit
for the physical mass ensembles at four lattice spacings
as well as the continuum extrapolation of the fit function.
The valence light mass $m_x$ is tuned to $m_s$; the graphs illustrate heavy-strange meson masses.
The meson mass or the difference of the meson mass and the $h$-quark MRS
mass are plotted versus the continuum limit of the $h$-quark MRS mass in Fig.~\ref{fig:mass-vs-mMRS}.
Data points with open symbols to the right of the dashed vertical line of the corresponding color in Fig.~\ref{fig:mass-vs-mMRS}
are not included in the fit because they have $am_h>0.9$.
At nonzero lattice spacing the masses of the sea quarks are set to their simulation values,
while in the continuum extrapolation the masses of sea quarks are tuned to the physical quark masses $m_l$, $m_s$ and
$m_c$.

The widths of the fit lines in Fig.~\ref{fig:mass-vs-mMRS} correspond to a part of the statistical error coming from the fit.
They do not include the statistical errors coming from scale setting and tuning light quark masses.
Moreover, they are sensitive to numerical errors in computing the fit parameters' covariance matrix.
A jackknife procedure, with 20 jackknife resamples, is used for a robust determination of the total statistical error of each output quantity.

The fit determines the 67 free parameters of $M_{H_x}$ including
$p_1 = m_{p4s,\msb}(2\,\GeV)$ and $p_2 = \Lambdabar_\MRS$.
Moreover, the fit function evaluated at zero lattice spacing, infinite volume and physical sea-quark masses, i.e.
\begin{align}
  & M_{H_x}\big(\{m_h, m_x\},\{m_l,m_l,m_s,m_c\}, \{0,\infty\}; \{p_1,\cdots,p_{67}\}\big)\, ,
  \label{eq:EFT:M_hx:cont}
\end{align}
yields the meson masses as a function of the valence heavy and light quark masses.
We now define the physical charm and bottom quarks
such that the $D_s$- and $B_s$-meson masses take their physical values,
albeit after subtracting the electromagnetic effects from the meson masses.
We use the phenomenological formula
\begin{equation}
     M^\text{expt}_{H_x} = M^\text{QCD}_{H_x} + A e_x e_h + B e_x^2,
     \label{eq:EFT:EM-model}
\end{equation}
to relate the experimental masses to the ``pure-QCD'' masses.
In Eq.~\eqref{eq:EFT:EM-model}, $e_h$ and $e_x$ are charges of the valence heavy and light
components, respectively, and the coefficients are $A=4.44$~MeV and $B = 2.4$~ MeV~\cite{Bazavov:2017lyh}.
For details of calculation one should consult Refs.~\cite{Bazavov:2018omf,Bazavov:2017lyh}.
Using the experimental meson masses $M^\text{expt}_{D_s}=1968.27(10)$~MeV and
$M^\text{expt}_{B_s}=5366.82(22)$~MeV~\cite{Olive:2016xmw},
Eq.~\eqref{eq:EFT:EM-model} yields the pure-QCD masses: $M^\text{QCD}_{D_s}=1967.01$~MeV and
$M^\text{QCD}_{B_s}=5367.04$~MeV.

\begin{figure}
    \includegraphics[width=1\textwidth]{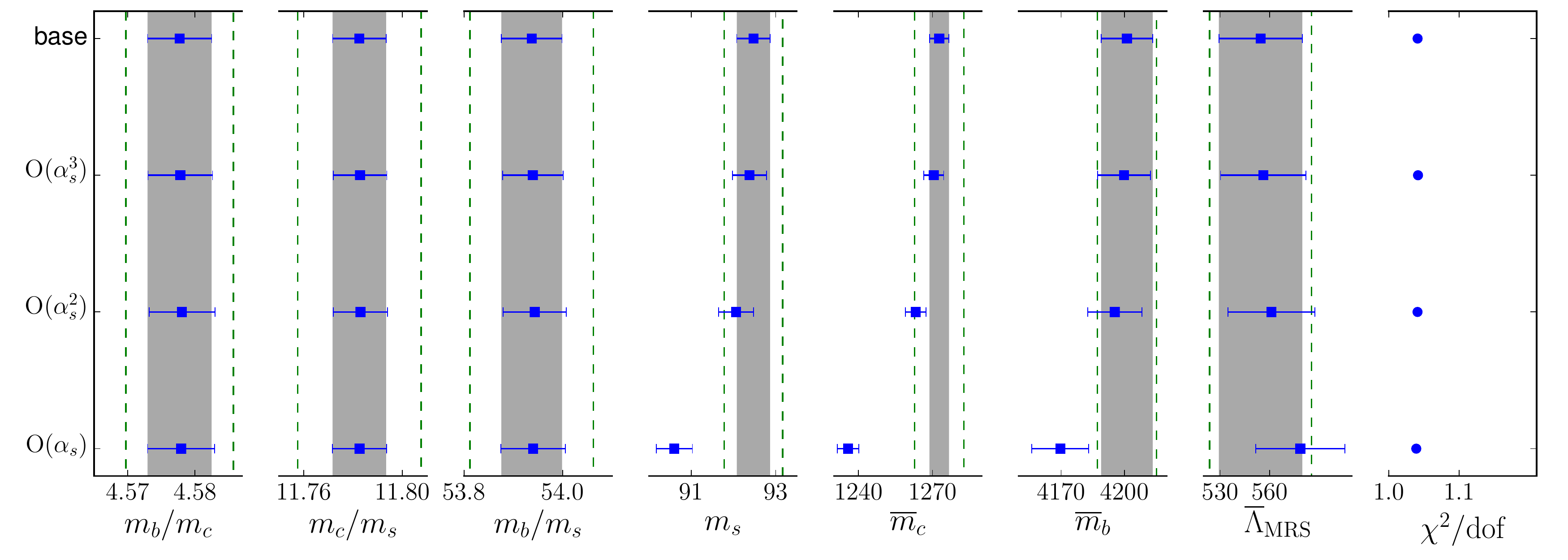}
    \caption{Stability plot showing the sensitivity to truncation error in perturbative-QCD relations.
        The perturbative series in the base fit are accurate through order~$\alpha_s^4$.
        Here the strange-quark mass is given at 2~GeV, and the charm- and bottom-quark masses given
        in their the self-consistent scales.
        The error bars show only the statistical errors, the gray error bands correspond to the statistical error of the base fit,
        and the dashed green lines correspond to total errors. From Ref.~\cite{Bazavov:2018omf}.}
   \label{fig:stability_loop}
\end{figure}

Before reporting the final results for quark masses,
we discuss the effects from truncating perturbative QCD
in the relation between quark-mass definitions and the beta function.
The analysis uses the fit parameter $m_{p4s,\msb}(2~\GeV)$
and $am_h/am_{p4s}$ to calculate $m_{h,\msb}(2~\GeV)$ for each heavy quark.
Then Eqs.~\eqref{eq:EFT:mass-anomalous} and~\eqref{eq:EFT:mass-anomalous_c} are used to calculate $\mbar_h$
and Eq.~\eqref{eq:mMRS} to calculate $m_{h,\MRS}$.
For these calculations we rely on known perturbative QCD results:
the beta function and quark-mass anomalous dimension at five loops~\cite{Baikov:2014qja,Baikov:2016tgj}
and the pole mass at four loops~\cite{Marquard:2015qpa,Marquard:2016dcn}.
By rerunning the analysis with fewer orders
in Eqs.~\eqref{eq:EFT:mass-anomalous_c} and~\eqref{eq:mMRS} and in the beta function
one can monitor the  truncation errors in perturbative QCD.
Figure~\ref{fig:stability_loop} shows the stability of the results as the order of perturbation theory is increased;
$\order(\alpha_s^n)$ denotes a fit that includes $n$ orders
beyond the leading terms in Eqs.~(\ref{eq:EFT:mass-anomalous_c}), (\ref{eq:mMRS}), and the beta function.
The quark mass ratios are not sensitive to the truncations in the perturbative-QCD relations.
This is expected because these ratios are essentially the continuum limit of the corresponding bare masses;
they do not need to get renormalized.
The masses of the strange, charm, and bottom quarks as well as the HQET matrix element $\Lambdabar_\MRS$
show good convergence as the order of $\alpha_s$ in the perturbative expressions is increased.
As a result there is no need to introduce any additional systematic error associated
with truncation in perturbative-QCD results. The fact that the truncation effects are negligible
is a consequence of using the MRS mass in the analysis
because the renormalon-subtracted perturbative coefficients in the MRS mass are all very small;
see Eq.~\eqref{eq:EFT:rn-Rn}.

We now present the final results of this analysis. As discussed above,
the first factor in Eq.~\eqref{eq:rabbit} is the mass of the reference quark, 
which is set to 0.4 times the mass of the strange quark (denoted by $m_{p4s}$)
and treated as a fit parameter. The analysis yields
\begin{equation}
    m_{s,\msb}(2~\GeV) = 92.47 (39)_\text{stat} (18)_\text{syst} (52)_{\alpha_s} (11)_{\fpiPDG}~\MeV .
    \label{eq:EFT:ms_MS}
\end{equation}
There are four uncertainties: statistical,
which includes uncertainties from statistics and EFT fit,
systematic, discussed below, the uncertainty from the parametric input $\alpha_s$,
given in Eq.~\eqref{eq:EFT:alpha5GeV},
and the uncertainty in the value of $f_\pi$ used for scale setting.
Note that the $\alpha_s$ uncertainty is the largest uncertainty in Eq.~\eqref{eq:EFT:ms_MS}
where the strange quark mass is reported at 2~GeV.
When running the strange quark mass, using Eqs.~\eqref{eq:EFT:mass-anomalous} and \eqref{eq:EFT:mass-anomalous_c},
from 2~GeV to higher scales, it turns out that the $\alpha_s$ uncertainty shrinks
much faster than the other sources of uncertainties because of the obvious correlation
with the value of $\alpha_s$ used in the running procedure.
This is discussed further below.

We now solve $M_{H_x}$ in Eq.~\eqref{eq:EFT:M_hx:cont} to obtain the points that give the pure-QCD masses
$M^\text{QCD}_{D_s}=1967.01$~MeV and $M^\text{QCD}_{B_s}=5367.04$~MeV.
This yields the mass ratios
\begin{align}
    m_c/m_s  &= 11.783 (11)_\text{stat} (21)_\text{syst} (00)_{\alpha_s} (08)_{\fpiPDG} ,
    \label{eq:EFT:mc_ms}\\
    m_b/m_s  &= 53.94   (6)_\text{stat} (10)_\text{syst}  (1)_{\alpha_s}  (5)_{\fpiPDG} .
    \label{eq:EFT:mb_ms}
\end{align}
Multiplying the strange quark mass in Eq.~\eqref{eq:EFT:ms_MS}
and the quark mass ratios in Eqs.~\eqref{eq:EFT:mc_ms} and~\eqref{eq:EFT:mb_ms}
then gives the charm- and bottom-quark masses in the \msb\ at scale 2~GeV.
For the heavy quarks, it is customary to present the quark masses at the self-consistent
scale $\mbar_h = m_{h,\msb}(m_{h,\msb})$, which can be iteratively calculated
using Eqs.~\eqref{eq:EFT:mass-anomalous} and~\eqref{eq:EFT:mass-anomalous_c}.
For the charm quark, we then obtain
\begin{align}
    \mbar_c  &= 1273 (4)_\text{stat} (1)_\text{syst} (10)_{\alpha_s}  (0)_{\fpiPDG}~\MeV .
    \label{eq:EFT:mc_SI}
\end{align}
For the bottom quark, because the lattice simulations used in this analysis contain
only the effects of four flavors in the sea, we should adjust the mass
to include effects of the bottom quark in the sea.
The adjustment can be calculated perturbatively by matching a theory
with four active quarks in the sea to a theory with five active quarks.
Specifically, we use~\cite{Liu:2015fxa}
\begin{align}
    m_b^{(n_l)}(\mu) &= \mbar_b^{(n_f)} \left[1 + 0.2060 \left(\frac{\alpha_s^{(n_f)}(\mu)}{\pi}\right)^2
        + (1.8476 + 0.0247 n_l) \left(\frac{\alpha_s^{(n_f)}(\mu)}{\pi}\right)^3 
    \right. \nonumber \\ \hspace{3em} & \left.
        + (6.850 - 1.466 n_l+0.05616 n_l^2)\left(\frac{\alpha_s^{(n_f)}(\mu)}{\pi}\right)^4 + \cdots \right] ,
        \nonumber 
\end{align}
where $n_l=n_f-1$ and $\mu=\mbar_b^{(n_f)}$,
and we obtain
\begin{align}
    \mbar_b^{(n_f=5)} &= 4195 (12)_\text{stat} (1)_\text{syst}  (8)_{\alpha_s}  (1)_{\fpiPDG}~\MeV 
    \label{eq:EFT:mb_SI_nf5}
\end{align}
in a theory with five active quarks in the sea.

In addition to the strange-, charm- and bottom-quark masses,
Ref.~\cite{Bazavov:2018omf} presents the up and down quark masses by combining their analysis
with a separate, but correlated, analysis of light mesons from Ref.~\cite{Bazavov:2017lyh}.
The light quark masses read
\begin{align}
    m_{l,\msb}(2~\GeV) &= 3.402 (15)_\text{stat} (05)_\text{syst} (19)_{\alpha_s} (04)_{\fpiPDG}~\MeV ,
    \label{eq:EFT:ml_MS} \\
    m_{u,\msb}(2~\GeV) &= 2.130 (18)_\text{stat} (35)_\text{syst} (12)_{\alpha_s} (03)_{\fpiPDG}~\MeV ,
    \label{eq:EFT:mu_MS} \\
    m_{d,\msb}(2~\GeV) &= 4.675 (30)_\text{stat} (39)_\text{syst} (26)_{\alpha_s} (06)_{\fpiPDG}~\MeV ,
    \label{eq:EFT:md_MS}
\end{align}
where $m_l$ is the average of the up- and down-quark masses.

This analysis directly yields results for the HQET matrix elements.
It is worthwhile to present the HQET matrix element $\Lambdabar$ in the MRS scheme:
\begin{equation}
    \Lambdabar_\MRS   = 555  (25)_\text{stat}  (8)_\text{syst} (16)_{\alpha_s}  (1)_{\fpiPDG}~\MeV .
    \label{eq:Lambadbar_MRS}
\end{equation}

In addition to computing the quark masses, mass ratios,
and HQET matrix elements, the fit function can be used to investigate flavor splittings
in the $D$- and $B$-meson systems as well as their SU(2) and SU(3) chiral limits.
We refer the readers to Ref.~\cite{Bazavov:2018omf} for the results and end this section with a few remarks.

A brief discussion on different sources of uncertainties in quark masses may be in order.
The systematic error of each quantity includes systematic effects that are not captured in the
EFT fit to lattice data.
These are systematic uncertainties in scale setting quantities, light quark mass tuning,
finite volume effects, topological charge distribution, the excited-state contamination
in two-point correlator fits, and finally electromagnetic effects.
For the full descriptions of the systematic error the reader should see Ref.~\cite{Bazavov:2018omf}.
In particular, Table~I in Ref.~\cite{Bazavov:2018omf} tabulates the contribution
of each source of uncertainty.
One can see that these systematic effects are relatively small for quark masses
especially for the heavy quark masses. They, however, get enhanced in ratios of quark masses;
see Eqs.~\eqref{eq:EFT:mc_ms} and~\eqref{eq:EFT:mb_ms}.
This is due to the fact that the statistical, $\alpha_s$,
and $\fpiPDG$ uncertainties in the quark masses are highly correlated
and thus they cancel in the quark mass ratios while this is not the case for the systematic errors.

The $\alpha_s$ uncertainty in the results for quark masses
in Eqs.~\eqref{eq:EFT:ms_MS} and~\eqref{eq:EFT:mc_SI}--\eqref{eq:EFT:md_MS}
is one of the largest uncertainties.
As mentioned above, when we run the quark masses to higher scales,
using \eqref{eq:EFT:mass-anomalous} and \eqref{eq:EFT:mass-anomalous_c},
the error bars corresponding to the $\alpha_s$ uncertainty shrinks faster than the other sources of uncertainties.
This happens because there is $100\%$ correlation between the $\alpha_s$ uncertainty
in the quark masses and the uncertainty in the value of $\alpha_s$ given in Eq.~\eqref{eq:EFT:alpha5GeV},
which is used for running the quark masses.
For convenience, we also present the up-, down-, strange-, and charm-quark masses
at 3~GeV and the bottom-quark mass at 10~GeV:
\begin{align}
  m_{l,\msb}(3~\GeV)  &= 3.072  (13)_\text{stat}  (04)_\text{syst}  (10)_{\alpha_s}  (04)_{\fpiPDG}~\MeV , \\
  m_{u,\msb}(3~\GeV)  &= 1.923  (16)_\text{stat}  (32)_\text{syst}  (06)_{\alpha_s}  (02)_{\fpiPDG}~\MeV , \\
  m_{d,\msb}(3~\GeV)  &= 4.221  (27)_\text{stat}  (35)_\text{syst}  (14)_{\alpha_s}  (05)_{\fpiPDG}~\MeV , \\
  m_{s,\msb}(3~\GeV)  &= 83.49  (36)_\text{stat}  (16)_\text{syst}  (28)_{\alpha_s}  (10)_{\fpiPDG}~\MeV , \\
  m_{c,\msb}(3~\GeV)  &= 983.7  (4.3)_\text{stat} (1.4)_\text{syst} (3.3)_{\alpha_s} (0.5)_{\fpiPDG}~\MeV , \\
  m_{b,\msb}(10~\GeV; n_f=5) &= 3665 (11)_\text{stat} (1)_\text{syst} (1)_{\alpha_s} (1)_{\fpiPDG}~\MeV .
\end{align}
Remarkably, the $\alpha_s$ uncertainty becomes negligible at 10~GeV
and higher scales relevant to the physics beyond the Standard Model.

The results presented here show that
the combined EFT and lattice QCD program developed in Refs.~\cite{Brambilla:2017mrs,Bazavov:2018omf}
and highlighted here is both qualitatively and quantitatively successful.
The qualitative success relies on the clean separation of scales provided by HQET
with the MRS definition of the heavy-quark mass.
The quantitative success relies on the high statistics of the MILC collaboration's HISQ
ensembles~\cite{Bazavov:2010ru,Bazavov:2012xda},
as well as the availability of high order coefficients in perturbation theory:
the order-$\alpha_s^5$ coefficient for the running of the
quark mass~\cite{Baikov:2014qja} and strong coupling~\cite{Baikov:2016tgj},
and the order-$\alpha_s^4$ coefficient for relating the \msb\ mass to the pole mass
and, consequently, the MRS mass~\cite{Marquard:2015qpa,Marquard:2016dcn}.


As the concluding remark, note that
this program can be considered, on the one hand, as an application of EFTs in extracting
real-world results (quark masses) from lattice simulations and, on the other hand,
as an application of lattice simulations in computing parameters of EFTs (such as $\Lambdabar_\MRS$).
For a review on different aspects of connection between EFTs and lattice QCD see, for example, Ref.~\cite{Bernard:2015wda}.

\section{Other determination of $\als$}
\label{sec:otherals}
In the previous sections we discussed $\als$ determinations using the static 
quark-antiquark energy and moments of quarkonium correlators. 
In this section we will briefly review other lattice determinations of the 
strong coupling constant. 
We will only consider $\als$ determinations that are based on $2+1$ flavor
and $2+1+1$ flavor calculations since these are the only phenomenologically 
relevant ones.
We note, however, that determinations of the strong coupling constant for 
smaller number of dynamical quark flavors, including no dynamical quarks, 
are interesting for understanding of systematic effects in the 
underlying calculations. 
These are discussed in detail in the FLAG review~\cite{Aoki:2019cca}.

In the following subsection we will discuss $\als$ determinations from 
step scaling analysis, small Wilson loops, hadronic vacuum polarization, 
QCD vertices and eigenvalues of the Dirac operators. 
We will present average values of $\als(M_Z)$ obtained from each of these
methods.

\subsection{$\als$ determination from step scaling}

The step scaling method allows to calculate $\als$ at large energy scales 
while avoiding the window problem. 
In this method one defines a coupling $g_S^2(1/L)$, where $L$ is the size 
of the lattice with the specified boundary conditions. 
The boundary conditions here define the scheme for the coupling constant. 
Then one decreases the box size, while keeping $L \gg a$ and follows the 
running of the coupling constant. 
For sufficiently small $L$ the coupling constant $g_S^2(1/L)$ can be related 
to the coupling constant in $\MS$ scheme,
$g_{\MS}^2(1/L)=g_S^2(1/L) (1+c^{1} g_S^2(1/L)+c^{2} g_S^4(1/L)+\ldots$.
One first starts at some $g_S^2(1/L_\mr{max})$ with $L_\mr{max} \sim 1$ fm. 
Using lattice calculations in large volumes and the same lattice action one 
determines $L_\mr{max}$ in the continuum limit for example by
performing calculations of $L_\mr{max}/r_0$ at several lattice spacings and 
then extrapolating to the continuum, $a \to 0$ limit.
Then one studies the change of the coupling constant as $\nu$ changes from $1/L_\mr{max}$ to $s^n/L_\mr{max}$ for
$n=2,~,3,~4,\ldots$ until $\nu$ reaches a large enough energy 
scale, where the resulting coupling can be converted to $\MS$ scheme.

In practice Schr\"odinger functional (SF) boundary conditions are used 
because the relation of the coupling constant in this scheme to the $\MS$ coupling is known to two-loop order.
In the SF scheme Dirichlet boundary conditions are imposed 
in the time direction~\cite{Luscher:1992an}
\begin{equation}
A_k(x)|_{x_0=0}=C_k,~A_k(x)|_{x_0=L} = C_k', ~k=1,2,3.
\end{equation}
Periodic boundary conditions are imposed on gluon and quark fields. 
Here $C_k$ and $C_k'$ are diagonal $3 \times 3$ matrices that depend on a 
real dimensionless parameter $\eta$ \cite{Luscher:1992an}.
The coupling constant in this scheme is defined in terms of the derivative 
of the effective action 
\begin{equation}
\partial_{\eta} \langle S \rangle |_{\eta=0}=\frac{12 \pi}{g_{SF}^2}.
\end{equation}
One could also use the gradient flow coupling for the step 
scaling method~\cite{Narayanan:2006rf, Fodor:2012td, Fritzsch:2013je}.
The first step scaling analysis in 2+1+1 flavor QCD was done by the PACS collaboration using the SF scheme~\cite{Aoki:2009tf}. 
They obtained $\als=0.1183$ \cite{Aoki:2009tf}. 
More recently the ALPHA collaboration determined $\als$ using a step scaling analysis of the gradient flow coupling and the 
SF coupling~\cite{Bruno:2017gxd}. 
For $2\,\mr{GeV} < \nu < 4\,\mr{GeV}$ the gradient flow coupling was used. 
They switched to SF coupling at $4$ GeV and determined the running of the 
SF coupling until $\nu=70$ GeV, where the two-loop relation 
to the $\MS$ coupling is sufficiently accurate. 
The use of the gradient flow coupling allowed the ALPHA collaboration to 
significantly reduce the statistical errors at small values of $\nu$ and 
obtain a precise determination~\cite{Bruno:2017gxd}
\begin{equation}
\als(M_Z)=0.11852(84).
\end{equation}
This result agrees well with the PACS determination but has much smaller 
error. 
Therefore we will use it as the $\als$ value from the step scaling method.
The experimental input of this result is the scale 
$f_{\pi K}=(2f_K+f_\pi)/3=147.6(5)\,\mr{MeV}$, which is sub-leading in 
the error budget.

\subsection{$\als$ from small Wilson loops}

One could try to determine the strong coupling constant from short distance 
quantities calculated on the lattice. 
In this case the relevant scale is $\mu \sim 1/a$. 
Examples of such short distance quantities are the small Wilson loops or quantities derived from them, \eg Creutz ratios. 
If the lattice spacing is sufficiently small the quantities can be calculated 
in lattice perturbation theory. 
However, it is well known that the convergence of lattice perturbation theory 
is poor. 
This problem can be cured if the perturbative series are expressed in terms 
of some renormalized coupling like $\alpha_V$~\cite{Lepage:1992xa}. 
For quantities $X$ obtained from small Wilson loops the perturbative series 
can be written as 
\begin{equation}
X=\sum_n c_n^X \alpha_V(q^*) .
\end{equation}
The coefficients $c_n^X$ are known through $n=3$ and the scale $q^*=d^X/a$ \cite{Mason:2005zx}, 
where $d^X$ slightly depends on the chosen quantity \cite{Hornbostel:2002af}, but
typically $d^X \sim \pi$. 
Comparing results from lattice QCD calculations with the above perturbative 
expression one obtains $\alpha_V(q^*)$, which then can be converted to $\als$ 
in $\MS$ scheme. 
In this approach there is no problem with the continuum extrapolations since 
the comparison to perturbation theory is performed at finite lattice spacing. 
The reliability of this method is limited by the accuracy of the perturbative 
result and the smallness of the lattice spacing. 
Typically relatively large scales can be achieved since $q^*$ is larger than 
the inverse lattice spacing. 

The strong coupling constant has been determined by the HPQCD 
collaboration~\cite{Mason:2005zx, Davies:2008sw, McNeile:2010ji}
as well as by Maltman et al~\cite{Maltman:2008bx} using this approach. 
The lattice data sets entering the above analyses largely overlap. 
The analysis of \mbox{Ref.}~\cite{McNeile:2010ji} results in 
$\als(M_Z)=0.1184(6)$ and supersedes the previous HPQCD
determinations since it uses finer lattices. 
Maltman et al use a different analysis strategy of the perturbative expansion 
and the condensates. 
They also use a subset of the data from \mbox{Ref.}~\cite{Davies:2008sw}
and obtain $\als(M_Z)=0.1192(11)$~\cite{Maltman:2008bx}. 
Performing the weighted average of the above values we estimate the central 
value and error of $\als(M_Z)$ to be $0.1185^{+0.0006}_{-0.0005}$. 
The uncertainty of the experimental input, $f_\pi$, which contributes to the 
uncertainty of the lattice scale $r_1$, is sub-leading in the error budget in 
both studies.

\subsection{The strong coupling constant from hadronic vacuum polarization}

The idea behind this approach is similar to the one based on the moments 
of quarkonium correlation functions but uses the vector or axial-vector 
current of light quarks. 
Let us consider the correlator 
\begin{equation}
\Pi_{\mu \nu}(q)=\int d^4 x e^{iq x} \langle j_{\mu}(x) j_{\nu}^{\dagger}(0) \rangle,
\end{equation}
where $j_{\mu}$ is the non-singlet vector or axial vector current. 
The correlator can be decomposed into transverse and longitudinal components 
\begin{equation}
\Pi_{\mu \nu}(q)=(q^2 g_{\mu \nu}-q_{\mu} q_{\nu}) \Pi^1(q^2)-q_{\mu} q_{\nu} \Pi^0(q^2).
\end{equation}
In the isospin symmetric limit $\Pi_{\mu \nu}(q)$ is purely transverse, 
\ie $\Pi^0=0$.
For Euclidean momenta $Q^2=-q^2>0$, $\Pi^1$ can be calculated in terms of OPE. 
Here the large scale is given by the momentum transfer not by the quark masses. 
The OPE is conveniently formulated in terms of Adler function, defined as
\begin{equation}
D(Q^2)=-q^2 \frac{d \Pi^1}{d q^2},
\end{equation}
which is regularization and scheme independent. 
The OPE in terms of Adler function is written as 
\begin{equation}
D(Q^2)=D^{(0)}(Q^2,\nu^2)+\frac{m_l^2}{Q^2} D^{(2)}(Q^2,\nu^2)+\sum_{n=2} \frac{c_{2n}}{Q^{2n}}.
\label{adler}
\end{equation}
$D^{(0)}$ is known to five-loop \cite{Chetyrkin:1979bj, Surguladze:1990tg, 
Gorishnii:1990vf, Baikov:2008jh}, 
\ie to order $\als^4$, $D^{(2)}$ is known only to two-loop, but the corresponding 
contribution being proportional to $m_l^2$ is numerically small. 
The last term in the above expression parametrizes the nonperturbative
contribution to the Adler function in terms of the local condensates. 
For sufficiently large $Q^2$ the nonperturbative contribution to $D(Q^2)$ is 
small, and comparing the Adler function obtained on the lattice to 
\Eqref{adler}, we can extract $\als(\nu)$ at some scale $\nu$ that is not too 
different from $Q$. 
A first analysis of this kind was performed by JLQCD/TWQCD for two flavors of 
light dynamical quarks~\cite{Shintani:2008ga} and then by the JLQCD collaboration 
in 2+1 flavor case~\cite{Shintani:2010ph}. 
While the perturbative calculation of the Alder function is available at high 
orders this method is very challenging because of the window problem. 
For small values of $a$ the contribution of the condensates is significant, 
while at large $Q^2$ we have to deal with large discretization effects. 
The determination in \mbox{Ref.}~\cite{Shintani:2010ph} resulted in a very 
low value of $\als$, $\als(M_Z)=0.1118(3)(+16)(-17)$. 
The systematic errors in this method have been carefully studied in 
\mbox{Ref.}~ \cite{Hudspith:2018bpz} involving the lead authors of the 
previous work. 
In particular, the values of $Q$ were limited to the one closest to 
the diagonal in order to minimize the discretization effects. 
This analysis resulted in $\als(M_Z)=0.1184(27)(+8)(-22)=0.1184(34)$. 
We added the errors in quadrature and symmetrized the final error in the 
last equation.
The experimental input for this study is the Omega baryon mass, 
which is sub-leading in the error budget.

\subsection{The strong coupling constant from QCD vertices}

A straightforward way to obtain the strong coupling constant is to calculate 
three- and four-gluon vertices or quark-gluon and ghost-gluon vertices in a 
fixed gauge on the lattice. 
The QCD coupling constant can be related to the renormalization constants
of these vertices. 
In practice one uses the Landau gauge and the renormalized coupling constant 
is defined in some intermediate MOM like scheme. 
For pioneering studies within this approach we refer the reader to 
\mbox{Refs.}~\cite{Alles:1996ka, Boucaud:2001qz}.
The ghost-gluon vertex offers an attractive possibility to extract $\als$ 
because it involves only the calculation of the ghost and gluon two-point 
functions~\cite{Boucaud:2008gn}. 
Furthermore, it is practical to use the MOM Taylor scheme in which
the renormalized running coupling constant is defined 
as~\cite{Blossier:2010ky}
\begin{equation}
\alpha_T(\nu)=\frac{g_T^2(\nu)}{4\pi}=Z_3(\nu^2,a^2) \tilde Z_3(\nu^2,a^2) \frac{g_0^2(a)}{4 \pi},
\end{equation}
where $Z_3(\nu^2,a^2)$ and $\tilde Z_3(\nu^2,a^2)$ are the gluon and ghost 
propagator renormalization constants and $g_0^2$ is the bare lattice gauge 
coupling. 
For a large value of $\nu$, the strong coupling constant in this scheme, $\alpha_T(\nu)$ can be written  as sum of
the perturbative result and a nonperturbative contribution due to the 
dimension-two gluon condensate $\langle A^2 \rangle$ 
in Landau gauge \cite{Blossier:2010ky}.
The running of $\alpha_T$ is known to four-loop \cite{Chetyrkin:2000dq}, 
and the relation between the $\Lambda$ parameter of the MOM Taylor scheme
and the $\MS$ scheme is also known~\cite{Chetyrkin:2000dq}. 
As many other methods this approach, too, suffers from the window problem: 
for small $\nu$ the nonperturbative contributions like the ones from the 
dimension-two gluon condensate are large, while for large $\nu$ 
discretization errors are important. 
Determinations of the strong coupling constant along these lines have been 
carried out by ETM collaboration for 2+1+1 
flavors using twisted mass fermions ~\cite{Blossier:2011tf, Blossier:2012ef, Blossier:2013ioa}. 
Discretization effects have been carefully studied in the above publications.
The latest ETM analysis gave 
$\als(M_Z)=0.1196(4)(8)(6)$~\cite{Blossier:2013ioa} 
or $0.1196(11)$ if all errors are combined in quadrature.
The experimental input for this study is the pion decay constant, 
which is sub-leading in the error budget. 
This result supersedes the previous ETM determinations. 
Very recently the determination of $\als$ using this approach but  
the DWF formulation and 2+1 flavors has been reported~\cite{Zafeiropoulos:2019flq}. 
This study finds $\als(M_Z)=0.1172(11)$ \cite{Zafeiropoulos:2019flq}.
The experimental input for this study is the Omega baryon mass, 
which is sub-leading in the error budget. 
This clearly deviates from the previous result by more than one sigma, 
possibly indicating that not all systematic errors in this method are 
under control. 
Taking the weighted average of the above result gives
$\als(M_Z)=0.1184$. 
We assign an error of $0.0012$ to this result to cover the spread in 
individual determinations.

\subsection{The strong coupling constant from eigenvalues of the Dirac operator}

One can also determine the strong coupling constant from the eigenvalues 
$\lambda$  of the Dirac operator.
For sufficiently large $\lambda$ values the density of eigenvalues can be calculated in perturbation theory
\begin{equation}
\rho(\lambda)=\frac{3}{4 \pi^3} \lambda^3 (1+\als \rho_1+\als^2 \rho_2+\als^3 \rho_3 + \ldots),
\end{equation}
with known coefficients $\rho_1, \rho_2$ and $\rho_3$ in 
the $\MS$ scheme~\cite{Chetyrkin:1994ex, Kneur:2015dda}. 
The eigenvalues of the Dirac operator using overlap fermions have been 
analyzed in \mbox{Ref.}~\cite{Nakayama:2018ubk}.
The analysis gives \cite{Nakayama:2018ubk}
\begin{equation}
\als(M_Z)=0.1226(36)\, .
\end{equation}
The experimental input for this study is the Omega baryon mass used in setting 
the gradient flow scale $\sqrt{t_0}$, which is sub-leading in the error budget.

\section{Other heavy quark mass determinations}
\label{sec:otherhqm}
In the sections 4 and 5 we discussed the quark mass determination using 
reduced moments of quarkonium correlators and the EFT approach. 
In the EFT approach the quark mass determination relies on HQET and
chiral perturbation theory to interpolate the hadron mass dependence on 
the quark masses. 
In other approaches one calculates the bare quark masses that correspond 
to physical values of a set of hadron masses and then performs the 
renormalization to obtain the quark masses in the $\MS$ scheme. 
The calculations of the reduced moments can be viewed as a special method
to calculate these renormalization constants. 
A more conventional approach to calculate the quark mass renormalization 
is to use some intermediate scheme like MOM or SF non-perturbatively on the 
lattice and then use the relation of these intermediate schemes to the $\MS$ 
scheme to finally obtain the $\MS$ scheme quark masses. 
In the following subsection we will briefly review the determination of 
the charm quark and bottom quark masses using this approach. 
As has already been pointed out, the determination of the bottom quark mass 
on the lattice is challenging because the bare bottom quark mass in lattice 
units is not small. 
In this case using some form of EFT approach and perturbation theory is 
necessary.
It is customary to quote the charm quark mass in the $\MS$ scheme at a scale 
equal to the charm quark mass, \ie $m_c=m_c(m_c)$ for four active 
flavors, see \eg \cite{PDG18}. 
Similarly, the bottom quark mass, $m_b$ is quoted at the scale of the 
bottom quark mass, $m_b=m_b(m_b)$ and five
active flavors. 
In what follows we will show the results for the charm and bottom quark 
masses using this convention unless stated otherwise.

\subsection{Charm quark mass determination}

In 2+1 flavor QCD there is a determination of the charm quark mass from 
the $\chi$QCD collaboration using overlap fermions in the valence sector and 
domain wall fermions for the sea quarks. 
They use two lattice spacings $a=0.087\,\mr{fm}$ and $a=0.11\,\mr{fm}$ and 
several light quark masses corresponding to a pion mass in the range of
$290\,\mr{MeV}$ to $420\,\mr{MeV}$. 
They obtain $1.304(5)(20)~\rm{GeV}$~\cite{Yang:2014sea}. 
We note, however, that using only two lattice spacings may be 
problematic for obtaining reliable continuum results for the charm quark mass. 
There are two charm quark mass determinations by the ETM collaboration using 
$N_f=2+1+1$ and twisted mass fermion 
action~\cite{Carrasco:2014cwa, Alexandrou:2014sha}.
In \mbox{Ref.}~\cite{Carrasco:2014cwa} the bare charm quark mass was 
determined using $D$ and $D_s$ meson masses resulting in 
$m_c=1.348(46)\,\mr{GeV}$, while in \mbox{Ref.}~\cite{Alexandrou:2014sha} 
the $\Omega_c^+$ was used to fix the bare charm quark mass leading to 
$m_c=1.3478(27)(195)\,\mr{GeV}$. 
The HPQCD collaboration used the same lattice setup as for the calculation 
of moments of quarkonium correlators to determine the renormalization factors
for the quark masses in RI-sMOM scheme~\cite{Lytle:2018evc}. 
Using these for the charm quark mass they obtain $m_c=1.2757(84)\,\mr{GeV}$. 
This value agrees well with the HPQCD determination using the moments of 
quarkonium correlators.

\subsection{Bottom quark mass determination}

To deal with the problem of the large bottom quark mass one can use NRQCD. 
The HPQCD collaboration calculated the $\Upsilon$ masses using NRQCD and 
$N_f=2+1$~\cite{Lee:2013mla}. 
The NRQCD mass parameter, $m_{b0}$, can be related to the $\MS$ bottom 
quark mass using three-loop perturbation theory~\cite{Lee:2013mla}. 
With this approach the HPQCD collaboration obtained $m_b=4.166(43)$ GeV using a 
single volume and pion mass of around $300\,\mr{MeV}$~\cite{Lee:2013mla}.
Using only one lattice volume and a single value of the pion mass that is 
larger than the physical value may be of some concern. 
However, we note that the $\Upsilon$ mass is largely insensitive to the pion mass 
and is not much affected by the volume.
One can also calculate the moments of bottomonium correlators also in 
NRQCD~\cite{Colquhoun:2014ica}.
Such calculations have been performed by the HPQCD collaboration using the HISQ
action for the sea quarks and $N_f=2+1+1$.
The ratio of the moments of bottomonium correlators together with the ratio 
of the lattice kinetic mass of bottomonium and the lattice NRQCD mass 
parameter, $m_{b0}$, can be related to the ratio of the experimental 
bottomonium mass and the $\MS$ b-quark mass. 
Using this relation the HPQCD obtained $m_b=4.196(23)$ GeV.

The ETM collaboration calculated the heavy-light and heavy-strange meson 
masses using the twisted mass action in (2+1+1)-flavor QCD~\cite{Bussone:2016iua}. 
Using HQET analysis these masses can be related to the pole mass in the static 
limit~\cite{Bussone:2016iua}, which then can be converted to the  $\MS$ 
b-quark mass. 
With this approach the ETM collaboration obtains
$m_b(\mu=m_b,N_f=4)=4.26(10)\,\mr{GeV}$. 
Finally Gambino et al used the same gauge configurations as the above ETM 
study and a very similar method to obtain $m_b$~\cite{Gambino:2017vkx}. 
The main difference in their analysis is that they relate the bottom quark
mass to the kinetic mass, and obtain 
$m_b(\mu=m_b,N_f=4)=4.26(18)$~\cite{Gambino:2017vkx}.

\section{Executive summary of the strong coupling 
constant and heavy quark mass determinations}
\label{sec:exec}

In this section we present an executive summary of the lattice 
determinations of the strong coupling constant and heavy quark masses. 

\begin{figure}[tb]
\begin{center}
\begin{minipage}[t]{16.5 cm}
\centering
\includegraphics[scale=1.0]{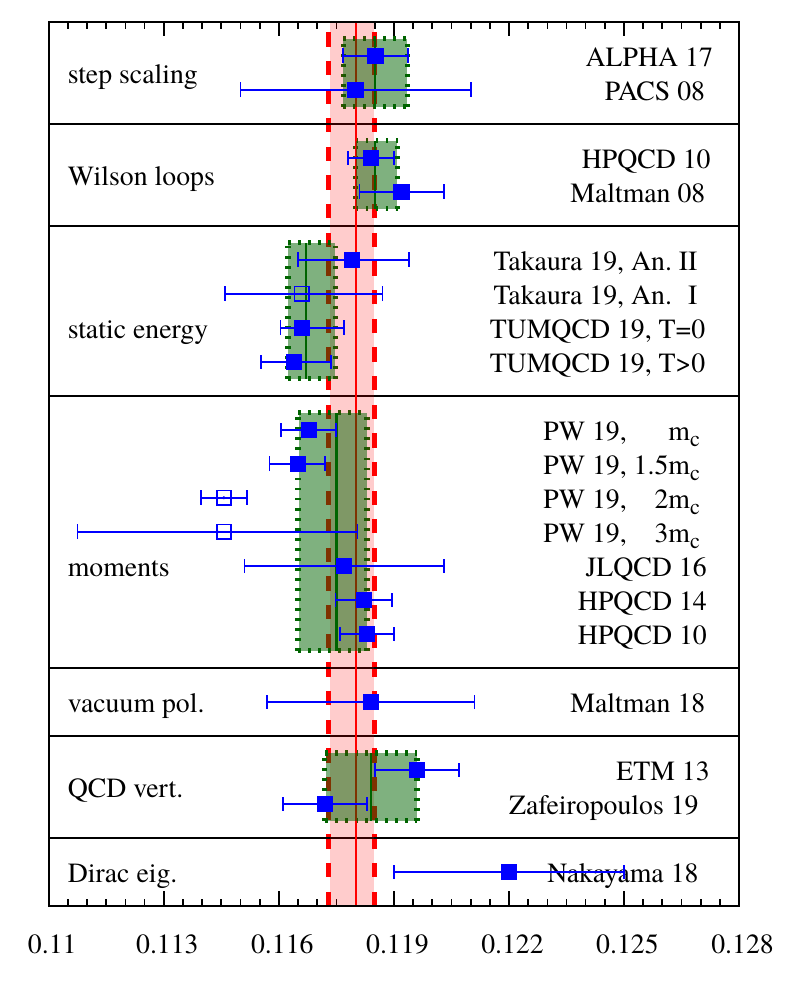}
\end{minipage}
\begin{minipage}[t]{16.5 cm}
\caption{
Summary of lattice determinations of $\als(M_Z)$ using different
methods. 
Green boxes show the pre-averages for the different observables. 
Open symbols do not enter the averaging procedure, see text.
\label{fig:exec_all_alphas}}
\end{minipage}
\end{center}
\end{figure}

We reviewed lattice determinations of $\als$ using different methods. 
For all the methods, with the exception of the Dirac operator eigenvalue 
approach, there are calculations performed by different groups. 
This allows to cross-check the error estimates within a given method. 
We performed pre-averages for each of the methods of $\als$ determination 
and assigned errors to these pre-averages to cover the spread in the 
individual determination within one method. 
The summary of different lattice determinations is shown in 
\Figref{fig:exec_all_alphas}.
The pre-averages are indicated by the green bands in the figure.
As one can see from the figure there is no large tension between different 
lattice determinations of $\als$. 
We note that different methods have quite different sources of systematic 
errors, so the lack of significant tension is non-trivial check of the 
lattice methods. 
As has been pointed out the major uncertainties are due to the perturbative 
errors and the continuum extrapolation for most of the methods. 
The step scaling method has control over these effects as large scales can 
be reached with controllable discretization errors. 
Because of the high energy scale the perturbative errors in this method are 
negligible. 
Since different lattice methods have different systematic errors it makes 
sense to perform a weighted average of $\als$ performed by different methods 
using the pre-averages. 
This procedure is similar to the one used by PDG and FLAG.

Performing the weighted averages of the pre-averages as well as the $\als$ 
determination from hadronic vacuum polarization and eigenvalues of the Dirac 
operators we obtain
\al{\label{exec:as_av}
 &\als(M_Z)=0.11803^{+0.00047}_{-0.00068},&&\cdf=6.54/6.
}
The combined lattice result has a smaller error than the step scaling method, 
$\als(M_Z)=0.11852(84)$, which can be considered as the most reliable.
On the one hand, if we perform the average without the static energy result, 
we obtain $\als(M_Z)=0.11838^{+0.00044}_{-0.00048}$, $\cdf=2.75/5$. 
Furthermore, if we perform the average without the eigenvalues of the Dirac 
operator, we obtain $\als(M_Z)=0.11802^{+0.00046}_{-0.00069}$, $\cdf=4.79/5$. 
Finally, if we perform the average without the step scaling result, we obtain 
$\als(M_Z)=0.11791^{+0.00054}_{-0.00074}$, $\cdf=6.39/5$, which agrees well 
with the other numbers. 
Thus, the combination of all lattice determinations other than the step 
scaling analysis leads to consistent and similarly accurate results. 
The consistency of different lattice averages confirms that the errors of
the lattice determinations are indeed realistic.

\begin{figure}[tb]
\begin{center}
\begin{minipage}[t]{16.5 cm}
\centering
\includegraphics[scale=1.0]{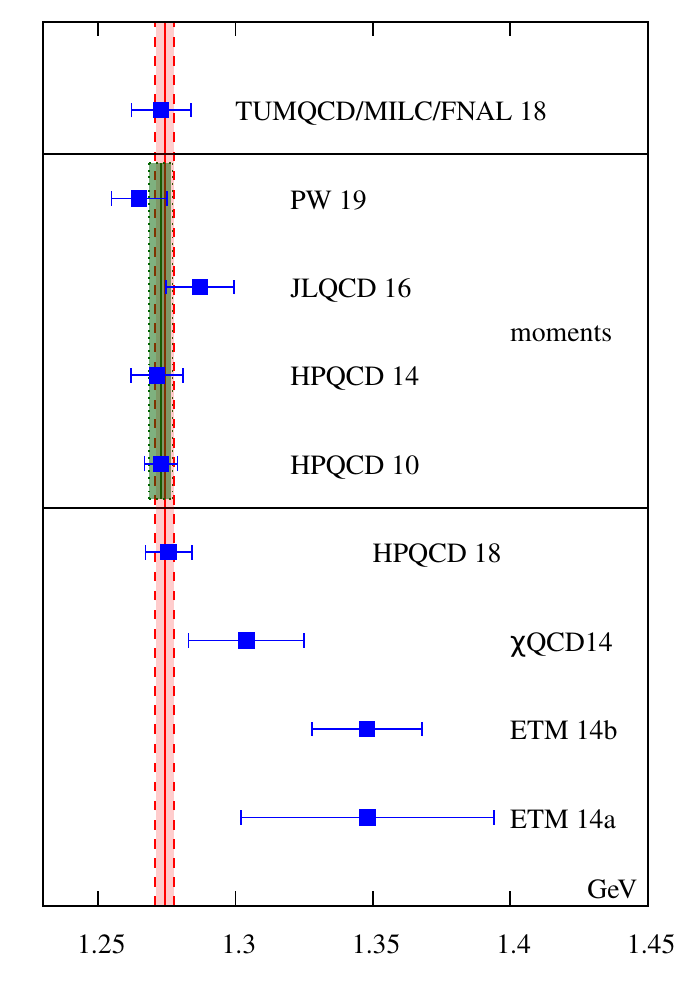}
\end{minipage}
\begin{minipage}[t]{16.5 cm}
\caption{The summary of charm quark mass determinations by different lattice groups.
The labels TUMQCD/MILC/FNAL~18, PW~19, JLQCD~16, HPQCD~14, HPQCD~10, HPQCD~18, $\chi$QCD~14,
ETM~14b, ETM~14a correspond to the results from Refs. \cite{Bazavov:2018omf}, \cite{Petreczky:2019ozv}, \cite{Nakayama:2016atf},
\cite{Chakraborty:2014aca}, \cite{McNeile:2010ji}, \cite{Lytle:2018evc}, \cite{Yang:2014sea}, \cite{Carrasco:2014cwa} and
\cite{Alexandrou:2014sha}, respectively. 
The vertical line and band correspond to the value of the charm quark mass 
given in \Eqref{exec:mc_av} and its uncertainty.
The green box shows the pre-average for the moments. 
\label{fig:exec_mc_sum}}
\end{minipage}
\end{center}
\end{figure}

There are several determinations of the charm quark mass. 
The determination of the charm quark mass first involves the determination
of the bare charm quark mass in the lattice scheme by matching the masses
of hadrons with charm quarks to the corresponding experimental values, and 
then performing the renormalization to obtain the charm quark mass in 
$\MS$ scheme.
Typically the renormalization involves the use of some intermediate scheme 
like RI-MOM.
The method of reduced moments can be considered as a special method of 
calculating the renormalization constant. 
The perturbative errors as well as the errors of the continuum extrapolation 
in the charm quark mass determination are mostly under control.
Perhaps the largest uncertainty is due to the determination of the lattice 
spacing.
In \Figref{fig:exec_mc_sum} we show the summary of the charm quark mass 
determinations on the lattice.
Different lattice determination of $m_c$ are consistent with each other, except 
for the ETM result.
There is a clear tension between the results from ETM collaboration and 
other lattice calculations.
There are several determinations of the charm quark mass using the 
reduced moments of quarkonium correlators. 
These determinations are grouped together in \Figref{fig:exec_mc_sum} and agree 
well with each other.
We performed a weighted average of the corresponding results, which gave 
$m_c^\mr{mom}=1.2729(42)\,\mr{GeV}$ and $\cdf=0.66$. 
Taking this value together with $m_c$ determinations from other methods
we calculated the weighted average of $m_c$. 
When results from ETM collaborations are included in the averaging we obtain 
$m_c=1.2768(64)$ and $\cdf=3.53$. 
The large value of $\cdf$ indicates the incompatibility of ETM results with 
other lattice determinations. 
If we perform the averaging excluding the ETM results we obtain
\begin{equation}
m_c(\mu=m_c,N_f=4)=1.2743(35), ~\cdf=0.72.
\label{exec:mc_av}
\end{equation}
The central value of $m_c$ did not change within errors but the resulting error is 
reduced almost by factor two.
We will use the above value as our final estimate for the charm quark mass.
The above value of the charm quark mass is considerably more precise than
the PDG value. This is partly due to the averaging procedure but also to some
quite precise individual lattice determinations. We note that 
the two most precise determination of $m_c$, the TUMQCD/MILC/FNAL determination
and the combined determination from the moment method agree very well with each
other. 
This suggests that the error in \Eqref{exec:mc_av} is realistic, since the two
methods have different systematic errors.

\begin{figure}[tb]
\begin{center}
\begin{minipage}[t]{16.5 cm}
\center
\includegraphics[scale=1.0]{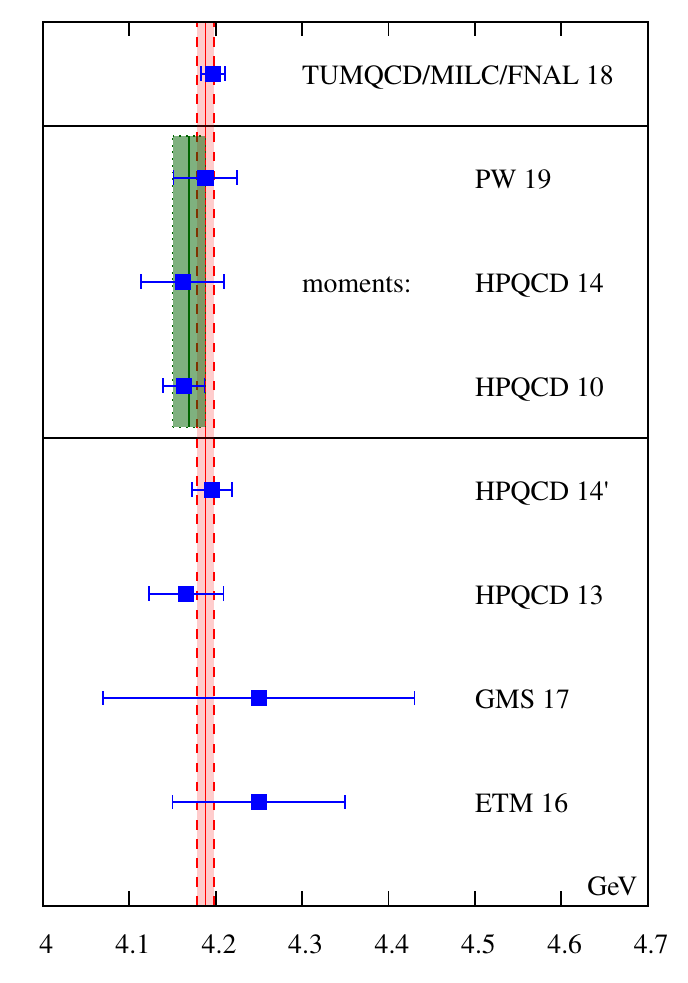}
\end{minipage}
\begin{minipage}[t]{16.5 cm}
\caption{
The summary of bottom quark mass determinations by different lattice groups.
The labels TUMQCD/MILC/FNAL~18, PW~19, HPQCD~14, HPQCD~10, HPQCD~14', HPQCD~13, 
GMS~17, ETM~16 refer to results from Refs.~\cite{Bazavov:2018omf}, 
\cite{Petreczky:2019ozv}, \cite{Chakraborty:2014aca}, \cite{McNeile:2010ji}, 
\cite{Colquhoun:2014ica}, \cite{Lee:2013mla}, \cite{Gambino:2017vkx} and 
\cite{Bussone:2016iua}, respectively.
The vertical line and band correspond to the value of the bottom quark mass 
given in \Eqref{exec:mb_av}) and its uncertainty.
The green box shows the pre-average for the moments. 
\label{fig:exec_mb_sum}
}
\end{minipage}
\end{center}
\end{figure}

The lattice determination of the bottom quark mass is challenging because
of the large discretization effects that are proportional to powers of $a m_b$.
In the moment method one has to use very fine lattices to ensure that the continuum
extrapolation is under control. The problem of the large bottom-quark mass causing
large discretization effects can be avoided by combining EFT and lattice
approaches as discussed in sections \ref{sec:heavylight} and \ref{sec:otherhqm}. The determination of the bottom
quark mass using such methods can be very precise.
The lattice determinations of the bottom quark mass is summarized 
in \Figref{fig:exec_mb_sum}.
The different lattice determinations agree well with each other. 
The bottom quark mass determination by 
TUMQCD/MILC/FNAL collaboration~\cite{Bazavov:2018omf} is the most precise one. 
The bottom quark mass in \mbox{Refs.}~\cite{Bussone:2016iua,Gambino:2017vkx} 
was given for four active flavors. 
We converted the results for the five flavor theory using perturbative 
decoupling at the value of $m_b(\mu=m_b)$ \cite{Chetyrkin:2000yt}.
As for the charm quarks we took a pre-average of the bottom quark mass 
determinations from the moments of quarkonium correlators which resulted 
in $m_b^\mr{mom}=4.169(19)\,\mr{GeV}$ and $\cdf=0.17$. 
Using this value together with other lattice determinations and performing 
the weighted average results in
\begin{equation}
m_b(\mu=m_b,N_f=5)=4.188(10), ~\cdf=0.46.
\label{exec:mb_av}
\end{equation}
This average value is shown in \Figref{fig:exec_mb_sum} as the vertical line with 
the corresponding error band. This value is more precise than the PDG value 
and the consistency of different approaches suggest that the error is indeed realistic.

\section{Conclusions}
\label{sec:concl}
In this review paper we discussed the lattice determinations of the 
strong coupling constant and the heavy quark masses. 
We emphasized the role of the EFT approach in the determinations
of these parameters. 
We explained how the use of combined lattice and EFT approaches helps
the determination of the quark masses and the strong coupling constant. 
The use of NRQCD and HQET is very important for the determination of 
the bottom quark masses. 
Combining lattice calculations with HMrAS$\chi$PT leads to the most 
accurate determination of strange, charm and bottom quark masses. 
We have found that with very few exceptions the lattice determinations 
are consistent with each other and therefore the lattice determination 
of the heavy quark masses and $\als$ can be considered as quite reliable. 
Our main results are given by Eqs.~\eqref{exec:as_av}, \eqref{exec:mc_av} and
\eqref{exec:mb_av}. 
These lattice results on the strong coupling constant, charm-quark mass and 
bottom-quark mass are considerably more precise than the ones obtained by PDG.

\section*{Acknowledgments}
JK would like to thank the Fermilab Lattice, MILC, and TUMQCD collaborations 
for such a fruitful collaboration. 
JHW would like to thank the TUMQCD collaboration for the productive work, and, 
in particular, Y. Kiyo for the useful discussions and the correspondence of the 
static energy results with domain-wall fermions. 
JHW was supported by the U.S. Department of Energy, Office of Science, Office 
of Nuclear Physics and Office of Advanced Scientific Computing Research within 
the framework of Scientific Discovery through Advance Computing (SciDAC) award 
Computing the Properties of Matter with Leadership Computing Resources. 
PP was supported by the U.S. Department of Energy under
Contract No. DE-SC0012704.

\appendix

\section{Coefficients of the force at $\mr{N^3LL}$}
\label{app:force}

In this appendix, we summarize the set of coefficients that appear 
in the various weak-coupling results that are used in this review.
\vskip1ex
For $\Nc$ being the number of colors, the $\mr{SU(\Nc)}$ color factors read
\al{
&\CF=\frac{\Nc^2-1}{2\Nc},&
& \CA=\Nc,&
& \TF=\frac{1}{2}.
}
\vskip1ex
The QCD beta function is
\al{
\frac{d\,\als(\nu)}{d\ln \nu}
 &=
\als\beta(\als)
  = -\frac{\als^2}{2\pi}\sum_{n=0}^\infty \left( \frac{\als}{4\pi} \right)^n \beta_n
  = -2\alpha_s\left[\beta_0\frac{\alpha_s}{4\pi}+\beta_1\left(\frac{\alpha_s}{4\pi}\right)^2+\cdots\right],
}
where the first three coefficients read~\cite{Herzog:2017ohr, Luthe:2017ttg}
\al{
\beta_0 & =  \frac{11}{3}\CA-\frac{4}{3}\TF \Nf,\\
\beta_1 & =  \frac{34}{3}\CA^2-\frac{20}{3} \CA \Nf \TF-4 \CF \Nf \TF,\\
\beta_2 & =  \frac{2857}{54}\CA^3+
   \left(-\frac{1415}{27}\CA^2-\frac{205}{9}\CA \CF+2\CF^2\right) \Nf
   \TF
   +\left(\frac{158}{27}\CA+\frac{44}{9}\CF\right)\Nf^2 \TF^2.
}
$\Nf$ is the number of active massless quarks. 
The higher-order coefficients \(\beta_3\), and \(\beta_4\) are also 
known~\cite{Herzog:2017ohr, Luthe:2017ttg}, but are not reproduced for 
brevity's sake. 

The singlet potential $V_s$ in \Eqref{eq:E3L} is given in terms of the coupling 
$\alpha_V(q^2)$, or in terms of the integral of the $\mr{N^3LO}$ force with 
or without the resummation of leading ultrasoft logarithms, \ie \Eqsref{eq:F3L} 
or \eqref{eq:FN3LL}, respectively.
In the following we collect the coefficients in these weak-coupling expressions.
We reproduce the coefficients \(\tilde{a}_i\) as given in \mbox{Ref.}~\cite{Tormo:2013tha} 
\al{ 
\tilde{a}_1 &= a_1+2\gamma_E\beta_0,
\\
\tilde{a}_2 &= a_2 +\left(\frac{\pi^2}{3}+4\gamma_E^2\right)\beta_0^2
  +\gamma_E\left(4a_1\beta_0+2\beta_1\right), \\
\tilde{a}_3 &= 
a_3+\Big(8\gamma_E^3+2\gamma_E\pi^2+16\zeta(3)\Big)
\beta_0^3+2\gamma_E\beta_2
\nn\\&
  +\Big[\left(12\gamma_E^2+\pi^2\right)\beta_0^2+4\gamma_E\beta_1\Big]
a_1
  +\left[6a_2\gamma_E+\frac{5}{2}\left(4\gamma_E^2+\frac{\pi^2}{3}\right)\beta_1\right]\beta_0, 
}
where the coefficients \(a_1\) and \(a_2\) 
read~\cite{Fischler:1977yf, Billoire:1979ih, Schroder:1998vy} 
\al{
a_1 &= \frac{31}{9}\CA-\frac{20}{9}\TF\Nf,
\\
a_2 &= \left(\frac{4343}{162}+4\pi^2-\frac{\pi^4}{4}+\frac{22}{3}\zeta(3)\right)\CA^2
\nn\\&
 -\left(\frac{1798}{81}+\frac{56}{3}\zeta(3)\right)\CA\TF\Nf
 -\left(\frac{55}{3}-16\zeta(3)\right)\CF\TF\Nf +\left(\frac{20}{9}\TF\Nf\right)^2. 
} 
The coefficient \(a_3\) reads~\cite{Anzai:2009tm, Smirnov:2009fh} 
\al{ 
a_3 &=  a_3^{(3)}\Nf^3+a_3^{(2)}\Nf^2+a_3^{(1)}\Nf+a_3^{(0)},\\
a_3^{(3)} &= - \left(\frac{20}{9}\right)^3 \TF^3
\nn\\
a_3^{(2)} &=
\left(\frac{12541}{243}
  + \frac{368}{3}\zeta(3)
  + \frac{64\pi^4}{135}
\right) \CA \TF^2
+
\left(\frac{14002}{81}
  - \frac{416}{3}\zeta(3)
\right) \CF \TF^2
\\
a_3^{(1)} &=
\left(-709.717
\right) \CA^2 \TF
 +
\left(-\frac{71281}{162}
  + 264 \zeta(3)
  + 80 \zeta(5)
\right) \CA\CF \TF
\nn\\&
 +
\left(\frac{286}{9}
  + \frac{296}{3}\zeta(3)
  - 160\zeta(5)
\right) \CF^2 \TF
+
\left(-56.83(1)
\right) 
\frac{18-6\Nc^2+\Nc^4}{96\Nc^2}
\\
a_3^{(0)} & =  502.24(1) \CA^3 -136.39(12)
\frac{\Nc^3+6\Nc}{48}
+
\frac{8}{3}\pi^2\CA^3\left(-\frac{5}{3}+2\gamma_E+2\log2\right).\label{eq:a30}
}
The three numerical values given above are nowadays known analytically~\cite{Lee:2016cgz, Lee:2016lvq}. 
Lastly, we reproduce the coefficient \(a_{3}^L\) as given in \mbox{Ref.}~\cite{Brambilla:1999qa}, 
which multiplies the leading finite term of the perturbative piece of the ultrasoft contribution in 
\Eqref{eq:E3L},
\al{
 a_3^L & = \frac{16\pi^2}{3}\CA^3.
}


\section{One-loop chiral corrections to heavy-light mesons masses}
\label{sec:app:heavylight}
In HMrAS\chpt, the mass of $H_x$ meson is described by Eq.~(4.2) of Ref.~\cite{Brambilla:2017mrs}
\begin{align}
    M_{H_x}(\{m_h,m_x\},\{m'_l,m'_l,m'_s\},\{a\}) &= m_{h,\MRS} + \Lambdabar_\MRS + \frac{\mu_\pi^2
        - \mu_G^2(m_h)}{2m_{h,\MRS}} + 2\lambda_1 B_0 m_x \label{eq:EFT:app:M_Hx} \\ &
   + 2\lambda'_1 B_0(2m'_l + m'_s) + \delta M_{H_x}(\{m_h,m_x\},\{m'_l,m'_l,m'_s\},\{a,V\}) - \mathcal{C} ,   \nonumber
\end{align}
where $B_0$ is the low energy constant (LEC) in the relation $m^2_\pi = B_0(m_u+m_d)$ between the pion mass and light quark masses,
$\lambda_1$ and $\lambda'_1$ are LECs independent of the heavy quark mass,
$\delta M_{H_x}$ is the one-loop corrections to the mass of the $H_x$ meson,
and $\mathcal{C}$ is a constant elaborated below.
The arguments of $M_{H_x}$ and $\delta M_{H_x}$ correspond to the heavy and light valence-quark mass,
the set of three light sea-quark masses, and the lattice spacing~$a$.
$\delta M_{H_x}$ contains effects of flavor, hyperfine, and order-$a^2$ (staggered) taste splittings
as well as finite lattice size.
For a theory with $(2+1)$ light flavors in the sea, we have
\begin{align}
    \delta M_{H_x} &= - \frac{3g_\pi^2}{16\pi^2f^2} \Biggl\{
        \frac{1}{16}\sum_{\mathscr{S},\Xi} K_1(m_{\mathscr{S}x_{\Xi}},\Delta^* + \delta_{\mathscr{S}x})
    \label{eq:chiral-form:2p1} \\
        & \qquad + \frac{1}{3}\sum_{j\in \mathcal{M}_I^{(2,x)}} \frac{\partial}{\partial m^2_{X_I}}
            \left[R^{[2,2]}_{j}(\mathcal{M}_I^{(2,x)};     \mu^{(2)}_I) K_1(m_{j},   \Delta^*) \right]
    \nonumber \\
        & \qquad + \biggl( a^2\delta'_V \sum_{j\in \hat{\mathcal{M}}_V^{(3,x)}} \frac{\partial}{\partial m^2_{X_V}}
            \left[R^{[3,2]}_{j}(\hat{\mathcal{M}}_V^{(3,x)}; \mu^{(2)}_V) K_1(m_{j},   \Delta^*) \right] + [V\to A]\biggr)  
            \Biggr\}
    \nonumber \\
        & \qquad +  a^2 \frac{3g_\pi^2}{16\pi^2f^2}
        \left[ \lambda'_{a^2} \bar{\Delta} \sum_\mathscr{S}\delta_{\mathscr{S}x} 
        + \lambda_{a^2} \Delta^* \left( 3\bar{\Delta} -\frac{1}{3}\Delta_I + \delta'_V + \delta'_A\right)\right].
    \nonumber
\end{align}
For the full description of each term one should see Refs.~\cite{Brambilla:2017mrs,Bazavov:2018omf}.
Here we give a brief description of the main indices and terms.
$g_\pi$ is the $H$-$H^*$-$\pi$ coupling,
and $\Delta^*$ is the lowest-order hyperfine splitting between pseudoscalar and vector mesons.
The indices $\mathscr{S}$ and $\Xi$ denote the light sea-quark flavors and meson tastes, respectively.
There are 16 mesonic tastes labeled with $\{I,V,T,A,P\}$ corresponding to multiplets of
$(1,4,6,4,1)$, respectively~\cite{Kronfeld:2007ek}.
$m_{\mathscr{S}x,\Xi}$ is the mass of the pseudoscalar meson with taste $\Xi$ and flavors $\mathscr{S}$ and $x$;
at the tree level we have~[\cite{Aubin:2003mg},~Eq.~(18)]
\begin{equation}
    m^2_{\mathscr{S}x,\Xi} = B_0 (m_\mathscr{S} + m_x) + a^2 \Delta_\Xi\, ,
    \label{eq:pion-mass-sq:Xi}
\end{equation}
where $\Delta_P=0$.
$\delta_{\mathscr{S}x}$ is the flavor splitting between a heavy-light meson with light quark of flavor
$\mathscr{S}$ and one of flavor $x$.
$\delta'_A$ and $\delta'_V$ are the taste-breaking hairpin parameters.
$\lambda_{a^2}$ and $\lambda'_{a^2}$ are parameters in S\chpt\, related to taste breaking in meson masses,
and
\begin{align}
 a^2 \bar\Delta &= \frac{1}{16} \sum_\Xi \left(m^2_{\mathscr{S}x,\Xi} - m^2_{\mathscr{S}x,P}\right) , \\
 a^2 \Delta_I   &= m^2_{\mathscr{S}x,I} - m^2_{\mathscr{S}x,P} \, .
\end{align}
Definitions of the chiral-log function $K_1$ at infinite and finite volumes,
the residue functions $R_j^{[n,k]}$, and the sets of masses in the residues
are given in Ref.~\cite{Brambilla:2017mrs} and references therein.

In Eq.~\eqref{eq:EFT:app:M_Hx}, $\mathcal{C}$ is a constant
that can be set to
\begin{equation}
    \mathcal{C} = 2\lambda_1 B_0 m_l + 2\lambda'_1 B_0(2m_l + m_s) + \delta M_{H_x}(\{m_h,m_l\}, \{m_l,m_l,m_s\}, \{0\})
    \label{eq:EFT:fit-function:C}
\end{equation}
so that in the continuum limit, for physical values of see-quark masses,
and $m_x=m_l$ one obtains
\begin{equation}
    M_{H_x}(\{m_h,m_l\},\{m_l,m_l,m_s\}, \{0\}) = m_{h,\MRS} + \Lambdabar_\MRS + \frac{\mu_\pi^2- \mu_G^2(m_h)}{2m_{h,\MRS}}\, .
\end{equation}

\bibliographystyle{unsrt}
\bibliography{ref}

\end{document}